\documentclass[11pt,a4paper]{article}
\pdfoutput=1
\usepackage{jheppub}
\usepackage[T1]{fontenc}
\usepackage[english]{babel}
\usepackage{amsmath,amssymb,stmaryrd}
\usepackage{amsthm,amsfonts,dsfont}
\usepackage[dvipsnames]{xcolor}
\usepackage[normalem]{ulem}
\usepackage{tikz}
\usepackage[mathscr]{euscript}
\usepackage{mathrsfs,hyperref,mathtools,graphicx,enumerate}
\usepackage{hhline}
\usepackage[shortlabels]{enumitem}
\usepackage{caption, multirow, makecell}
\usepackage{siunitx}
\usepackage{mhchem}
\usepackage{float}
\usepackage{adjustbox}

\DeclareMathAlphabet{\mathpzc}{OT1}{pzc}{m}{it}

\usepackage{enumitem}
\setlist[enumerate,1]{ref=\theenumv}

\usepackage{cancel}

\usepackage{bbold} 
\DeclareMathAlphabet{\mathpzc}{OT1}{pzc}{m}{it}

\usepackage{hyperref}
\hypersetup{
    colorlinks=true,
    linkcolor=blue,
    filecolor=magenta,      
    urlcolor=cyan} 

\numberwithin{equation}{section}								

\begin{document}
\begin{flushright}
	MPP-2023-275\\
    LMU-ASC 36/23
\end{flushright}
	$$\quad  $$
	{\LARGE\center \bfseries Topology change and non-geometry\\ at infinite distance  \par } 
	\vspace{10pt}
	\begin{center}
	{\center Saskia Demulder$^a$,\hspace{2pt} Dieter L\"ust$^{b,c}$ and Thomas Raml$^b$}
	\end{center}
	\begin{center}
	{\small
 	\textit{${}^a$Ben Gurion University of the Negev,\\ David Ben Gurion Blvd 1,  84105 Be'er Sheva, Israel}\\ %7Q6X+HR
  \vspace{8pt}
		\textit{${}^b$Max-Planck-Institut f\"ur Physik (Werner-Heisenberg-Institut),\\ Boltzmannstr. 8, 85748 Garching, Germany}	\\
			\vspace{8pt}
   \textit{${}^c$Arnold-Sommerfeld-Center for Theoretical Physics,\\
   Ludwig-Maximilian-Universit\"at, 80333 M\"unchen, Germany}	\\
			\vspace{10pt}
	E-mail: \texttt{saskia.demulder@gmail.com, luest@mpp.mpg.de, raml@mpp.mpg.de}}
		\end{center}
	\vspace{15pt}
	\noindent

 \begin{changemargin}{1.5cm}{1.5cm} 
	{\sc  Abstract.}  The distance conjecture diagnoses viable low-energy effective realisations of consistent theories of quantum gravity by examining their breakdown at infinite distance in their parameter space. At the same time, infinite distance points in parameter space are naturally intertwined with string dualities.  
 We explore the implications of the distance conjecture when T-duality is applied to curved compact manifolds and in presence of (non-)geometric fluxes. We provide evidence of how divergent potentials signal pathological infinite distance points in the scalar field space where towers of light states cannot be sustained by the curved background. This leads us to  suggest an extension to the current statement of the Swampland distance conjecture in curved spaces or in presence of non-trivial fluxes supporting the background. \\ 
\end{changemargin}

\newpage
\tableofcontents

\newpage
%%%%%%%%%%%%%%%%%%%%%%%%%%%%%%%%%%%%%%%%%%%%%%%%%%%%%
%%%%%%%%%%%%%%       Introduction      %%%%%%%%%%%%%%
%%%%%%%%%%%%%%%%%%%%%%%%%%%%%%%%%%%%%%%%%%%%%%%%%%%%%
\section{Introduction}
The Swampland program \cite{Vafa:2005ui} sets out to find a sufficiently exhaustive list of conditions, called conjectures, to rule out effective field theories which cannot arise from a low-energy realisation of a consistent theory of quantum gravity. Theories which do not go past one of these tests are declared to be in the `Swampland', while theories which cannot be ruled out are left in the purgatory of the `Landscape', potentially lifting to a valid formulation of quantum gravity in the UV. In recent years, the Swampland program has let to a growing set of tightly interwoven conjectures. The different  conjectures are born out of a mix of principles we believe should be realised in quantum gravity together with a thorough understanding of string theoretical examples which hints us towards defining properties of a theory of quantum gravity. By now multiple pedagogical reviews are available covering the different Swampland conjectures \cite{Palti:2019pca,Grana:2021zvf,vanBeest:2021lhn,Agmon:2022thq}.

Amongst these conjectures, the Swampland Distance Conjecture (SDC)  \cite{Ooguri:2006in} describes the phenomenology of deforming the metric of the internal manifold and other fields of the effective theory.
This particular conjecture predicts that in a quantum theory of gravity, a large excursion in the expectation value of a massless scalar or modulus of the theory and its fields inevitably leads to an exponentially light tower of states.
At  heart, the SDC is a claim about either (de-)compactification or (de-)coupling limits: taking the radius of a compact dimension or a coupling constant to infinity or zero. It comes then as no surprise that the SDC is intimately intertwined with string dualities: T- or S-duality. Applying a duality, the original theory and its dual share the same moduli space. Looking at \linebreak{T-duality} and from  a lower dimensional point of view (that is when the internal geometry is compactified), the different backgrounds lie on a T-duality orbit and describe the same physics.   In its mildest incarnation, T-duality of the free boson on a compact circle, the duality inverts the compactification radius while simultaneously exchanging winding and momentum modes. This is in perfect harmony with the expectations set by the distance conjecture. The two infinite points of the moduli space, here the radius, are related by duality, which consistently swaps the zero modes furnishing the tower of light states.

In the present paper we would like to scrutinise the SDC when T-dualities are applied to curved internal manifolds or geometries supported by fluxes. Introducing curvature and fluxes leads to a collection of new phenomena that shines a new light on the relation between T-duality and the SDC.  We will consider in particular the topology changing properties of \linebreak{T-duality} and the appearance of non-geometry.  In both cases, the spacetime and background fields of dual models may wildly differ. Indeed, T-duality can be applied to non-trivial circle-fibration, generically inducing a non-trivial topology change in the dual background \cite{Giveon:1993ph,Alvarez:1993qi}. The latter phenomenon is also known to take place under the closely related mirror symmetry in Calabi-Yaus \cite{Witten:1993yc,Aspinwall:1993yb}. 

\begin{figure}[t]
    \centering
    \includegraphics[scale=0.24]{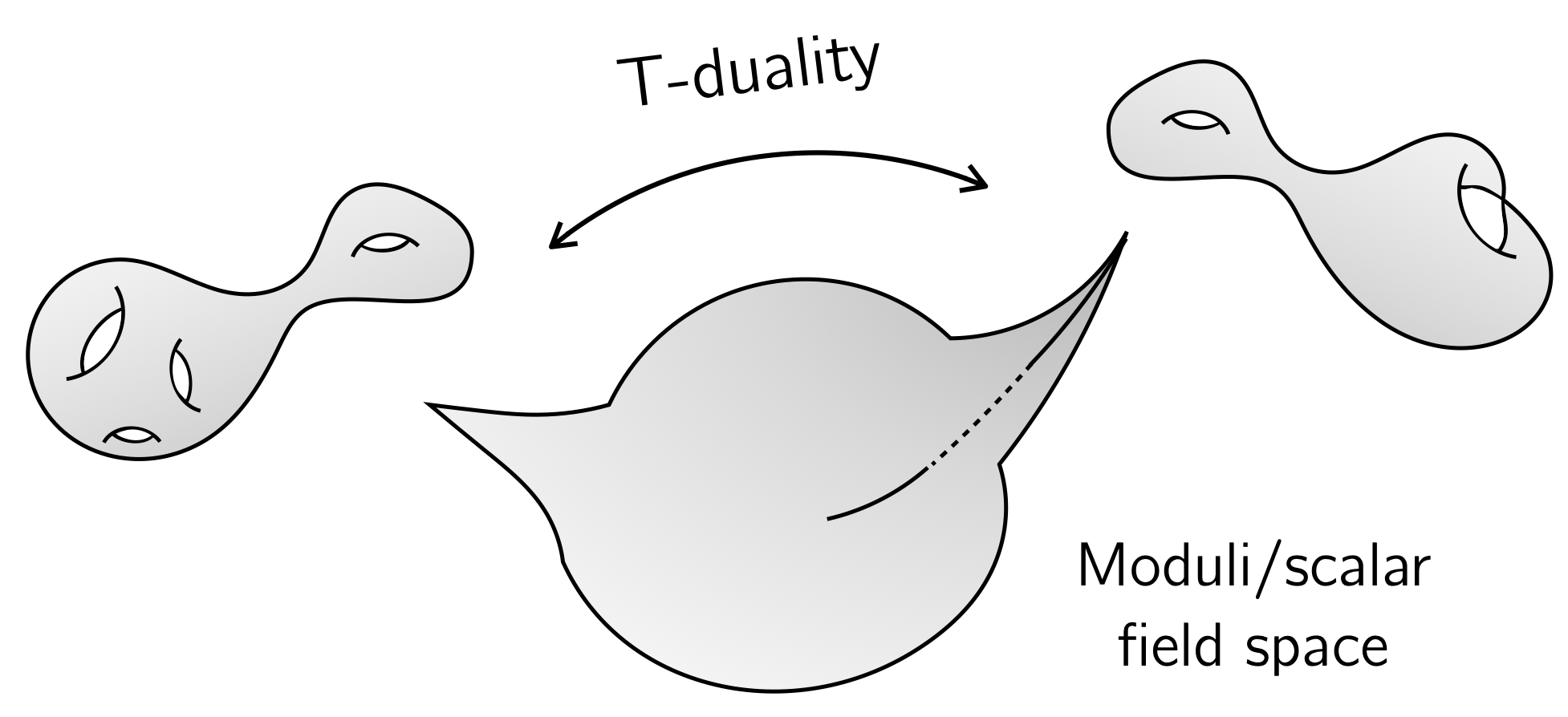}
    \caption{When considering curved manifolds, T-duality generates an array of distinct phenomena: change of topology including the trivialisation of homology cycles or appearance of torsion-full cycles and non-geometry with its associated fluxes. }
    \label{fig:T-duals}
\end{figure}

At the level of the zero-modes accounting for the tower of infinite states this raises the question: What happens when topology precludes for the existence (or alternatively entails the instability) of zero-modes which should, by living at infinite distance in the scalar field space, furnish the infinite light tower? This idea is illustrated in figure \ref{fig:T-duals}.  To reconcile the absence of certain modes and the existence of infinite distance in the scalar field space, we will have to require for an additional refinement of the SDC.  We will see  how a self-protecting mechanism arising by the inclusion of a non-trivial potential enters the stage, signalling  the problematic infinite distance points.

%%%%%%%%%%%%%%%%%%%%%%%%%%%%%%%%%%%%%%%%%%%
%%%%%%%%%%%%%%%%%%%%%%%%%%%%%%%%%%%%%%%%%%%
\subsection{Summary of the results}\label{sec:summary_results}
Our main goal is to explore the relationship between the distance conjecture and (generalised) T-duality for geometries with fluxes and non-zero curvature. We consider a number of prototypical examples of backgrounds featuring a large array of exotic behaviours T-duality is known to generate: topology change, exotic zero-mode exchange, and non-geometric fluxes. For the sake of simplicity we will restrict the discussion to backgrounds characterised by (essentially) one-dimensional scalar field spaces. In the eye of collecting sufficiently many elements to be in the position of addressing the distance conjecture and simultaneously keeping computational control, none of the examples consider here have the ambition to be a realistic candidate of a full-fledged theory of quantum gravity. Instead, we will mainly restrict our attention to the internal part of backgrounds and we will restrict the discussion to towers arising from winding or momentum modes. The examples considered here have to be considered as lab-experiments enabling us to garner clues and intuition behind the interplay between the SDC and T-duality in generality. 

\noindent
The two main results of this paper can be summarised as follows
\begin{enumerate}[(i)]
    \item Curved background generically display complicated topological and  symmetry properties which oftentimes render these spaces unable to host either winding or momentum modes. This is obviously at odds with the expectations originating from the SDC, as zero modes like the string winding and momentum modes usually lie at the origin of the tower of infinitely light states. On the other hand, to be part of a supergravity solutions, curved backgrounds generically necessitate supporting fluxes. Ricci curvature and fluxes will generically generate a non-trivial potential on the scalar field space. Exploring the behaviour in several exemplifying cases leads us to resolve the aforementioned problem by proposing the following modification to the SDC:

    \noindent
    \textit{In effective field theories that can be consistently lifted to a theory of quantum gravity in the UV, a divergence in the scalar potential emerges when approaching an infinite locus point for which the target space geometry cannot give rise to a light tower of states. That is, the potential signals pathological infinite distance loci in the scalar field space.}

    \item Backgrounds supported by non-trivial three-form flux naturally lead, upon T-dualising, to non-geometric spaces.  We consider non-geometric backgrounds within the distance conjecture and their consistency across different duality frames. We will see when considering the three-sphere and its generalised T-dual that a consistent picture only materialises through a number of delicate cancellations which can only be taken into account by moving to an alternative formulation of supergravity, known as $\beta$-supergravity \cite{Andriot12,Andriot:2013xca}, in which frame the geometry is smooth. 
\end{enumerate}
This paper is structured as follows.
In section \ref{sec:T-duality_non_geometry}, we review several aspects of T-duality and non-geometry. The aim is to highlight certain features which will appear to challenge the currents form of the SDC. Having set the stage, the remaining sections turn towards examples treating systematically these issues and study their possible resolutions with the predictions of the SDC.
In section \ref{sec:S3_H-flux}, we first investigate the simplest example of a \linebreak{T-duality} in a curved space: the thee-sphere  and associated Lens spaces. In section \ref{sec:NATD_S3}, we will consider the non-Abelian T-dual of the three-sphere and have a first encounter with a T-dual which is globally non-geometric. In section \ref{sec:T-duality_chain}, we tackle the so-called T-duality chain, which forms the basic example of how non-geometric backgrounds can be generated from successive applications of T-duality. Finally, in section \ref{sec:discussion}, we will conclude with future directions suggested by the results presented here.

%%%%%%%%%%%%%%%%%%%%%%%%%%%%%%%%%%%%%%%%%%%
%%%%%%%%%%%%%%%%%%%%%%%%%%%%%%%%%%%%%%%%%%%
%%%%%%%%%%%%%%%%%%%%%%%%%%%%%%%%%%%%%%%%%%%
\section{T-duality and the distance conjecture: setting the stage}\label{sec:T-duality_non_geometry}
The SDC specifies how, when taking large excursion in their moduli space, the low-energy effective actions arising from a theory of quantum gravity break down. Approaching these infinite distance point in moduli space, new light states with masses that become exponentially light will appear which were not part of the effective field theory and will lower the cut-off of the EFT, eventually invalidating its low energy effective description.

The SDC not only narrows down the characteristics of a would-be low-energy realisation of a quantum gravity, but it is also closely related to string dualities. Common-wisdom build from known examples  and solidified by the closely related emergent string conjecture \cite{Lee:2019wij} indicates that infinite points in moduli space are closely associated to string dualities. Indeed, at infinite points of the moduli space coupling constants vanish/blow up or generate a compactification/decompactification limit of the geometry. While the former is related to a strong/weak duality, the second is related to T-duality. In each case, new modes have to be included: for the latter KK-modes and in the former non-perturbative modes. Inclusion of those modes dramatically alters the infrared description at these infinite distance points. Infinite distance points in moduli spaces of field theories which cannot be coupled consistently to gravity are not expected to display any of these particularities.

In this section we will first review the Swampland Distance Conjecture. We will subsequently revisit the canonical example of T-duality of a single free boson which exemplifies the relation between the SDC and T-duality. Although beautiful in its simplicity this example fails to tackle many of the more challenging features of T-duality once more complicated backgrounds are considered. The rest of the section will then cover these precise properties of T-duality (topology change, zero mode exchange, non-geometric backgrounds and fluxes) which will come to confront the current understanding of the relationship between the SDC and T-duality.

%%%%%%%%%%%%%%%%%%%%%%%%%%%%%%%%%%%%%%%%%%%
%%%%%%%%%%%%%%%%%%%%%%%%%%%%%%%%%%%%%%%%%%%
\subsection{The SDC and dualities}
Consider a $D$-dimensional spacetime $M_D=M_d\times K_n$ where $K_n$ is an $n$-dimensional compact subspace of the total space $M_D$ and $M_d$ is the $d$-dimensional external space which completes the geometry. We will denote by $G$ the metric of the total space $\mathcal M$, and $g,h$ for the metrics of the external and internal manifolds, respectively. After compactifying, the geometry on the compact space $K_n$, one obtains an effective field theory coupled to Einstein gravity
\begin{align}\label{eq:DC_EH_action_no_pot}
       S_\mathrm{EH}=\frac{M_p^{D-2}}{2} \int \mathrm d^Dx \sqrt{-g}\left( \mathcal R(g)-\gamma_{ij}\partial_\mu\phi^i\partial^\mu\phi^j - V(\phi^i)\right)\,,
\end{align}
and where $V(\phi^i)$ accounts for a possible scalar potential.
The $\phi^i$ are scalar fields pinning down the shape, couplings and other parameters of the compactified part of the geometry. In absence of a scalar potential, these massless scalar fields span a space called the moduli space\footnote{ We will however, in abuse of the nomenclature, sometimes refer to massive and massless scalar fields equally as ``moduli'' or ``scalar fields'', interchangeably. When important, the massless or massive characters of the fields will be highlighted.}. The matrix $\gamma_{ij}$ controls their kinetic term and will soon be interpreted as the metric on  moduli space.

Denote by $\mathcal M$ the moduli space parametrised by the massless scalar fields. The Swampland distance conjecture then states that \cite{Ooguri:2006in} 
\begin{itemize}
	\item[1.] The moduli space  $\mathcal M$ is non-compact. This moduli space is parametrised by the expectation values of some field $\phi^i$, the moduli, which are not subjected to a potential. There are always two points at infinite distance from each other. 
\end{itemize} 
Distances are measured by means of geodesic paths and with respect to the metric $\gamma_{ij}$ as appearing in the kinetic term for the scalar fields $\phi_i$, i.e.
\begin{align}\label{eq:metric_moduli_space}
    \Delta \phi=\int \mathrm d t \sqrt{\gamma_{ij}\partial_t \phi^i \partial_t \phi^j}\,.
\end{align}
 Having postulated the existence of at least two infinitely separated points in moduli space, a central part of the conjecture is to characterise what happens when one tries to approach either of them:
 \begin{itemize}
	\item[2.] When approaching an infinite distance point, there is an infinite tower of states that becomes exponentially light as follows 
	\begin{align*}
		m(Q)\sim m(P)e^{-\lambda \Delta\phi}\quad \text{when}\quad\Delta\phi\rightarrow\infty\,,
	\end{align*}
	where $\lambda$ is a constant, $m$ denotes the mass of the corresponding states as a function of the points $P,Q\in \mathcal M$ on the moduli space and $\Delta\phi\equiv d(P,Q)$ is the geodesic field distance.
\end{itemize}
See the reviews \cite{Palti:2019pca,Grana:2021zvf,vanBeest:2021lhn,Agmon:2022thq} for additional details and relation to other Swampland conjectures.

As mentioned earlier, the SDC espouses a close connection to dualities in quantum gravity. Two dual theories, being an equivalent representation of the same theory, must have equivalent moduli spaces, featuring the same number of infinite distances, albeit reshuffled  at dual values of the moduli. Under a duality transformation,  the corresponding tower of light state is mapped to a dual tower which becomes exponentially light when approaching the dual distance point. This observation has the deep implication that studying the phenomena arising at infinite distance points as predicted by the distance conjecture and our understanding of dualities in quantum gravity are intimately intertwined. This phenomenon is succinctly demonstrated by compactification on a circle, which we will review in the following subsection, but before we need a slight generalisation of the original SDC.

The statement of the SDC given above applies for moduli, i.e. massless scalar fields underpinning the geometry. Generically however the parameters of the theory will be subjected to scalar potentials. These scalar potentials are the result of the presence of non-trivial fluxes and/or curvature in the original ten-dimensional spacetime. Upon compactication these will source scalar potential for the scalar fields of the compact space. This will be reviewed in detail in section \ref{sec:reduction}. In presence of a potential, the scalar fields become massive and their ``motion'' within field space will be subjected to the particular shape of that potential. The inclusion of a non-vanishing scalar potential was first considered in \cite{Baume:2016psm}, see also \cite{Klaewer:2016kiy,Valenzuela:2016yny,Blumenhagen:2017cxt}. There, it was established that when the scalar parameters are subjected to a potential, the conjecture remains true when now formulated on the field space of the effective theory rather than a moduli space. See also \cite{Calderon-Infante:2020dhm,Freigang:2023ogu} for other works investigating the implications of including a scalar potential in the context of the SDC.

%%%%%%%%%%%%%%%%%%%%%%%%%%%%%%%%%%%%%%%%%%%%%%%%%%%%%
\subsubsection{A simple  example: circle compactification}\label{sec:cpt_free_boson}
In this section, we will revisit the relation of T-duality in the free boson system and the SDC.
Take a $D=(d+1)$-dimensional space time $M_D=M_d\times S^1$ where we denote the coordinate along the compact circle $S^1$ by $y$. We then consider the effective field theory that results from compactifying on that circle down to $d$ dimensions. In the Einstein frame, the $D$-dimensional metric takes on the form 
\begin{align}\label{eq:metric_Ansatz_circle}
ds^2&=G_{IJ}\mathrm{d}X^I\mathrm{d}X^J=R^{\frac{-2}{d-2}}g_{\mu \nu}(x) \mathrm{d}x^\mu \mathrm{d}x^\nu +  R^{2} \mathrm{d} y^i \mathrm{d} y^j\,, 
\end{align}
where $g_{\mu\nu}(x)$ is the external space metric and $x$ collectively stands for the $d$ spacetime coordinates parametrising the external space $M_d$. The scalar factor in front of the external metric has been introduced for later convenience.
The radius $R$ of the circle is a dynamical field\footnote{As pointed out in the recent \cite{Li:2023gtt}, promoting a parameter to a dynamical field is not an innocuous step, as the background will no longer verify the equation of motion, here the supergravity equations. We will return to this point in section \ref{sec:promote_dyn} in context of the T-dual of the three-sphere with $H$-flux presented in section \ref{sec:S3_H-flux}.} in $d$ dimension and forms the moduli space of the effective field theory. Performing a reduction on the compact circle $S^1$ parametrised by $y$ (see e.g. relevant discussion in the recent reviews \cite{Palti:2019pca,Grana:2021zvf,vanBeest:2021lhn,Agmon:2022thq}), 
\begin{align*}
       S_\mathrm{EH}=\frac{M_p^{d-2}}{2} \int \mathrm d^dx \sqrt{-g}\left( \mathcal R(g)-\frac{d-1}{d-2}\frac{1}{R^2}(\partial R)^2 \right)\,.
\end{align*}
The general reduction procedure will be reviewed in detail in section \ref{sec:reduction_potential}.

Reading off the canonically normalised kinetic term yields a distance metric \eqref{eq:metric_moduli_space} on the modulus space
\begin{align*}
    \Delta R=\sqrt{2}\sqrt{\frac{d-1}{d-2}}\log R\,.
\end{align*}
It follows that the modulus space features two infinite distances: $R=0$ and $R\rightarrow \infty$.  From the SDC we expect the appearance of towers of exponentially light states when approaching either of these points. Since these points are related to compactification and decompactification limits, the towers will come from the winding zero-modes and Kaluza-Klein zero-modes of the string, respectively. The bosonic mode expansion for the free boson on a compact circle yields the familiar masses for the zero modes 
\begin{align*}
M_{n,w}^2=\left(\frac{1}{2\pi R}\right)^\frac{2}{d-2}\left(\frac{n}{R}\right)^2+(2\pi R)^\frac{2}{d-2}\left(\frac{wR}{\alpha'}\right)^2\,,
\end{align*}
where $n$ and $w$ are the zero modes for the center of mass moment and the winding number of the string, respectively.
Considering first the infinite distance point $R\rightarrow\infty$, the Kaluza-Klein states have masses
\begin{align*}
    M_\mathrm{KK}^{(n)}=\left(\frac{n}{(2\pi)^{d/{d-2}}}\right)\frac{1}{R^{(d-1)/(d-2)}}= \left(\frac{n}{(2\pi)^{d/{d-2}}}\right)e^{-\sqrt{\frac{d-1}{d-2}}\Delta R}\,,
\end{align*}
realising the expected tower of light states.
In a very similar manner one encounters an infinite tower of states that become exponentially light when approaching $R\rightarrow 0$, but now arising from the winding zero modes.

 The existence of these two infinite distance points and the fact that the spectrum of the compact free boson is T-dual symmetric are thus inexorably linked. Extrapolating from this example, one could hope to use T-duality as a tool to explore the different infinite distance points in moduli space together with their corresponding towers of exponentially light states.
One of the main observations here is that this relation between duality and infinite distance points featured by this example strongly relies on the exchange of string zero-modes under the T-duality. This exchange is simple and clear cut when T-dualising on a circle or with respect to any trivially fibered $U(1)$ over an external manifold. The nature of zero-mode exchange under T-duality when applied on more general curved spaces becomes, as we will see in the next sections, somewhat more thorny. Indeed, slightly generalising the simple set-up of a free boson, we will see that applying T-duality to explore different points on the space of parameters of the effective field theory will dramatically change the narrative relating T-duality and the SDC. Investigating different exemplifying scenarios and reconciling certain of these exotic phenomena with the SDC will be the main matter of this paper.

Let us sketch the main ways in which we will explore this relation beyond the free boson example. The first obvious step is to T-dualise with respect to a non-trivially fibrered $U(1)$ over a curved base manifold. The resulting T-dual background will display a change in topology, often implying a less symmetric exchange of zero modes. The second generalisation is that in higher dimensions or in curved space we can include fluxes. These can come under the form of geometric fluxes (as for example a Kalb-Ramond three-form flux) or non-geometric fluxes. While the former will essentially bring us back to a change in topology, the second will force us to consider a class of drastically different  backgrounds known as non-geometric backgrounds.

%%%%%%%%%%%%%%%%%%%%%%%%%%%%%%%%%%%%%%%%%%%%%%%%%%%%%
%%%%%%%%%%%%%%%%%%%%%%%%%%%%%%%%%%%%%%%%%%%%%%%%%%%%%
\subsection{Computing the  reduction and scalar potential}\label{sec:reduction_potential}\label{sec:reduction}
In this section we will review in detail the reduction procedure of an  $D=(d+n)$-dimensional theory defined on space $M_D$, to a lower dimensional, effective, theory in $d$-dimensions on $M_d$, obtained by integrating out an internal compact submanifold $K_n$ of dimension $n$. We are thus considering a curved space $M_D= M_d\bowtie K_n $, where the subscript denotes the corresponding dimensionality and the product $\bowtie$ is meant to convey that the product is not necessarily direct. Allowing for non-trivial fiber bundle together with the possibility of (non-)geometric fluxes threading the spacetime will lead to a rich phenomenology for the SDC. The coordinates parametrising the internal space $K_n$ will be collectively referred to by $y$ and the coordinates for the external space by $x$. We will denote by $(G,B)$ the background fields of the $D$-dimensional theory consisting of a metric and a Kalb-Ramond two-form. The metric of the $n$-dimensional compact submanifold will be denoted by $h$ and metric of the remaining part by $\hat g$. The line element for the total space $\mathcal M$ will be taken to take on the general form
\begin{align}\label{eq:metric_Ansatz}
ds^2&=G_{IJ}\mathrm{d}X^I\mathrm{d}X^J=e^{\frac{4\Phi_x}{d-2}}R^{\frac{-2\alpha}{d-2}}g_{\mu \nu}(x) \mathrm{d}x^\mu \mathrm{d}x^\nu + h_{ij}(x,y) \mathrm{d} y^i \mathrm{d} y^j\,, 
\end{align}
where $\hat{g}_{\mu \nu}(x)\equiv e^{\frac{4\Phi_x}{d-2}}R^{\frac{-2\alpha}{d-2}}g_{\mu \nu}(x) $ denotes the metric on the external space $M_d$. We have added a ($x$-dependent) conformal factor for later convenience, the precise form of which we will explain shortly. On the compact part $h_{ij}(x,y)$ and $\{y^i\}$ are the metric and coordinates and we allow for a dependence on the external coordinates $x^\mu$. In the examples under study here, this dependence   will enter implicitly by promoting one (or several) constant parameter to  moduli fields of the theory, i.e functions of the external space.  These constants will have a notion of radius associated to the compact space and will be denoted by $R=R(x)$ (or $R_i(x)$ in the case of multiple moduli).

 After obtaining the effective field theory, we will explore the infinite distance points in the corresponding scalar field space in different duality frames. Note in particular that we will be mostly ignoring the additional external coordinates needed to embed the solution into string theory (and thus also the corresponding moduli). In addition, we will restrict the analysis to backgrounds featuring the NSNS-sector and we will leave the completion or addition of the backgrounds by RR-fluxes for future work.

%%%%%%%%%%%%%%%%%%%%%%%%%%%%%%%%%%%%%%%%%%%
\subsubsection{Scalar potentials for geometric backgrounds }\label{sec:pots_geom}
In order to perform the dimensional reduction we start from the $D$-dimensional Einstein-Hilbert action for the metric $G$ including an $H$-flux $\mathpzc{H}$ and dilaton $\Phi$ in the string frame
\begin{align}
S = \frac{1}{2 \kappa^2_0} \int \mathrm{d}^D X \sqrt{-G} e^{-2 \Phi} \left( \mathcal{R}(G) - \frac{1}{12}\mathpzc{H}_{IJK}\mathpzc{H}^{IJK} + 4 \partial_I \Phi \partial^I \Phi \right)\,.
\end{align}
Here $\mathpzc{H}_{IJK}=3\partial_{[I}B_{JK]}$ and we assume in the following that the Kalb-Ramond field and dilaton respect the block-diagonal structure of our geometry in the following way 
\begin{gather}\label{eq:product_strucutre_B}
B_{\mu \nu}= B_{\mu \nu}(x)\,, \quad B_{ij}=B_{ij}(x,y)\,, \quad B_{\mu j}=0\,,\nonumber\\
\Phi(X)= \Phi_0 + \Phi_x(x)+\Phi_y(x,y)\,,
\end{gather}
where $\Phi_0$ is constant and we allow for a $x$-dependence of the internal quantities. In the present case, the product structure above warrants for the Ricci scalar to split into three parts
\begin{align}
    \mathcal{R}(G) = \mathcal{R}(\hat{g}) + \mathcal{R}(h) + \mathcal{J}(\hat{g},h)\,,
\end{align}
where the explicit from of the third term rest on the $x$-dependence of the metric $h$. For the sake of pedagogy we will restrict the derivation to a single moduli here, while the multiparameter case will be detailed in  appendix \ref{App:reduction_formula}. When the only $x$-dependence is through a scalar field $R=R(x)$ we get
\begin{align}\label{eq:formula_ricciscalar}
\mathcal{R}(G) = \mathcal{R}(\hat{g}) + \mathcal{R}(h) -  \mathrm{tr}(h^{-1} \square_{\hat{g}} h) + f(h) \partial_\mu R \partial^{\hat{\mu}} R\,,
\end{align}
where $\square_{\hat{g}}$ is the d'Alembertian with respect to the metric $\hat g$ given below eq. \eqref{eq:metric_Ansatz} and we have defined 
\begin{align}
   f(h) = \frac{3}{4} \mathrm{tr}(h^{-1}\partial_R h)^2 -\frac{1}{4} \mathrm{tr}(h^{-1}\partial_R hh^{-1}\partial_R h) \,.
\end{align}
Taken all together, it results into the action
\begin{multline}\label{equ:10D_NSNS_action_split}
S= \frac{1}{2 \kappa^2} \int \mathrm{d}^d x \mathrm{d}^n y \sqrt{-g} \sqrt{h}  e^{-2 \Phi_y} e^{\frac{4 \Phi_x}{d-2}} R^{\frac{-d\alpha}{d-2}}\Bigl( \mathcal{R}(\hat{g}) + \mathcal{R}(h)  +  \mathcal{J}(\hat{g},h)\\
 -\frac{1}{12} \mathpzc{H}_{\mu\nu\lambda} \mathpzc{H}^{\hat{\mu}\hat{\nu}\hat{\lambda}} -\frac{1}{12} \mathpzc{H}_{ijk} \mathpzc{H}^{ijk} -\frac{1}{4} \mathpzc{H}_{\mu jk} \mathpzc{H}^{\hat{\mu} jk} + 4 \partial_\mu \Phi \partial^{\hat{\mu}} \Phi + 4 \partial_i \Phi_y \partial^i \Phi_y  \Bigr)\,,
\end{multline}
where a hat denotes contraction with $\hat{g}^{\mu \nu}$ as defined below eq. \eqref{eq:metric_Ansatz}.
Keeping in mind the transformation rules of the Ricci scalar $\mathcal R$ under a Weyl rescaling, the above can be further reduced to 
\begin{multline}\label{equ:reduced_NSNS}
    S= \frac{1}{2 \kappa^2} \int \mathrm{d}^d x \mathrm{d}^n y \sqrt{-g} \sqrt{h} e^{-2 \Phi_y}R^{-\alpha} \Bigl(  \mathcal{R}(g) - \frac{1}{12}\mathpzc{H}_{\mu\nu\lambda} \mathpzc{H}^{\mu\nu\lambda}e^{\frac{-8 \Phi_x}{d-2}}R^{\frac{4\alpha}{d-2}}+4\partial_\mu \Phi_x \partial^\mu \Phi_x\\
    + \left(\mathcal{R}(h)-\frac{1}{12} \mathpzc{H}_{ijk} \mathpzc{H}^{ijk} + 4 \partial_i \Phi_y \partial^i \Phi_y  \right) e^{\frac{4\Phi_x}{d-2}}R^{\frac{-2\alpha}{d-2}}\\
    + \mathcal{J}(\hat{g},h)e^{\frac{4\Phi_x}{d-2}}R^{\frac{-2\alpha}{d-2}} -\frac{1}{4} \mathpzc{H}_{\mu jk} \mathpzc{H}^{\mu jk}+ 8 \partial_\mu \Phi_x \partial^\mu \Phi_y+ 4 \partial_\mu \Phi_y \partial^\mu \Phi_y + \Sigma(\Phi_x,R)\Bigr)\,,
\end{multline}
with $\Sigma(\Phi_x,R)$ arising form the transformation of $\mathcal{R}(\hat{g})$ and the explicit from can be found in \eqref{eq:Sigma} in the appendix.
From this it is clear that we introduced the factor of $R^{\frac{-2\alpha}{d-2}}$ in the metric $\hat{g}$ in order to be able to bring the action to the Einstein frame. Since $R$ is a modulus of the internal manifold it also enters the action through the factor $\sqrt{h}$ in the above action as well as possibly through the dilaton $\Phi_y$. Hence we will always choose the parameter $\alpha$ such that $\sqrt{h} e^{-2 \Phi_y}R^{-\alpha}\propto 1$. We can further simplify the above expression if we assume the dilaton to be of the form\footnote{In fact this assumption is  just for convenience here since all our examples will be of this form and we again refer to the appendix \ref{App:reduction_formula} for the gerneric case.}
\begin{align}\label{eq:ansatz_dilaton}
    \Phi_y(x,y) = 1/2 \log(R^\beta) + \Phi_y^0 \,, \quad \Phi_x \equiv \phi\,, 
\end{align}
where $\Phi_y^0$ denotes any expression that  is at most a function of the internal coordinate $y$. Furthermore we assume that the only $x$-dependence of $B_{ij}$ is through $R$. Assuming that $h\equiv \det(h_{ij})= R^{2\gamma} \Omega^2(y)$ the above condition reads
\begin{align}\label{eq:relation_alpha_beta_gamma}
\sqrt{h} e^{-2 \Phi_y}R^{-\alpha}\propto 1 \Longleftrightarrow \gamma-\beta-\alpha = 0\,.
\end{align}
Under the above assumptions, and as detailed in \ref{App:reduction_formula}, one can simplify the action and finally arrive at the reduced $d=D-n$ dimensional action
\begin{multline}\label{equ:reduced_NSNS_2}
    S= \frac{\widehat{\mathcal{V}}_{int}}{2 \kappa^2} \int \mathrm{d}^d x \sqrt{-g} \, \Biggl( \mathcal{R}(g) - \frac{1}{12}\mathpzc{H}_{\mu\nu\lambda}\mathpzc{H}^{\mu\nu\lambda}e^{\frac{-8 \phi}{d-2}}R^{\frac{4\alpha}{d-2}}-\frac{4}{d-2}\partial_\mu \phi \partial^\mu \phi\\
    +\frac{4 \alpha}{d-2}  R^{-1} (\partial_\mu \phi \partial^\mu R) - \gamma_{RR}(\partial_\mu R \partial^\mu R) - V(\phi,R) \Biggr)\,,
\end{multline}
where $ V(\phi,R)$ is the potential and $\gamma_{RR}$ the kinetic term, the latter of which determines the metric $\gamma$ on scalar field space, are given by
\begin{align}
    \gamma_{RR}&=\widehat{\mathcal{V}}_{int}^{-1} \int    \mathrm{d}^n y \,\Omega \Bigl\{ \frac{1}{4}\mathrm{tr}\left((h^{-1}\partial_R h)^2\right) -\frac{1}{4}\mathrm{tr} \left((h^{-1}\partial_R B )^2\right) + \frac{\alpha^2}{d-2}R^{-2}\Bigr\}\,,\label{eq:met_conv_red}\\
    V(\phi,R)&=-\widehat{\mathcal{V}}_{int}^{-1}e^{\frac{4\phi}{d-2}}R^{\frac{-2\alpha}{d-2}} \int    \mathrm{d}^n y \,\Omega \Bigl\{\mathcal{R}(h)-\frac{1}{12} \mathpzc{H}_{ijk} \mathpzc{H}^{ijk} + 4 \partial_i \Phi_y \partial^i \Phi_y  \Bigr\} \,.\label{eq:pot_conv_red}
\end{align}
We defined 
\begin{align}\label{eq:V_int}
    \widehat{\mathcal{V}}_{int} = \int  \mathrm{d}^n y e^{2 \Phi_y^0} \Omega\,, 
\end{align}
which for $\Phi_y^0=0$ is just the $R=1$ unit volume of the internal space.
The first expression is nothing but the well-known DeWitt-like metric for scalar field variations \cite{DeWitt:1967yk,Gil-Medrano:1991ncm}. As can be seen from the reduced action in eq. \eqref{equ:reduced_NSNS_2}, after compactification, the scalar fields $\phi,R$ will be massive, controlled by a potential, as given in \eqref{eq:pot_conv_red}, that is being sourced by the curvature of the compactification space, the corresponding components of the ambient three-form flux and the dilaton.

In the next section, we will review how this is not the end of the story and we will be forced to complete the ellipses with new contributions, building on its application to the SDC in later sections. These new contributions to the potential will need to be considered for the spaces which are no longer Riemannian. These so-called non-geometric backgrounds feature new fluxes contributing to the non-trivial scalar potential. Before introducing non-geometric backgrounds, we first review another particularity of applying T-dualities on curved backgrounds: change of topology.

We will close this review by making a simple but critical observation: one can easily show that the metric on the moduli space is Abelian T-duality invariant using the tools from Double Field Theory. Indeed, one can straightforwardly check that 
\begin{align}\label{eq:Odd_expr_metric}
    \frac{1}{2}\mathrm{tr}[(\mathcal H^{-1}\partial_R \mathcal H)^2]&=  \mathrm{tr}\left((h^{-1}\partial_R h)^2\right)- \mathrm{tr}\left((h^{-1}\partial_R B)^2 \right)\,,
\end{align}
where $\mathcal H$ is the generalised metric for a curved internal background with metric $h$ and NSNS two-form $B$.
In particular, using the rewriting in eq. \eqref{eq:Odd_expr_metric} together with the invariance under T-duality of the combination given in \eqref{eq:relation_alpha_beta_gamma},
we see that the metric of the scalar field $R$ in eq. \eqref{eq:met_conv_red} is an $O(d,d)$ invariant object and remains invariant under T-duality. A short review of Double Field Theory included the generalised metric and the realisation of T-duality as an $O(d,d)$ rotation is provided in section \ref{sec:beta-gravity}.

%%%%%%%%%%%%%%%%%%%%%%%%%%%%%%%%%%%%%%%%%%%
%%%%%%%%%%%%%%%%%%%%%%%%%%%%%%%%%%%%%%%%%%%
\subsection{T-duality, topology change and zero modes}
T-duality does not only affect the geometry locally but can affect the global structure of the original manifold by changing its topology \cite{Giveon:1993ph,Alvarez:1994wj}. Change of topology with respect to the original manifold can for example take place when the circle undergoing T-duality is non-trivially fibered over some (curved) base manifold. This phenomenon was later formalised using topological T-duality \cite{Bouwknegt:2003vb}, see also \cite{Bunke:2005um}. This approach to T-duality is a purely topological reformulation and made it explicit that under T-duality a non-trivial $H$-flux and the first Chern class of the dual circle bundle are exchanged. Denote by $k$ the vector field generating an isometry of the background with respect to which one T-dualises. The space is then a, possibly non-trivial, fibration $E_k$. The non-trivialness of the fibration is measured by the first Chern class $c_1(E_k)$. T-duality precisely interchanges the two-form $\imath_k\mathpzc{H}$ and the first Chern class of the circle fibration:
\begin{align}\label{eq:top_T-duality}
\mathrm{fluxes:}\;\imath_k \mathpzc{H}  \quad \longleftrightarrow \quad \mathrm{topology:}\; c_1(E_k)\,.
\end{align}
The simplest possible example illustrating this striking property is when T-dualising the three sphere. The three-sphere, seen for example as the Hopf-fibration, has first Chern number 1. As a result, T-dual background is the trivially fibered $S^2\times S^1$ together with a non-trivial $H$ three-form with a single unit of flux. While the former has no non-contractible cycles, since its fundamental group vanishes $\pi_1(S^3)=0$, the latter has infinitely many since $\pi_1(S^1\times S^2)=\mathbb Z$ coming from the trivial $S^1$-fiber. We will study this example at length in section \ref{sec:S3_H-flux}. 

In the context of the SDC, the topology changing nature of T-duality inevitably leads us to conclude that the duality can ``teleport'' us to infinite corners of the moduli space with very different global properties. In some cases, the change of topology may even lead to map to a background not supporting any winding modes. This happens when the fundamental group is trivial or as will argue later, is pure torsion. In view of the example of the compact free boson given in section \ref{sec:cpt_free_boson}, one may then wonder what, if anything, can provide the tower of KK-states and what happens in the T-duality frames.  

%%%%%%%%%%%%%%%%%%%%%%%%%%%%%%%%%%%%%%%%%%%%%%%%%%%%%
%%%%%%%%%%%%%%%%%%%%%%%%%%%%%%%%%%%%%%%%%%%%%%%%%%%%%
\subsection{Non-geometry and non-geometric fluxes}
Strings perceive spacetime differently than point particles. One of the most striking implication of their stringy nature is that one is forced to drastically revise basic assumptions on the global structure of string theory compactifications. In particular one has to step away from the usual expectation  that identifies spacetimes with smooth, Riemannian manifolds, i.e. a space that consists out of patches locally sewn together by diffeomorphisms and local gauge transformations. Indeed, string dualities leave the action invariant up to a total derivative and can be thus put on similar footing as diffeomorphisms and $B$-field transformations. In particular, strings can propagate on spaces whose coordinate patches are not only glued by diffeomorphisms or gauge transformations but also by T- and S-dualities.  This observation was originally made in  \cite{Dabholkar:2002sy,Hull:2004in,Dabholkar:2005ve}. Such spacetimes are called globally non-geometric. Locally these spaces can be treated as Riemannian manifolds, globally however the geometry will, when going around a non-contractible loop in the geometry, not be single valued and will pick up non-trivial monodromies accounting for the stringy symmetry. More radical steps away from Riemannian manifolds also exists where the manifold depends  on additional coordinates that are canonically conjugate to the winding modes of the string. In this case the geometry is not even locally Riemannian and one speaks of locally non-geometric background. 

Although non-geometric backgrounds may seem quite distant from phenomenological expectations, the opposite is in fact true.   The scalar fields or moduli describing the compactified internal space have to, lacking experimental observation, to be stabilised to a constant expectation value. Often these moduli can be stabilised by the addition of fluxes in the higher dimensional theory, such as a three-form flux or Ramond-Ramond flux. Upon compactification, these automatically generate a scalar potential in the corresponding lower-dimensional, effective field theory. Since these are part of the geometric field content of supergravity, these fluxes are called geometric. In some instances however geometric fluxes are known to not be sufficient in order to fix all moduli. Instead it was soon realised \cite{Shelton:2005cf,Shelton:2006fd,Becker:2006ks,Hellerman:2006tx,Palti:2007pm} that the moduli could be stabilised by the introduction of fluxes which do not result from the conventional NSNS and RR-sector of ten-dimensional string theory \cite{Shelton:2006fd}, see also \cite{Aldazabal:2006up,Palti:2007pm}. Although bringing about the desired the stabilisation, the ten-dimensional origin of these new fluxes remained a mystery. In \cite{Andriot:2011uh,Andriot:2012wx,Andriot12} by exploiting Double Field Theory, the authors resolved this mystery by writing down a ten-dimensional effective action involving ten-dimensional fluxes on a non-geometric space, which upon compactification yielding the exotic fluxes observed in four dimensional effective field theories. These new ten dimensional fluxes in turn characterise the global and local features  of non-geometric space. The action describing them will play a crucial role in what follows and will be review in more depth in section \ref{sec:beta-gravity}. First however we would like to review a simple example of non-geometry.

%%%%%%%%%%%%%%%%%%%%%%%%%%%%%%%%%%%%%%%%%%%%%%%%%%%%%
\subsubsection{A simple example of non-geometry}\label{sec:intro_nongeomchain}
The prototypical incarnation of non-geometry can be found in the, so-called, non-geometric T-duality chain \cite{Kachru02,Shelton06}. We will briefly discuss the main features of this chain of backgrounds as it introduces in an succinct way the four types of non-geometric fluxes. A more detailed treatment is delayed until section \ref{sec:T-duality_chain}.  The starting point is a three-torus with a non-trivial $H$-flux. At each step, the T-duality will convert the original $H$-flux to new fluxes 
\begin{align}\label{eq:tdualitychain}
	\mathpzc{H}_{ijk}\xrightarrow[]{T_i} \mathpzc{f}^i{}_{jk}\xrightarrow[]{T_j} \mathpzc{Q}_{\,k}{}^{ij}\xrightarrow[]{T_k} \mathpzc{R}^{ijk}\,.
\end{align}
For now, we remark that the first two slots of the chain correspond to well-known fluxes: the field strength of the Kalb-Ramond field and the geometric flux which, as the name suggests, is related to the derivative of the vielbeins and the spin connection of the background. The two subsequent fluxes do not have a geometric meaning and are instances of non-geometric backgrounds. They do however lead to the two distinct types of non-geometry. The first flux, the $Q$-flux backgrounds, leads to a geometry which requires to be glued together by T-dualities. As a result, $Q$-flux backgrounds are globally non-geometric. The last arrow of the diagram leading to the $R$-flux is a little unorthodox as the corresponding T-duality transformation is performed with respect to a non-isometric directions. The $R$-flux is in addition even more exotic, and by far less understood, and is even locally non-geometric and was shown to feature non-commutative properties \cite{Blumenhagen:2010hj,Lust:2010iy,Blumenhagen:2011ph,Condeescu:2012sp,Andriot12a}. Even more $R$-flux backgrounds are notoriously hard to probe, lacking a target space description altogether and generically even feature non-associative coordinates \cite{Bouwknegt:2004ap,Blumenhagen:2011ph}.

Anticipating on the discussion in section \ref{sec:potential_nongeom} on reductions on non-geometric internal manifolds and the analysis of the T-duality chain in section \ref{sec:T-duality_chain}, one can already expect that, like their geometric counterpart, the non-geometric part will contribute to a scalar potential. This is indeed that case and the potential admits two additional sources which can be schematically summarised as
\begin{align*}
	V\supset \mathpzc{Q}^2\,, \mathpzc{R}^2\,,
\end{align*}
sourced by the (square of the) non-geometric fluxes $\mathpzc{Q}_{\,k}{}^{ij}$ and $\mathpzc{R}^{ijk}$. Details will be provided in section \ref{sec:potential_nongeom}, together with the exact expression of the potential in \eqref{eq:potential_Q-flux_bkrgd} which is applicable for the cases considered here.

Arguably, since any of these spaces in the chain in eq. \eqref{eq:tdualitychain} are dual to one another, one could argue that the non-geometry of the last two spaces is but an artefact. Indeed, one can always move, by applying multiple T-duality, back to the familiar geometric setting. Some non-geometric backgrounds are however not equivalent to geometric ones. Examples of truly non-geometric backgrounds have been realised by using asymmetric orbifold constructions \cite{Condeescu:2012sp,Condeescu:2013yma} and lack as of yet a target space description. We will distinguish the two possibilities by referring to the latter as ``truly non-geometric'' and the former as ``non-geometric''. From a point of view of the SDC, non-geometric backgrounds, even only `up to T-duality' display the same fluxes as truly non-geometric backgrounds, and offer a tractable vantage point to capture a similar phenomenology as those not on a T-duality orbit of a geometric background. We will thus restrict the present discussion to non-geometric background.

%%%%%%%%%%%%%%%%%%%%%%%%%%%%%%%%%%%%%%%%%%%
\subsubsection{Non-geometry and \texorpdfstring{$\beta$}{beta}-gravity}\label{sec:beta-gravity}
Since non-geometric spaces are closely related to T-duality and, in the case of locally non-geometric spaces, even involve explicit dependence on winding coordinates, it will come as no surprise that Double Field Theory (DFT) as a $O(D,D)$-covariant formalism which puts winding modes on the same footing as momentum mode, forms a crucial tool. 
Using the framework of DFT,  the exact functional form with which the different fluxes contribute has been derived in  \cite{Andriot:2011uh} with a simplifying assumption and later in full generality in \cite{Andriot12}. In addition, as we will see in later sections, DFT will be a crucial tool to track down additional terms, materialised by total derivative terms in the action, whose role was underscored in \cite{Andriot:2011uh,Andriot12}. These new contributions will be critical to reconcile the scalar field space on both side of the duality whenever one of the sides is (globally) non-geometric.

To lay the ground for the following sections discussing non-geometric backgrounds, we first need to introduce $\beta$-supergravity. For more details see \cite{Andriot13,Andriot:2013xca,Andriot:2014uda}.  $\beta$-supergravity (and the corresponding $\beta$-frame) is a reformulation of the standard supergravity equations for the NSNS-sector of the ten-dimensional Lagrangian. 
We start from the familiar supergravity Lagrangian capturing the string NSNS sector  
\begin{align}\label{eq:SUGRA_NSNS_Lagr}
    \mathcal L_\mathrm{NSNS}=e^{-2\Phi}\sqrt{|G|}\left(\mathcal R(G)-\frac{1}{12}\mathpzc{H}^2+4(\partial\Phi)^2\right)\,,
\end{align}
where $\mathpzc{H}_{ijk}=3\partial_{[i}B_{jk]}$. The NSNS data can be repackaged into the $O(d,d)$-covariant generalised metric
\begin{align}\label{eq:generalised_M}
	\mathcal H=\begin{pmatrix}
		G-BG^{-1}B & BG^{-1}\\ -G^{-1}B& G^{-1}
	\end{pmatrix}
\end{align}
called the generalised metric. We can invoke a generalised vielbein $\mathcal E$ such that
\begin{align}
    \mathcal H=\mathcal E^T\mathds I\mathcal E\,,\qquad \mathds I=\begin{pmatrix}
        \eta_d &0\\
        0 & \eta_d^{-1}
    \end{pmatrix}\,,
\end{align}
where $\eta_d$ is the flat metric, which we will take to be the identity $\mathds 1_d$ in what follows.
The vielbein is not unique and is only fixed up to an $O(2d)$-transformation. A distinguished choice is fixed by taking 
\begin{align}\label{eq:vlb_geom}
	\mathcal E_{(e,B)}=\begin{pmatrix}
		e & 0\\
		-e^T B& e^{-T}
	\end{pmatrix}\,,
\end{align}
where $G=e^T\mathds 1_d e$ is the conventional $d$-dimensional vielbein decomposition of the (positive definite) metric $G$ and $B$ is the field appearing in the generalised metric decomposition given in eq. \eqref{eq:generalised_M}. 
In what follows, we will use beta-gravity exclusively to describe non-geometric internal manifolds. As a consequence we will be using the notation $h$ for the metric instead of $G$, consistent with the notation chosen earlier to distinguish the total from the internal metric. 

The choice of generalised vielbein in eq. \eqref{eq:vlb_geom} is however born from an expectation for the background manifold to be geometric. As discussed earlier this is not a given for backgrounds in which strings propagate. These new backgrounds can be in fact better described by new fields which are motivated by the fact that  globally non-geometric backgrounds display monodromies which are neither diffeomorphisms or gauge transforamtions. Instead the generalised NSNS-vielbein in eq. \eqref{eq:vlb_geom}  is traded for  a vielbein ${\mathcal{E}_{\tilde e,\beta}}$ in terms of new fields $\tilde h$ and $\beta$. This  field redefinition of the standard formulation of the NSNS sector of supergravity \cite{Andriot:2011uh,Andriot:2012wx,Andriot12,Andriot:2014uda} (but see also earlier use of this parametrisation in generalised geometry in \cite{Grana:2008yw,Grange:2006es,Grange:2007bp}) where we trade the usual metric $h$, Kalb-Ramond field $B$ and dilation $\Phi_y$ for new fields $(\tilde h, \beta, \tilde \Phi_y)$ by the map
\begin{align}\label{eq:SNSNS_to_beta_fields}
    (h+B)^{-1}=(\tilde h^{-1}+\beta)\,,\qquad e^{-2 \Phi_y}\sqrt{| h|}=e^{-2\tilde \Phi_y}\sqrt{|\tilde h|}\,.
\end{align}
The matrix $\beta$ is a $d$-dimensional square anti-symmetric matrix.  The generalised metric now reads
\begin{align}
	\mathcal H=\begin{pmatrix}
		\tilde h & \tilde h\beta\\
  -\beta\tilde h& \ \ \tilde h^{-1}\!-\beta \tilde h\beta
	\end{pmatrix}\,.
\end{align}
In terms of these fields the decomposition of the generalised metric is realised through the vielbein \cite{Andriot:2011uh}  
\begin{align}
	{\mathcal{E}_{(\tilde e,\beta)}} =\begin{pmatrix}
		\tilde e& \tilde e\beta \\ 0 & \tilde e^{-T}
	\end{pmatrix}\,.
\end{align}
Note that this generalised vielbein still describes the same generalised metric and hence, i.e. $\mathcal H=\mathcal H=\mathcal E_{(e,B)}^T\mathds 1_{2d}\mathcal E_{(e,B)}=\mathcal E_{(\tilde e,\beta)}^T\mathds 1_{2d}\mathcal E_{(\tilde e,\beta)}$, is related through an $O(d,d)$-transformation\footnote{More precisely an $O(d-1,1)\times O(1,d-1)$-transformation, see appendix C of \cite{Andriot:2013xca} for details.} to the NSNS-generalised vielbein in eq. \eqref{eq:vlb_geom}.

The Lagrangian capturing the theory through the fields $\tilde h$ and $\beta$ requires some care \cite{Andriot:2013xca}. The underlying reason is that due to the different properties of the fields $\tilde h$ and $\beta$ compared to the more familiar metric $h$ and $B$-field, it is more challenging to define analogues to the conventional covariant derivative and Ricci scalar.  In \cite{Andriot:2014uda}, it was however shown to define a suitable connection $\tilde \Gamma$, which in terms defines an associated ``Ricci tensor''
\begin{align*}
    \check{\mathcal{R}}=\check{\mathcal{R}}(\beta, \tilde \Gamma)\,,
\end{align*}
which can be shown to indeed behave as a tensor. The reason to invoke this new tensor is that it geometrises the obscure form of the NSNS-sector of the $\beta$-supergravity Lagrangian.  $\check{\mathcal{R}}$ is a scalar under diffeomorphism and therefore manifestly diffeomorphism invariant as are all the other quantities in \eqref{eq:beta_Lagr}.  The new `Ricci' tensor $\check{\mathcal{R}}$ is determined by the $\mathpzc{Q}$-flux (see \eqref{eq:Qflux_ito_beta} below) together with the $\beta$-frame metric \cite{Andriot12}
\begin{align}\label{eq:checkRicci}
    \check{\mathcal{R}}=-\frac{1}{4}\tilde h_{ij}\tilde h_{mn}\tilde h^{kl}\mathpzc{Q}_{\;k}{}^{\ell j}\mathpzc{Q}_{\;\ell}{}^{\ell j}-\frac{1}{2}\tilde h_{ij}\mathpzc{Q}_{\;k}{}^{\ell j}\mathpzc{Q}_{\;j}{}^{ki}-\tilde h_{ij}\mathpzc{Q}_{\;k}{}^{ki}\mathpzc{Q}_{\;\ell}{}^{\ell j}+\dots\,,
\end{align}
where the elipsis contains terms linear in $\mathpzc{Q}$ and (covariant) derivatives of $\beta$, the complete expression is given in eq. (A.8) of \cite{Andriot12}. For our purposes, these contributions will vanish and we will only consider the quadratic contribution in the non-geometric flux  $\mathpzc{Q}$.
With this new tensor the $\beta$-frame Lagrangian for a non-geometric background characterised by non-geometric fluxes $\mathpzc{Q}$ and  $\mathpzc{R}$
reads \cite{Andriot12,Andriot:2013xca}
\begin{multline}\label{eq:beta_Lagra_w_check_R}
\mathcal{L}_{\beta} = e^{-2\tilde{\Phi}_y}\sqrt{|\tilde h|}\bigl( \mathcal{R}(\tilde{h}) + \check{\mathcal{R}}(\tilde{h}) - \frac{1}{12}\mathpzc{R}^{ijk}\mathpzc{R}_{\,ijk}+ 4 (\partial \tilde{\Phi}_y)^2\\
+ 4(\beta^{ij} \partial_j \tilde{\Phi}_y -\frac{1}{2} \tilde{h}_{pq}\beta^{ij}\partial_j\tilde{h}^{pq}+ \mathpzc{Q}_{\,k}^{\ ki})^2 \Bigr)\,.
\end{multline}
As will be shown in the next section and examples treated later, this form for the \linebreak{$\beta$-supergravity} action is instrumental in assembling the ingredients sourcing the potential when the background is non-geometric.  This new Lagrangian  $ \mathcal L_{\beta}$ contains explicitly the non-geometric fluxes \cite{Andriot12}
\begin{align}\label{eq:Qflux_ito_beta}
    \mathpzc{Q}_{\,k}{}^{ij}=\partial_k \beta^{ij}\,,\qquad \mathpzc{R}^{ijk}=3\beta^{\ell[i}\partial_\ell\beta^{jk]}\,.
\end{align}
Let us remark in passing that, while $\mathpzc{R}^{ijk}=3\beta^{\ell[i}\nabla_\ell\beta^{jk]}$ is a tensor, $\mathpzc{Q}_{\,k}{}^{ij}$ as defined above is not! It was however shown in \cite{Andriot:2013xca} that by defining the non-geometric flux instead by $\hat{\mathpzc{Q}}_{\,k}{}^{ij}=\partial_k \beta^{ij}-2\beta^{\ell[i}f^{j]}{}_{k\ell}$ and rewriting $\check{\mathcal{R}}$ in terms of this new flux $\hat{\mathpzc{Q}}$, that in turn $\hat{\mathpzc{Q}}$ should be interpreted  as a connection. For us this will not play any role\footnote{See however appendix \ref{App:totdal_deriv} for some comments on the importance of working with the fully covariant Lagrangian $\mathcal{L}_\beta$.} in the following and we will adopt the definition in \eqref{eq:Qflux_ito_beta} for the non-geometric $Q$-flux.

Critically, the $\beta$-gravity Lagrangian $\mathcal L_\beta$ is equivalent to the NSNS-Lagrangian $\mathcal L_\mathrm{NSNS}$ only up to a total derivative \cite{Andriot12,Andriot:2013xca}
\begin{align}\label{eq:up_to_total_der}
    \mathcal L_\mathrm{NSNS} = \mathcal L_\beta +   \partial \left( \dots \right)\,.
\end{align}
In particular, differing only by a total derivation, the two Lagrangian a) share the same set of symmetries%\footnote{This is since a symmetry will in general leave the Lagrangian invariant up to a total derivative. As a result a symmetry of the $NSNS$-Lagrangian will automatically be a symmetry of the $\beta$-Lagrangian.}
, b) up to a field redefinition, the two Lagrangian share the same equations of motion. The latter implies that finding a vacuum of one theory will automatically yield a (local) vacuum for the other theory. We provide some details on the explicit form of the total derivative in eq. \eqref{eq:up_to_total_der} in appendix \ref{App:totdal_deriv} where we demonstrate the matching on the example of the NATD of $S^3$ discussed in section \ref{sec:NATD_S3}. We will see in that section that contributions from the total derivative will be critical when examining this non-geometric background through the lens of the SDC.

%%%%%%%%%%%%%%%%%%%%%%%%%%%%%%%%%%%%%%%%%%%
\subsubsection{Monodromies and non-geometry} \label{sec:monodromies}
As discussed earlier, non-geometry manifests itself either as T-duality monodromy around certain cycles or through the dependence of the non-geometric background on winding coordinates.
In this section we will give a minimalist review of how globally non-geometric backgrounds can by detected through their non-trivial monodromies, for a more in depth and sophisticated discussion see e.g. appendix C in \cite{Andriot12a}.  Globally non-geometric space cannot be detected locally, instead one has to rely on the studying the appearance of non-trivial monodromies when revolving around non-contractible cycles present in the geometries. In general, monodromies arise when the compact manifold is a non-trivial fibration. Going around a closed loop in the fiber direction of the total space, the metric is not single-valued upon closing the loop. That is, denoting the coordinate along this loop direction by $z$ with periodicity $z \sim z + 2 \pi$,  the generalised metric $\mathcal{H}$ will not respect the periodicity of the loop but changes according to 
\begin{align*}
    \mathcal{H}_{IJ}(z+2\pi) = M_I^{\ L} \mathcal{H}_{LK}(z)M^K_{\ J}\,.
\end{align*}
Demanding the background to be a valid string background, requires that the monodromy matrix $M$ is a symmetry of string theory, i.e. lies in the T-duality group $O(d, d;\mathbb Z)$ \cite{Dabholkar:2002sy}.

A simple example is provided by the Q-flux backgrounds obtained by performing two T-dualities starting from the torus with $H$-flux indicated on the third rug of the T-duality chain in eq. \eqref{eq:tdualitychain}. The corresponding background is given by 
\begin{align*}
    \mathrm d s^2= \frac{1}{1+h^2x_3^2}\left(\mathrm d x_1^2+\mathrm d x_2^2\right) +\mathrm d x_3^2\,,\quad B=\frac{hx_3}{1+h^2x_3^2}\mathrm d x_1\wedge \mathrm d x_2\,.
\end{align*}
Going once round the $x_3$-fiber it is clear that, due to the denominator, the metric and B-fields are not periodic modulo a possible combination of diffeomorphism or gauge field transformations. Instead the background is periodic up to a transformation of the T-duality group $O(2,2,\mathbb Z)$. This can be easily infered from the corresponding generalised metric
\begin{align*}
\mathcal H=
	\begin{pmatrix}
		\delta^i_j &0 \\
		-2hx_3 \delta^{[j}_{x_1}\delta^{i]} _{x_2} & \delta^j_i
	\end{pmatrix}\begin{pmatrix}
		\delta_{ik}& 0\\
		0& \delta^{ik}
	\end{pmatrix}\begin{pmatrix}
		\delta^k_r & 2hx_3 \delta^{[k}_{x_1}\delta^{r]} _{x_2}\delta_m^l  \\
		0& \delta^r_j
	\end{pmatrix}\,,
\end{align*}
which under a jump $x_3\rightarrow x_3+1$ picks up a monodromy 
\begin{align*}
	M^I{}_J=\begin{pmatrix}
		\delta^i_j & 2hx_3 \delta^{[i}_{x_1}\delta^{j]} _{x_2}\delta_m^l  \\
		0& \delta^j_i
	\end{pmatrix}\in O(2,2;\mathbb Z)\,.
\end{align*}
Since the monodromy takes the form of an upper-triangular matrix, it does not correspond to a diffeomorphism or gauge transformation.

%%%%%%%%%%%%%%%%%%%%%%%%%%%%%%%%%%%%%%%%%%%
\subsubsection{Scalar potentials for non-geometric backgrounds}\label{sec:potential_nongeom}
In this section we will review how the scalar potential is sourced by non-geometric fluxes \cite{Andriot:2011uh,Andriot:2012wx}.
Consider again a total space $M_D=M_d\bowtie K_n$ where this time the internal space $K_n$ is taken to be non-geometric and the product $\bowtie$ indicates again the possibly non-trivial fibration. 
 The initial step for the reduction is exactly the same as in the geometric setting detailed in section \ref{sec:pots_geom}. We start from the NSNS action of the total space and split the Ricci scalar $\mathcal R$, the H-flux three-form $\mathpzc{H}$ and dilaton $\Phi$ contribution according to the splitting of the space into an internal and external space, leading  to the splitting of the the NSNS data as in  eq. \eqref{equ:10D_NSNS_action_split}. When the background is non-geometric however, the characterising monodromies plaguing the NSNS-background fields forestall us to integrate out this compact part of the geometry. To circumvent that problem, we will adopt the strategy sketched above by switching to the non-geometric frame, i.e. the $\beta$-frame in\linebreak eq. \eqref{eq:SNSNS_to_beta_fields}. As discussed earlier, switching to $\beta$-gravity preserves the original symmetries and the equations of motion of the conventional supergravity formulation. In particular, the $\beta$-supergravity Lagrangian $\mathcal L_\beta$ describes the same vacuum solution and no new point in the moduli space is visited. One is merely picking a field redefinition which  simplifies and clarifies the analysis of the solution considering the unusual, non-Riemannian geometry.

Again using the letter $h$ to denote the internal metric characterising $K_n$, we thus trade the background fields $(h,B,\Phi_y)$ for their non-geometric equivalents $(\tilde{h},\beta,\tilde{\Phi}_y)$. In terms of these new fields the geometry is perfectly smooth and the integration can be carried out. The conventional NSNS-Lagrangian for the compact part is now replaced by the $\beta$-gravity Lagrangian given in  \eqref{eq:beta_Lagra_w_check_R}, now specialised to globally non-geometric backgrounds (i.e no $R$-flux) reads \cite{Andriot:2013xca}
\begin{align}\label{eq:beta_Lagr}
\mathcal{L}_{\beta} = e^{-2\tilde{\Phi}_y}\sqrt{|\tilde h|}\left( \mathcal{R}(\tilde{h}) + \check{\mathcal{R}}(\tilde{h})+ 4 (\partial \tilde{\Phi}_y)^2 + 4(\beta^{ij} \partial_j \tilde{\Phi}_y + \tilde{h}_{pq}\beta^{ij}\partial_j\tilde{h}^{pq}+ \mathpzc{Q}_{\,k}^{\ ki})^2 \right)\,,
\end{align}
where the $\beta$-frame `Ricci' scalar was introduced in eq. \eqref{eq:checkRicci}. Let us stress that, although the construction of this action allows for an explicit dependence on the dual coordinates, all the expressions here are ``on section'' and only depend on the physical coordinates.
Note however already, that this Lagrangian $\mathcal L_\beta$ contains terms that upon compactifiying and integrating over the internal space will give contributions to the potential, similar to the one coming from the Ricci scalar, $H$-flux and dilaton in the NSNS-frame. 

Anticipating on the examples discussed in what follows and the fact that we will not consider $R$-fluxes by comparing  to the reduction of the NSNS-frame Lagrangian result in \eqref{eq:SUGRA_NSNS_Lagr}, we infer that 
\begin{align}\label{eq:potential_Q-flux_bkrgd}
    \widetilde{V}(\tilde{\phi},R) &=- \hat{\mathcal{V}}_{\beta, int}^{-1} e^{\frac{4\phi}{d-2}}R^{\frac{-2\alpha}{d-2}} \int \mathrm{d}^ny\, \tilde{\Omega} \Bigl\{\mathcal{R}(\tilde{h})-\frac{1}{4}\mathpzc Q^2 - \frac{1}{2} \tilde{h}_{ij} \mathpzc Q_{\;k}^{\ lj}\mathpzc Q_{\;l}^{ki}+ 4 (\partial \tilde{\Phi}_y)^2 + \dots \Bigr\}\,.
\end{align}
where $\mathpzc Q^2=\tilde{h}_{ij}\tilde{h}_{mn}\tilde{h}^{kl}\mathpzc  Q_{\;k}^{\ mi}\mathpzc  Q_{\;l}^{\ nj}$ and $\dots$ denotes terms that will vanish in our explicit examples. $\hat{\mathcal{V}}_{\beta, int}^{-1}$ is the analogue of \eqref{eq:V_int} but now calculated with $\tilde{\Omega}$ defined with respect to the $\beta$-frame.
Let us note that under the simplifying assumption\footnote{This holds for simple examples like the Q-flux background in the T-duality chain. The example of the NATD of $S^3$ which we discuss in section \ref{sec:NATD_S3} violates this assumption and requires the more general formula.} that $\beta^{km}\partial_m(\cdot)=0$ this reduces to 
\begin{align}
    \widetilde{V}(\tilde{\phi},R) &= -\hat{\mathcal{V}}_{\beta, int}^{-1}  e^{\frac{4\phi}{d-2}}R^{\frac{-2\alpha}{d-2}}\int \mathrm{d}^ny \, \tilde{\Omega} \Bigl\{\mathcal{R}(\tilde{h})-\frac{1}{4}\mathpzc Q^2 + 4 (\partial \tilde{\Phi}_y)^2\Bigr\}\,,
\end{align}
which is of a particular similar for as the NSNS-frame expression \eqref{eq:pot_conv_red}.

Before closing this section let use tie on one of the central results of \cite{Andriot:2011uh}:
in some case one can write a general form for the dependence of the potential of the reduced theory including all the fluxes.
Assuming the dimensional reduction depends solely on two scalar fields: the volume scalar $\rho$ and the four-dimensional dilaton $\sigma$. The authors of \cite{Andriot:2011uh}, argued that for this specific case a general formula for the potential arising after compactification on these non-geometric spaces from 10d to 4d reads
\begin{align}\label{eq:pot_scaling}
V(\sigma,\rho)=\sigma^{-2}\left( -\rho^{-1}V_{\mathpzc{f}}^0+\rho^{-3}V_{\mathpzc{H}}^0  + \rho V_{\mathpzc{Q}}^0 + \rho^{3}V_{\mathpzc{R}}^0 \right)\,.
\end{align}
We refer to \cite{Andriot:2011uh,Andriot:2012wx} for details and conventions.  One of the aspect of this result of relevance here, is the scaling with respect to the volume modulus, which is defined as  
\begin{align}\label{eq:ansatz_fluct_dep_Pot_form}
    h_{ij} = \rho\, h_{ij}^{(0)}\,,\quad e^{-\Phi}=e^{-\Phi^{(0)}}e^{-\varphi}\,,\quad \sigma=\rho^{3/2}e^{-\varphi}\,,
\end{align}
where $\varphi$ captures the fluctuations of the dilaton around the vacuum expectation value $\Phi^{(0)}$ of the internal manifold, the expectation value for the metric is likewise indicated by the superscript $(0)$ and $\sigma$ is the 4d dilaton. Most of the examples we are discussing in this work  however have  a three dimensional internal manifold. The above formula can however still be applied with some mild modification to the external dilaton.
We will see that, in the cases in which we can identify a volume modulus to satisfy the particular Ansatz for the dependence on the volume scalar given in eq. \eqref{eq:ansatz_fluct_dep_Pot_form}, the potential indeed obeys the behaviour of equation \eqref{eq:pot_scaling}.

%%%%%%%%%%%%%%%%%%%%%%%%%%%%%%%%%%%%%%%%%%%
%%%%%%%%%%%%%%%%%%%%%%%%%%%%%%%%%%%%%%%%%%%
\subsection{Generalised T-dualities}\label{sec:generalised_Tdualities}
In the previous section, we reviewed how string dualities can be seen as a way to explore the different infinite distance points in moduli or scalar field space starting from a single infinite distance point. When considering curved spacetimes one can contemplate the possibility to generate yet new  relations between infinite distinct points in the moduli space. The potential tool to achieve precisely that is provided by generalised T-dualities \cite{delaOssa:1992vci,Klimcik:1995ux}.
Indeed, generalised dualities drastically enlarge the scope of the familiar (Abelian) T-duality by performing a duality transformation with respect to a (compact) non-Abelian part of the background geometry rather than just a single $U(1)$. In what follows, we will use the umbrella term ``generalised T-duality'' to encompass both non-Abelian T-duality and Poisson-Lie T-duality. We provide a lightning introduction to generalised T-duality in appendix \ref{App:gen_T-duality}. One of the basic example is to T-dualise with respect to $SU(2)$ of a geometry featuring a three-sphere, which we will consider in detail in section \ref{sec:S3_H-flux}. 
 Below, we would like to address several questions and issues around generalised T-duality in turn, and by doing so lay out the working assumptions used throughout the remainder of the text and finally sketch the potential role generalised T-dualities can play in the Swampland program.

\vspace{10pt}
\noindent
\textit{Solution generating technique vs quantum symmetry}

\vspace{3pt}
\noindent
It remains unclear if and which subsector of generalised T-duality can be lifted to the quantum level and could thus be seen as true symmetries of string theory. In the recent pre-print \cite{Sakatani:2023mcy}, the authors showed that for certain WZW models Poisson-Lie T-duality is valid quantum mechanically at all orders in $\alpha'$ and at any genus.  Indeed, although generalised \linebreak{T-dualities} drastically relaxes some of the starting assumptions of the conventional \linebreak{T-duality}, they do remain classical canonical equivalences \cite{Alvarez:1994wj,Lozano:1995jx,Klimcik:1995ux}: mapping one theory to another theory while still sharing the same phase space. The equivalence is as of yet only proven classically%\footnote{Note as well that the canonical equivalence only reproduces the transformation of the NSNS sector. The dilaton shift can be extracted by modifying the measure, see \cite{Lozano:1996sc}.} 
and it remains doubtful that the equivalence holds quantum mechanically without modifications. It was already put into question long ago whether non-Abelian \linebreak{T-duality} could be a full perturbative symmetry of string theory \cite{Giveon:1993ai}. Indeed it is believed that at the level of the worldsheet, non-Abelian T-duality (and by extension more general forms of T-duality) are not expected to hold at all order in string genus perturbation theory.  It does however preserve the structure of low energy regime of string theory, by mapping CFTs to (different) CFTs, at least to quadratic order in $\alpha'$ \cite{Borsato:2020wwk,Hassler:2020tvz,Codina:2020yma}. In that regard, generalised T-dualities have to be considered as a solution generating techniques. 
    
\vspace{10pt}
\noindent
\textit{Map between scalar field spaces}

\vspace{3pt}
\noindent  
The scalar field space determining the internal manifolds is determined by three individual pieces of data: the set of scalar fields, the metric describing the local geometry they span and their scalar potential.
Taking generalised T-dualities to be no more than solution generating, we are only guaranteed to map the given set of scalar fields of the original background to the same set of scalar fields. These fields may however potentially span a T-dual scalar field space described by a different metric and potential on scalar field space.  Indeed, even for the former, the argument applied for (Abelian) T-duality in eq. \eqref{eq:Odd_expr_metric} no longer carries over when going to generalised T-dualities on curved spaces.  Below will show however that  for the non-Abelian T-dual of $SU(2)$, T-dual space will  display the same potential and metric as the original background. As will me explained in detail in section \ref{sec:NATD_S3}, the non-trivial way to matching occurs hints towards a more general principle rather than a coincidental matching. At the same time, the subtle way the two scalar field spaces matches may point that a possible proof of matching metric and/or potentials is far from straightforward.

\vspace{10pt} 
\noindent
\textit{Global structure} 

\vspace{3pt}
\noindent     
Having discussed the nature of the duality and the relation between dual scalar field spaces, we are left to discuss the spacetime geometry of the dualisable backgrounds. At a technical level, it is indeed critical to understand the global structure of the internal manifold to be able to perform the reduction procedure reviewed in section \ref{sec:reduction}. Foremost, as we would like the internal space to remain compact. This is however not straightforward when applying a non-Abelian T-duality, as one does not have any global information about the T-dual manifold \cite{Alvarez:1994wj}. In fact, without any modifications the T-dual coordinates are valued on the real line and as such generate a non-compact manifold.

In \cite{Lozano:2019ywa}, considering the non-Abelian T-dual of the $AdS_3\times S^3\times T^4$-background with respect to the $SU(2)$ isometry group of the three-sphere, the authors proposed a completion for the geometry resolving the lack of global information. The driving motivation there was to demand that holographically dual quantum field theory to the non-Abelian T-dual background should be a well-defined conformal theory. The completion proposed in \cite{Lozano:2019ywa} effectively segments the problematic non-compact direction generated by the Lagrange multiplier into an infinite series of segments which are periodically identified, rendering the completed geometry effectively compact. 

Here we will choose to take a different path. In an attempt to bypass the problem of our lack of knowledge about the global properties of the dual background, we will choose to perform a non-Abelian T-duality utilising the more general Poisson-Lie T-duality. The main advantage will be that it will provide is with natural way to interpret the dual geometry is being again compact.

\vspace{10pt}
Concluding, the power of generalised T-duality lies in its ability to generate a slew of backgrounds with properties that challenge and enlarge known solutions. Generic supergravity solutions are often by design, but  also to their detriment, very symmetric. Solutions obtained by applying a generalised dualisable models are generically less symmetric, often forming multiparametric solutions, while remaining computationally tractable. A prime example is provided by the non-Abelian generalisation of TsT deformations of \cite{Borsato:2016pas,Hoare:2016wsk}.
These in turn are closely related to $J\bar J$-deformations, which can be seen as $O(d,d;\mathbb R)$ rotations \cite{Kiritsis93,Giveon:1993ph,Forste:2003km}, which are integrable deformations of conformal CFTs and form a large class of generalised T-dualisable backgrounds. These models are also known to feature many interesting global properties, displaying changes in topology when varying over their moduli space \cite{Kiritsis93}.

%%%%%%%%%%%%%%%%%%%%%%%%%%%%%%%%%%%%%%%%%%%%%%%%%%%%%
%%%%%%%%     Three-sphere and Lens spaces     %%%%%%%
%%%%%%%%%%%%%%%%%%%%%%%%%%%%%%%%%%%%%%%%%%%%%%%%%%%%%
\section{T-duality and infinite distance in three-spheres and Lens spaces}\label{sec:S3_H-flux}
In this section, we will consider the simplest family of backgrounds that departs from the example of the compact free boson in several significant ways. The first difference is that these backgrounds are curved and as a result the scalar field characterising the geometry (here the radius) is subjected to a potential. In addition, due to the topology of these curved spaces we will have to reconsider the existence of corresponding winding zero-modes. Thirdly, T-duality will act on circles that are non-trivially fibered over some base manifold. As a result the T-dual backgrounds will generically display a different topology compared to the original space and we will see that this brings interesting implications concerning the zero-modes associated to a would-be tower of exponentially light states. In the last part of this section we will, inspired by the recent \cite{Li:2023gtt}, discuss how T-duality can help understand the subtle interplay between on-shell and off-shell arguments when studying the SDC.

%%%%%%%%%%%%%%%%%%%%%%%%%%%%%%%%%%%%%%%%%%%%%%%%%%%%%
%%%%%%%%%%%%%%%%%%%%%%%%%%%%%%%%%%%%%%%%%%%%%%%%%%%%%
\subsection{Backgrounds and topology change}\label{sec:S3_lens_geom}
We will first introduce the details of the backgrounds, the three-sphere and Lens spaces, and review how, upon T-dualising along a $U(1)$, the T-dual background displays a change of topology.

Since the three-sphere $S^3$ is a group manifold, when allowing for a three-form flux, it can be realised as a $SU(2)$-WZW model at radius $R^2=k$, where $k$ is the level. Then the metric and the $H$-flux of the $SU(2)$ WZW-model at level $k$ are, following the parametrisation in \cite{Plauschinn:2017ism}, given by
\begin{align}
    \mathrm d s^2 &=R^2(\mathrm{d}\eta^2 + \mathrm{d}\xi_1^2 + \mathrm{d}\xi_2^2 - 2 \cos(\eta)\mathrm{d}\xi_1\mathrm{d}\xi_2 )\,,\nonumber\\
    \mathpzc{H} &= k \sin(\eta) \mathrm{d}\eta \wedge \mathrm{d}\xi_1 \wedge \mathrm{d}\xi_2\,,
\end{align}
with $\eta \in  [0,\pi]\,, \xi_1 \in [0, 2\pi)\,, \xi_2 \in [0, 4\pi)$. The three-sphere is a non-trivial $U(1)$-fibration of the two-sphere $S^2$ via the Hopf fibration, it has first Chern number given by $c_1=1$. Topologically, it features a trivial fundamental groups $\pi_1(S^3)=0$, and cannot carry any non-contractible loops.

We now consider a natural generalisation: the orbifold of the three-sphere $S^3$ by a discrete group $\mathbb Z_p$ leading to Lens spaces \cite{Maldacena:2001ky,Plauschinn:2017ism,Gaberdiel:1995mx,Schulz:2011ye}.
Identifying the three-sphere as
\begin{align}
    S^3=\{(z_1,z_2)\in \mathbb C^2\mid |z_1|^2+|z_2|^2=1\}\,,
\end{align}
by identifying the action of $\mathbb Z_p$ on $S^3$ given by
\begin{align}
    \exp(2\pi ik/p)\cdot (z_1,z_2)=(z_1,\exp(2\pi ik/p)z_2)\,,\quad k=0,1,\dots,p-1\,,
\end{align}
the resulting geometry is called the Lens space $L(p,1)$ with metric
\begin{align}\label{eq:Lens_background}
    \mathrm d s^2&= R^2\left(\mathrm d \eta^2+\mathrm d \xi_1^2+\frac{\mathrm d \xi_2^2}{p^2}-\frac{2}{p}\cos(\eta)\mathrm d\xi_1\mathrm d\xi_2\right)\,,\nonumber\\
   \mathpzc{H}&=q \sin (\eta) \mathrm d\eta\wedge \mathrm d \xi_1\wedge \mathrm d \xi_2\,,\qquad q=\frac{k}{p}\,.
\end{align}
This is an orbifold geometry $S^3/\mathbb Z_p$ with a total space that is a circle bundle over a base two-sphere with bundle characterised by a Chern class determined by the integer $p$. Lens spaces have a torsion-full fundamental group $\pi_1(L(p,1))=\mathbb Z_p$, where the torsion-full cycle is the $U(1)$-fiber of the non-trivial bundle fibration. In order to be be conformal and hence a valid string background, the radius has to satisfy $R^2=k$.

For two unequal integers $p$ and $q$, the spaces $L(p,1)$ and $L(q,1)$ are not homeomorphic. These two space are however related by T-duality when allowing for a non-trivial $H$-flux. Indeed, one can show that applying T-duality on the fiber of the fibration of $L(p,1)$ with $q$ units of $H$-flux leads to the geometry where the roles of $p$ and $q$ are interchanged, i.e. \cite{Alvarez:1993qi,Bouwknegt:2003vb}
\begin{align}
	L(p,1)\quad \mathrm{and}\quad [\mathpzc{H}]=q\quad \leftrightarrow \quad L(q,1)\quad \mathrm{and}\quad [\mathpzc{H}]=p\,.
\end{align}
The change in topology is again a consequence of the non-trivial fibration of Lens spaces. Taking $p=0$ and $q=1$, we retrieve the example mentioned in the introduction on how \linebreak{T-duality} changes topology. In this specific example the trivially fibered Lens space \linebreak ${L(0,1)=S^2\times S^1}$ (a torus) with a single unit of $H$-flux is mapped by T-duality to the three-sphere $L(1,1)=S^3$ with zero $H$-flux. Lens spaces thus provide an example of a background carrying non-trivial $H$-flux together with a cycle, strings unwind after winding $(p-1)$ times around the cycle: the 1st homology group of a lens space is torsion-full. 

%%%%%%%%%%%%%%%%%%%%%%%%%%%%%%%%%%%%%%%%%%%%%%%%%%%%%
%%%%%%%%%%%%%%%%%%%%%%%%%%%%%%%%%%%%%%%%%%%%%%%%%%%%%
\subsection{Zero-modes, mass spectrum and scalar potentials}\label{sec:threesphereH_zeromodes_spectrum_pot}
Having introduced the backgrounds, we now move on to identify the zero-modes which can potentially provide a tower of infinite states and associated mass spectrum. 

\vspace{10pt}
\noindent
\textit{Existence and conservation of winding and momentum zero modes}

\vspace{3pt}
\noindent
For the three-sphere, the Lens spaces and their T-duals, winding modes are not a reliable source for a tower of states. In the former case, since the fundamental group of the three-sphere is trivial $\pi_1(S^3)=0$, all loops are contractible and winding modes are simply absent. Lens spaces on the other hand have a non-trivial fundamental group, which is however pure torsion:  $\pi_1(L(p,1))=\mathbb Z_p$. Physically, this means that a string can wind $p-1$ times after which it will unwind and becomes contractible. In particular, although winding modes are allowed, there are only a finite number of them, precisely $p-1$. We are led to conclude that winding modes on Lens spaces, at least for this case, will not provide an \textit{infinite} tower of (light) states. Turning to the momentum modes, we do not expect for their existence to be obstructed. This is summarise in the last column of table \ref{table:spheres_ab_T}\footnotemark .

\begingroup
    \renewcommand{\arraystretch}{1.7}
\begin{table}[t]
\centering
  \begin{tabular}{c|c|c}
  {Background and T-dual}  & Scalar potential $V$ &   Obstructed modes \\ \hline \hline
     $S^3_R$ : $c_1=1$\,, $[\mathpzc{H}]=0$ &$-\frac{3}{2 R^2}$ &  absence of winding \\
   $S^2_R\times S^1_{1/R}$: $c_1=0$\,, $[\mathpzc{H}]=1$ & $-\frac{3}{2 R^2}$ & \ $\begin{cases} S^2: \text{absence of winding}\\S^1:  \text{non-conserved momentum}\end{cases}$  \\  \hline 
     $S^3_R$ : $c_1=1$\,, $[\mathpzc{H}]=q$  &$-\frac{3}{2 R^2}+\frac{q^2}{2R^6}$ & absence of winding \\
   $S^3/\mathbb Z_q$: $c_1=q$\,, $[\mathpzc{H}]=1$ & $-\frac{3}{2 R^2}+\frac{q^2}{2R^6}$ &  non-conserved winding/momentum \\  \hline
     $S^3_R/\mathbb Z_p$ : $c_1=p$\,, $[\mathpzc{H}]=q$&$-\frac{3}{2 R^2}+\frac{p^2q^2}{2R^6}$  & non-conserved winding   \\
  $S^3/\mathbb Z_q$: $c_1=q$\,, $[\mathpzc{H}]=p$ &$-\frac{3}{2 R^2}+\frac{p^2q^2}{2R^6}$ & non-conserved winding/momentum
  \end{tabular}
  \caption{Scalar potential and obstructed modes for the three-sphere variations and their T-dual. Each block forms a T-dual pair, which share the same value  for the scalar potential. The sphere as well as the sphere-bundles (lens spaces) feature no or non-conserved winding modes due to the triviality $\pi_1(M)=0$ and torsion $\pi_1(M)=\mathbb Z_p$, $p\in \mathbb Z$, of their respective first fundamental groups.In the rightmost column we only list the obstructed modes that would usually provide the light tower when moving towards the infinite distance point where the potential diverges, i.e in the above notation for $R \to 0$.} \label{table:spheres_ab_T}
\end{table}
\endgroup
\footnotetext{Note that, in every first line, $R$ denotes the ``natural'' radius of the geometry: applying T-duality will generate a fibre with radius $\tilde{R}=1/R$ in the dual geometry, as explicitly stressed in the first row.}

Tracing back the momentum zero modes is a bit more subtle. On the three-sphere with H-flux, the momentum zero modes would be \`a priori\footnote{Anticipating on a related discussion in section \ref{sec:T-duality_chain} and appendix \ref{App:dilute_flux}, one could imagine that the $H$-flux might similarly spoil the conservation of the momentum zero modes. Even if a similar reasoning could be applied to the curved three-sphere, the fact that it is controlled by as single rather than three parameters together with the absence of any winding zero modes, would render the possible obstruction in eq. \eqref{eq:obstruction_conser_mom} void. In particular one could imagine always being able to find at least a tower of conserved momentum zero modes along one isometry direction.} conserved and could provide a tower of states. Turning to the T-dual spaces, whenever the original space displays a torsionful fundamental group, this will in turn control the non-conservation of the momentum zero-modes on its dual background. Therefore we would expect that at least along one direction of the dual space, the corresponding momentum mode will not to be conserved.

\pagebreak
\noindent
\textit{Mass spectrum of light states}

\vspace{3pt}
\noindent
To be able to discuss the place of a low effective field theory within the distance conjecture requires understanding the dependence of the tower of states that results from compactifying the theory on the internal manifold. Tracing their origin back to the zero modes carried by the internal manifold, the momentum zero modes will label the KK-tower. 

When it comes to determining the spectrum of the lights modes and zero modes in particular, curved backgrounds are particularly challenging. One cannot perform the flat space mode expansion and  identifying the relation between the mass spectrum and the zero-mode of the theory becomes much harder. Instead, one has to perform a Hodge decomposition and determine the mass spectrum of the KK-tower for each zero-mode by solving for the corresponding scalar spectrum of the Laplacian problem\footnote{Before moving on, we need to add some words of caution. As pointed out in e.g. \cite{Andriot:2018tmb}, the Laplacian spectrum may not necessarily coincide with mass spectrum of the the effective theory, either due to the appearance due to certain massive modes combining into massless ones  \cite{Duff:1986hr} (known as a ``space-invaders''-scenario) or, by completing the solution to a ten dimensional background, additional flux entering the solution may affect the analysis.}. 
Although, unsolved for most backgrounds, the KK-spectrum is very well-understood for the three-sphere and takes on the form
\begin{align*}
m^2=\frac{l(l+2)}{R^2}\,, \quad l=0,1,2\dots \,.
\end{align*} 
The Kaluza-Klein spectrum for Lens spaces is obtained from that of the three-sphere by projecting onto the $\mathbb Z_m$-invariant states. In particular, the corresponding KK-spectrum will feature the same dependence on the radius $R$ and still contain an infinite number of states, see e.g. \cite{Grimm:2018weo}.
 It is well-known that for backgrounds supported by an H-flux, one has to turn to H-twisted cohomology \cite{Gualtieri04} determined by the nilpotent differential $\mathrm{d}_{\mathpzc H}=\mathrm d-\mathpzc H\wedge$. Accordingly the corresponding Hodge decomposition and Laplace problem should be built from the differential $\mathrm d_{\mathpzc H}$. For the matter of the SDC however, only the KK-spectrum approaching the infinite distance point $R\rightarrow \infty$ is relevant. Noting that the $H$-flux goes as $\mathpzc H\sim R^{-6}$, plus an additional power of the inverse metric entering the definition of the Laplacian, any contribution from the $H$-flux will be heavily suppressed as $R\rightarrow \infty$. As a result, we can simply use the KK-spectrum computed without taking the $H$-flux into account.

Finally, the winding zero-modes are more straightforward, as the corresponding towers and their moduli-dependence cannot be identified using a similar recipe. In curved spaces however, we can rely on the familiar intuition that winding modes wrapping a cycle of the internal will become heavier with the volume of the corresponding cycle.  In the limit of small non-contractible cycles, the corresponding winding states becomes infinitesimally light.

\begin{figure}[t]
    \centering
    \includegraphics[scale=0.23]{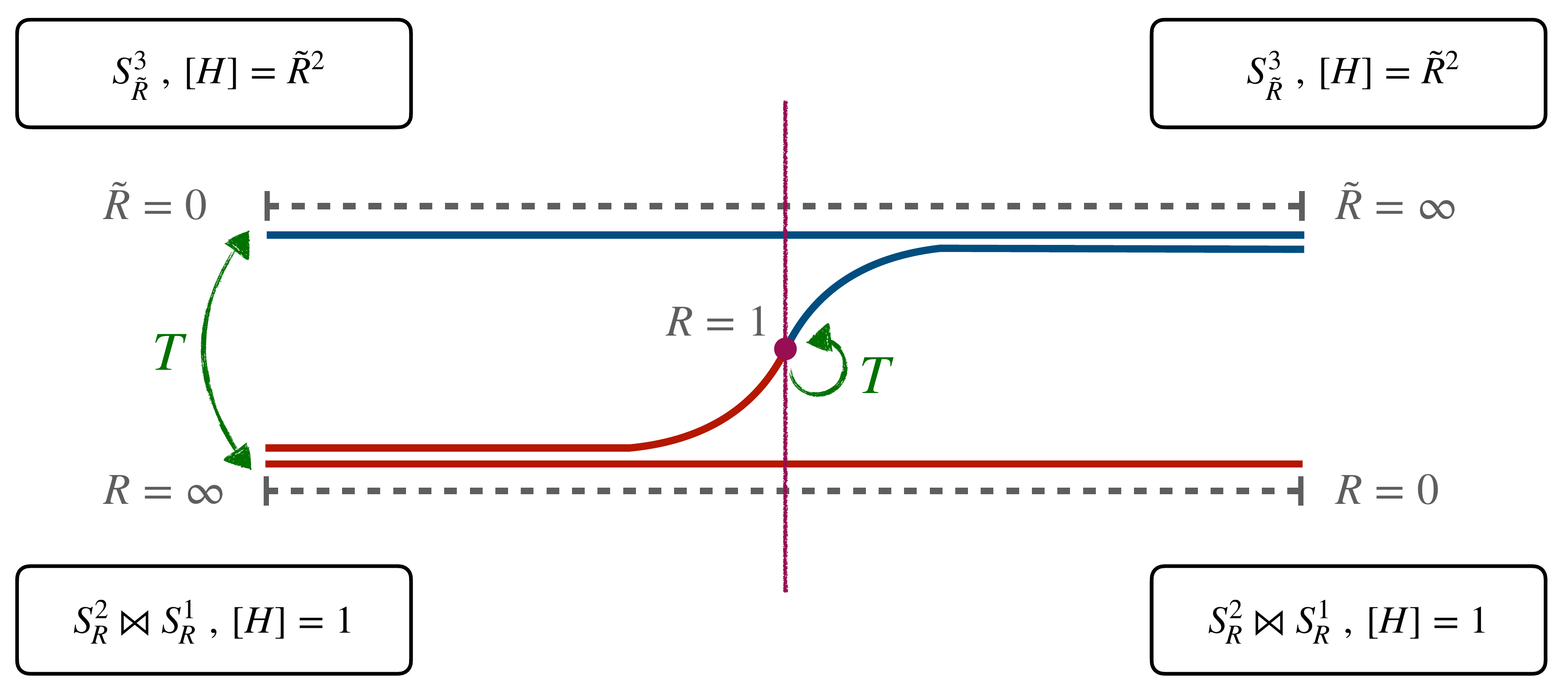}
    \caption{The three sphere with H-flux and its T-dual are a non-trivial example of backgrounds that share a self-duality point (the unit $S^3$). This allows for a special interpretation of the associated light towers. Starting from $R=\infty$ on the circle fibration (red line) one can either go to large $R$ within the same background and find the light states at small $R$ in terms of winding modes. These in turn can then be understood as KK-modes of the dual theory of $S^3$ at infinite radius. Equally well one can think of this process of changing frame as  a ``dynamical'' process, moving along the curved line, changing to the more natural frame of the $S^3$ (red line) at $R=1$,  in which we find the light KK modes as we move towards $R=0$.}
    \label{fig:duality_S3}
\end{figure}

\vspace{10pt}
\noindent
\textit{Computing the scalar potentials}

\vspace{3pt}
\noindent
Taking the the three-sphere or more generally a Lens space as the compact space $K_n$, we can apply the machinery introduced in section \ref{sec:pots_geom}. We subsequently perform a dimensional reduction on the Lens space $K_n$ and obtain the reduced Einstein-Hilbert action following the procedure reviewed in section \ref{sec:reduction_potential}.
Let us start from the background given in \eqref{eq:Lens_background}. The background is characterised by having first Chern class $c_1=p$ and $q$ units of $H$-flux, $[\mathpzc{H}]=q$. We will see that it is interesting to distinguish the cases where this relation is imposed or not and we will thus, in a first stage, step outside of the supergravity regime. Stepping outside of the supergravity regime by not imposing the relation  $R^2=k$ or equivalently $R^2/p=q$, we can perform\footnote{Note that $\xi_1$ is not an isometry and one cannot T-dualise with respect to that direction. Indeed, by computing the associated connection and reading off the corresponding charge the fiber spanned by $\xi_1$ is isomorphic with $U(1)/\mathbb Z_{p}$. This is nicely explained in \cite{Plauschinn:2017ism}.} a T-duality along $\xi_2$ and obtain a background of the form
\begin{align}\label{eq:dual_metric_Lens}
   \widetilde{ \mathrm{d}s^2} &= R^2 \left( \mathrm{d}\eta^2 + \left(1-\cos(\eta)^2\left(1-\frac{p^2q^2}{R^4}\right) \right) \mathrm{d}\xi_1^2 + \frac{p^2}{R^4}\mathrm{d}\tilde{\xi}_2^2 - 2 \frac{p^2q}{R^4} \cos(\eta)\mathrm{d}\xi_1\mathrm{d}\xi_2 \right)\,,\nonumber\\
    \widetilde{B} &= - p \cos(\eta) \mathrm{d}\xi_1 \wedge \mathrm{d}\xi_2\,.
\end{align}
It can be checked explicitly that this is indeed again a circle fibration with $c_1=q$ and $[\mathpzc H]=p$ and upon imposing the relation among $p,q,R$ stated above reduces to the Lens space action \eqref{eq:Lens_background} above. In particular this background describes a family of solution that share a selfduality point. This allows for a particular interpretation in view of the SDC, see figure \ref{fig:duality_S3}.
The generic expression for the reduced action was given in eq. \eqref{equ:reduced_NSNS}. For both the three-sphere and the Lens spaces the expression of the determinant of the compact space will be simply given by some power $\beta$ of the radius times the surface element of the sphere (or the appropriate analog for the Lens spaces), i.e. we will choose the dilaton to be given by
\begin{align}\label{eq:dilaton_lensspaces}
    \Phi_y=\frac{1}{2}\log(R^{\beta})\,, \quad \sqrt{h} = R^\beta \Omega\,.
\end{align}
With this choice we can simply set $\alpha=0$ in eq. \eqref{equ:reduced_NSNS_2} and obtain more explicitly for the action
\begin{multline}\label{equ:action_Lensspaces}
    S= \frac{\widehat{\mathcal{V}}_{int}}{2 \kappa^2} \int \mathrm{d}^d x \sqrt{-g}\, \Bigl(  \mathcal{R}(g) - \frac{1}{12} \mathpzc{H}_{\mu\nu\lambda} \mathpzc{H}^{\mu\nu\lambda}e^{\frac{-8 \phi}{d-2}}-\frac{4}{d-2}(\partial \phi)^2 - \gamma_{RR}(\partial R)^2 - V(\phi,R)\Bigr)\,.
\end{multline}
with 
\begin{align}
\gamma_{RR}&=\widehat{\mathcal{V}}_{int}^{-1}\int \mathrm{d}^3y\,\Omega\, \Bigl\{ \frac{1}{4}\mathrm{tr}\left((h^{-1}\partial_R h)^2\right) \Bigr\}\,,\nonumber\\
V(\phi,R) &= -\widehat{\mathcal{V}}_{int}^{-1} e^{\frac{4\phi}{d-2}}\int \mathrm{d}^3y\,\Omega\, \Bigl\{ \mathcal{R}(h)-\frac{1}{12} \mathpzc{H}_{ijk} \mathpzc{H}^{ijk} \Bigr\}\,,\nonumber\\
\widehat{\mathcal{V}}_{int}&=\int \mathrm{d}^3y\,\Omega\,.
\end{align}
From this we can easily determine that the metric on the moduli space by plugging in the explicit expression for the metric and $B$-field and respectively for the T-dual. Note that the dilaton of the dual space has to be computed by the usual Buscher rules and can not simply be set according to the above choice. However since $\sqrt{h}e^{-2\Phi_y}$ is invariant under (Abelian) T-duality we are guaranteed that also for the T-dual model $\tilde{\gamma}-\tilde{\beta}=0$ and we can use relation \eqref{equ:action_Lensspaces}.

The careful reader might wonder why we dropped the contribution $-\frac{1}{4}\mathpzc{H}_{\mu jk}\mathpzc{H}^{\mu jk}$ to the metric on the scalar field space. As mentioned earlier, when solving the supergravity equations for the Lens space, one is led to impose a relation between the flux quanta, the first Chern class and the radius of the circle of the fibration, and this contribution would not vanish. On the other hand, relaxing this requirement, one treats $q$ as an independent (quantised) quantity. The metric on moduli space for the backgrounds given by the family of metrics \eqref{eq:Lens_background}, is then given as
\begin{align}\label{eq:Lens_GRR_R}
   \gamma_{RR}= 3 R^{-2}\,,
\end{align}
while the scalar potential for generic $p$ and taking $q$ are summarised in table \ref{table:spheres_ab_T}. The supergravity case with the additional contribution coming from the flux will be the matter of section \ref{sec:promote_dyn}.

%%%%%%%%%%%%%%%%%%%%%%%%%%%%%%%%%%%%%%%%%%%%%%%%%%%%%
%%%%%%%%%%%%%%%%%%%%%%%%%%%%%%%%%%%%%%%%%%%%%%%%%%%%%
\subsection{Distance conjecture}
Turning finally to the distance conjecture, similarly to the simple circle compactification reviewed in the introduction, we have two infinite distance points: $R\rightarrow 0$ and $R\rightarrow \infty$. The different modes in the respective limits are summarised in table \ref{fig:HSphere_Lens_DC}. Note that we distinguish two cases: while the first two rows summarise the three-sphere with $H$-flux and its T-dual with the supergravity-imposed relation between the radius and the level, i.e. $k^2=R$, while the two next rows covers the case where the relation is not imposed.

\begingroup
    \begin{figure}[t]
        \centering
        \includegraphics[scale=0.289]{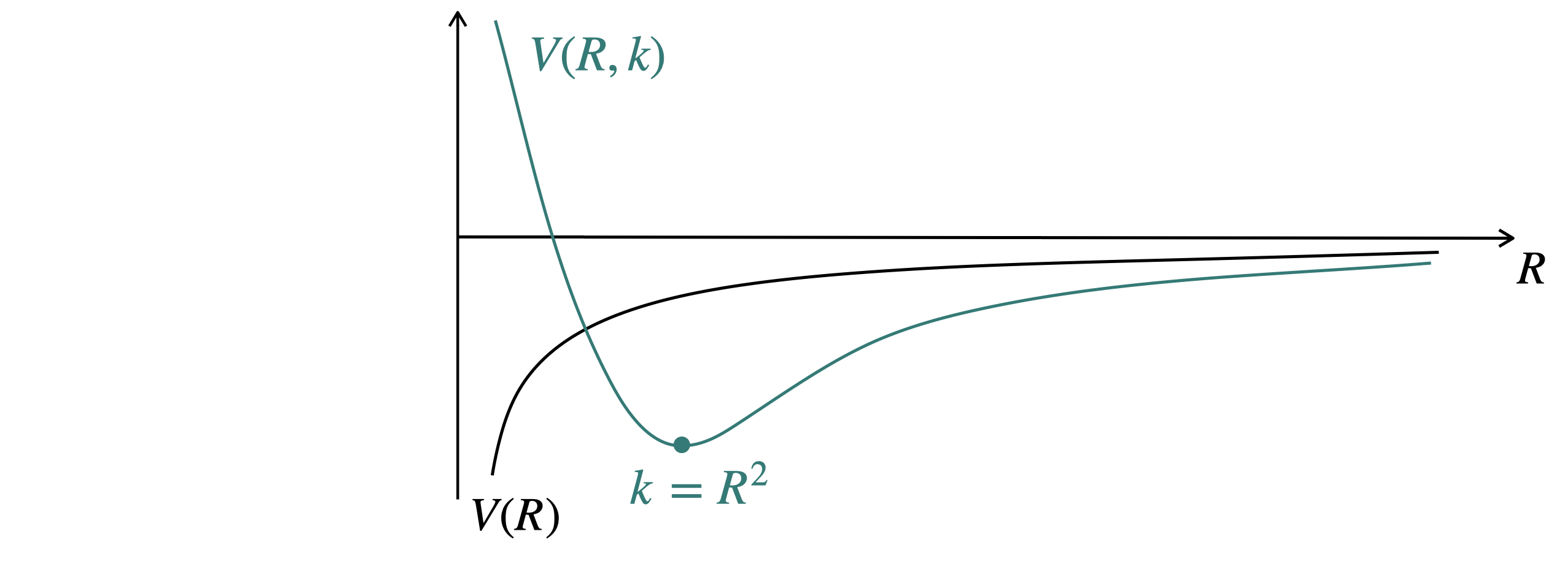}
    \end{figure}
    \setlength{\tabcolsep}{9pt}
    \renewcommand{\arraystretch}{1.3}
\begin{table}[t]
\centering
\begin{tabular}{c|c|c|c}
                  & $R\rightarrow 0$ & $R\rightarrow \infty$ &        Potential $V$           \\ \hline \hline
\multirow{2}{*}{$S^3_{R}$,\quad $[\mathpzc H]=R^2$} & $w:$ none& $w:$ none & \multirow{4}{*}{$-\frac{3}{2 R^2}$} \\ %\cline{2-3}
                  & $p:$ heavy  &  $p:$ light&                 \\ \cline{1-3}
\multirow{2}{*}{$S^3/\mathbb Z_{R^2}$,\quad $[\mathpzc H]=1$} &  $w:$  none&  $w:$  $\mathbb Z_{R^2\rightarrow \infty}$, heavy  &                   \\ %\cline{2-3}
                  &  $p:$ heavy &  $p:$  light &                   \\ \hline \hline
\multirow{2}{*}{$S^3_{R}$,\quad $[\mathpzc H]=k$} &$w:$  none & $w:$ none & \multirow{4}{*}{$-\frac{3}{2R^2}+\frac{k^2}{2 R^6}$} \\ %\cline{2-3}
                  &$p:$  heavy &  $p:$ light   &                   \\ \cline{1-3} 
\multirow{2}{*}{$S^3/\mathbb Z_k$,\quad $[\mathpzc H]=1$} &$w:$ $\mathbb Z_k$, light & $w:$ $\mathbb Z_k$, heavy &                   \\ %\cline{2-3}
                  & $p:$ heavy & $p:$ light &                  
\end{tabular} \caption{The table briefly summarises some of the salient aspects related to the zero modes for the three-sphere and Lens spaces with $H$-flux together with their duals. In the first two columns, the level and the radius are correlated via the relation $k=R^2$. Note that it is useful to remember that Lens space also allow for a description in terms of a non-trivial fiber bundle $S^1\hookrightarrow S^3 \rightarrow S^2$. The descriptive `light' for the momentum states only refer to the existence of only on such tower, as explained in the main text. For reference, the green plot follows the potential where the level $k$ is not promoted to a modulus but kept constant. This curve displays the characteristic minimum at $k=R^2$. }\label{fig:HSphere_Lens_DC}
\end{table}
\endgroup

The $R\rightarrow \infty$ corner of the scalar field space meets the prediction of SDC: one will encounter an infinite tower of exponentially light states arising from the momentum zero modes of the string. This is both true whether the relation $R^2=k$ imposed or not, but in a slightly different way. Note that for the supergravity solutions, the `cardinality' $k=\sqrt{R}$ of the torsion $\mathbb Z_{\sqrt{R}}$ grows as the radius. As a result, as one approaching $R\rightarrow \infty$, more and more, but \textit{only a finite} number of winding modes become available to give more and more states that become exponentially light. An infinite tower is however only strictly present at infinity when $R=\infty$ and not along a path approaching that point. This raises the question if this situation is still compatible with the statement of the SDC \cite{Ooguri:2006in}, which would require on \textit{infinite} tower of exponentially light states along a path leading to the infinite distance point.

Finally, let us return to the comment made earlier concerning the non-conservation of momentum zero modes in spaces dual to backgrounds with non-trivial torsion. As we remarked there, the torsion would only endanger, upon T-dualising, a single direction and thus single possible momentum tower. As the other momentum towers remain unaffected, the SDC in that limit can still be realised. This is a contrast the situation discussion later in section \ref{sec:T-duality_chain}. There we discuss a three-torus which has three moduli for each of its cycles, and therefore also have three independent infinite distance limits.

Going to $R\rightarrow 0$ limit of the scalar field space and contrary to the compact free boson, one runs now into an apparent contradiction. In absence of any winding modes (be it for the sphere with $H$-flux or the Lens spaces), no tower of exponentially light states can be produced to align with the requirements of the distance conjecture.
Taking now the non-trivial scalar potential sourced by the curvature and non-trivial fluxes, the first observation is that the potential displays a divergence. In particular, we see that a divergence develops in the potential when approaching $R\rightarrow 0$. This is the first example of a behaviour that we would like to argue in the following is much more general. Whenever certain zero modes cannot be carried by the internal geometry of an effective field theory living in Landscape, the SDC predicts the existence of a divergent scalar potential. 

The idea that, under certain circumstances, certain infinite distance points in moduli space cannot be accessed is not new. In \cite{Bonnefoy:2020uef}, a similar, but otherwise not obviously related obstruction was put forward under the name of Negative Distance Conjecture. 

Finally let us highlight once more that at the quantum level, we should correlate the radius with the level of the WZ term through the relation $k=R^2$. However, when the $H$-flux is being quantised, it also turns the scalar field space into a discrete space. The distance conjecture and its implications when moduli space is discrete have recently been investigated in \cite{Basile:2023rvm}. Before closing this example, we will discuss yet another way how T-duality may help understanding certain subtle aspects of the SDC.

%%%%%%%%%%%%%%%%%%%%%%%%%%%%%%%%%%%%%%%%%%%%%%%%%%%%%
%%%%%%%%%%%%%%%%%%%%%%%%%%%%%%%%%%%%%%%%%%%%%%%%%%%%%
\subsection{Flux variations and their metric contributions}\label{sec:promote_dyn}
As already pointed out several times, in order to have a valid string background (without the addition of further fluxes etc...) one has to set $k=R^2$, i.e. impose a relation between the radius $R$ of the sphere and the H-flux supported on it  by setting $[\mathpzc{H}]\equiv q = \frac{R^2}{p}$. This step is however not innocuous.
The compactification procedure, requires to promote the scalar field $R$ to an $x$-dependent field $R(x)$. In the present class of backgrounds, the supergravity imposes relations between the parameters given by $k=R^2$, effectively forcing one to take into account also variations of the three-form flux around its constant vacuum expectation value. This additional variation in the action inevitably sources a contribution to the metric on scalar field space which was not there before.

Very recently, a related problem was considered in \cite{Li:2023gtt} in the context of AdS spaces and the AdS Distance Conjecture \cite{Lust:2019zwm}. Indeed following the arguments of \cite{Lust:2019zwm} suggesting to consider not only variations coming form the (internal) metric but the full set of all possible field variations lead the authors to introduce an ``action metric'' providing a prescription for obtaining  variations of RR-fluxes. While the procedure faces some conceptual challenges, it was shown in \cite{Li:2023gtt} that for certain backgrounds the flux variations are crucial in order to obtain a positive metric on scalar field space.

In what follows, using  T-duality as a critical tool and with the example at hand, we will see that the additional variations are indeed crucial for leaving invariant the metric on scalar field space, providing an explicit  example on how the invariance of moduli space under T-duality can only be realised when taking into account the variations from the non-vanishing H-flux.  

We will this consider the example of the three-sphere with H-flux and correlate the level and radius through the relation $k=R^2$, i.e $(c_1=1, [\mathpzc{H}]=R^2)$. Since an explicit embedding into a 10d solution of supergravity is not important for our purposes we do not specify the external part in the following. We will merely assume that the full Kalb-Ramond field $B$ is of the general form given in section \ref{sec:pots_geom} such that the only flux contribution to the metric can arise from
\begin{align}
    -\frac{1}{4}\mathrm{tr} \left((h^{-1}\partial_R B)^2\right)= \frac{1}{4}  \mathpzc{H}_{\mu jk}\mathpzc{H}^{\mu jk} =   \frac{2 \cot(\eta)^2}{ R^2}(\partial R)^2\,.
\end{align}
This expression for the metric is however problematic. First of all, it depends on the angular coordinate $\eta$ in a way that when integrating over the sphere will give a divergent contribution. In particular we have to perform the integral
\begin{align}
    \widehat{\mathcal{V}}_{int}^{-1} \int    \mathrm{d}^3 y \,\Omega\,\Bigl\{ -\frac{1}{4}\mathrm{tr} \left((h^{-1}\partial_R B)^2\right) \Bigr\} =  \int_0^\pi \mathrm{d} \eta \ \sin(\eta) \frac{ \cot(\eta)^2}{R^2}(\partial R)^2\,,\label{eq:fluxvar_div_intrg}
\end{align}
which is non-convergent due to the divergent behaviour of the integrand. Before trying to resolve this puzzle, it will prove insightful to write down the total metric on moduli space as
\begin{align}\label{eq:modulispace_S3wH}
    \gamma_{RR} &= \widehat{\mathcal{V}}_{int}^{-1} \int    \mathrm{d}^3 y \,\Omega\,\Bigl\{ \frac{1}{4}\mathrm{tr}\left((h^{-1}\partial_R h)^2 \right) -\frac{1}{4}\mathrm{tr} \left((h^{-1}\partial_R B)^2\right) \Bigr\}\nonumber\\
    &= \frac{1}{2 R^2}\int_0^\pi \mathrm{d}\eta  \sin(\eta) \left(3+2\cot(\eta)^2\right)\,.
\end{align}
As was outlined above, the  background under consideration gets mapped under T-duality into the $(c_1=R^2, [\mathpzc{H}]=1)$-background, hence it becomes a not-trivially fibered circle bundle with one unit of H-flux. In particular the metric and $B$-field read
\begin{align}
    \tilde{G} = \begin{pmatrix}
        R^2 & 0 & 0\\
        0 & R^2 & -\cos(\eta)\\
        0 & -\cos(\eta) & \frac{1}{R^2}
    \end{pmatrix}\,, \qquad \tilde{B}= - \cos(\eta) \mathrm{d}\xi_1 \wedge \mathrm{d}\xi_2\,.
\end{align}
Clearly in the T-dual space, there will be no flux contribution to the metric. As a result, the metric on the scalar field space can be obtained in the ``standard way'' from the variations of the internal metric (i.e. without having to worry about on- and off-shell variations). What we find for $\tilde{\gamma}_{RR}$ (for example by plugging $\tilde{\gamma}$ into eq. \eqref{equ:reduced_NSNS}) exactly agrees with $\gamma_{RR}$ as given in \eqref{eq:modulispace_S3wH}. We see that this simple case exemplifies how the metric on scalar field space only remains invariant under T-duality when taking into account the simple interplay between on- and off-shell contributions. The contributions coming from the three-form flux are crucial in order to match the metric on the moduli spaces of T-dual internal manifolds.

Let us now return to the problem of the divergent integrand in the contribution to the scalar metric in eq. \eqref{eq:fluxvar_div_intrg}. In order to resolve this issue, inspired by \cite{Li:2023gtt}, we observe that
unlike the contribution to the potential $V_\mathpzc{H}$ that arises from the term $\mathpzc{H}_{\mu \nu \lambda}\mathpzc{H}^{\mu \nu \lambda}$ the flux contribution to the metric on moduli space is not gauge invariant: it is sensitive to the explicit choice of Kalb-Ramond field $B$. Using the freedom to choose the gauge field patchwise, we will indeed obtain a well-defined flux contribution. 
Recall that the flux  contribution comes from the term $\mathpzc{H}_{\mu jk}\mathpzc{H}^{\mu jk}$ where, due to our block-diagonal form of $B$, we have $\mathpzc{H}_{\mu jk} = \partial_\mu B_{ij}$. Adding subsequently a closed term $\hat{B}=B+\zeta \mathrm{d}\alpha(y)$ with $\zeta$ a constant and $\alpha$ a one-form. However consider the situation of $\zeta$ being (a function of) some modulus, i.e promoting $\zeta$ to a field $\zeta(x)$. This means we have
\begin{align}
    \hat{\mathpzc{H}} = \mathrm{d}\hat{B} = \mathrm{d}B + \mathrm{d}\zeta(x) \wedge \mathrm{d}\alpha(y)
\end{align}
or explicitly
\begin{align}
    \hat{\mathpzc{H}}_{\mu \nu \lambda} = \mathpzc{H}_{\mu \nu \lambda}\,, \qquad \hat{\mathpzc{H}}_{\mu jk} = \mathpzc{H}_{\mu jk} + \partial_\mu \zeta(x)(\mathrm{d}\alpha)_{jk}\,.
\end{align}
This is true on-shell, i.e considering the moduli to be constant this describes the same supergravity solution. However when we allow for the moduli to go off-shell in order to compute the metric on scalar field space and following \cite{Li:2023gtt}, this term will have a non-zero contribution to the moduli space metric. As already mentioned above we can use this freedom now to make the flux contribution well-defined.

We define the B-field now patch-wise
\begin{align}
    \hat B_N &= B + R^2 \mathrm{d}(\xi_1 \mathrm{d}\xi_2)\,,\qquad \hat{B}_S = B - R^2 \mathrm{d}(\xi_1 \mathrm{d}\xi_2)\,,
\end{align}
where $B_{N/S}$ denotes the gauge choice on the norther ($\eta \in [0,\pi)$ ) and southern ($\eta \in (0,\pi])$  hemisphere of $S^3$. They clearly differ by only an exact term and therefore the two fields are gauge equivalent. The resulting flux contribution is easily calculated to be
\begin{align}
     \frac{1}{4}  \mathpzc{H}^N_{\mu jk}\mathpzc{H}_N^{\mu jk} = \frac{ 2 \tan(\eta/2)^2}{R^2}(\partial R)^2\,.
\end{align}
for $B_N$ and a similar expression for $B_S$. The full flux contribution to the metric \eqref{eq:modulispace_S3wH} therefore reads 
\begin{align}
   \gamma_{RR}^\mathpzc{H}&= -\frac{1}{8} \left(\int_0^{\pi/2}  \mathrm{d}\eta  \sin(\eta)  \, \mathrm{tr} \left((h^{-1}\partial_R B_N)^2 \right) + \int_{\pi/2}^\pi   \mathrm{d}\eta  \sin(\eta)   \, \mathrm{tr} \left((h^{-1}\partial_R B_S)^2 \right)\right)\nonumber\\
   &= \frac{2 \log(4)-2}{R^2}\,,
\end{align}
and is therefore finite.
Taken altogether, the metric becomes
\begin{align}
\gamma_{RR}=\gamma_{RR}^\mathcal{R}+\gamma_{RR}^\mathpzc{H}= \frac{1+2 \log(4)}{R^2}\,,
\end{align}
where $\gamma_{RR}^\mathcal{R}$ is given by \eqref{eq:Lens_GRR_R}.
Now analogous to the situation above one may apply the Buscher rules in order to obtain the T-dual background. Since the B-field is defined only locally, we have to apply the Buscher rules patch-wise. The different patches glue together to form a smooth metric describing a non-trivial circle bundle of $S^2$.  The moduli space for the geometry can be calculated patch-wise using a similar calculation for the flux contribution above and the scalar field spaces can be seen to agree. 

%%%%%%%%%%%%%%%%%%%%%%%%%%%%%%%%%%%%%%%%%%%%%%%%%%%%%
%%%%%%%%%%%    NATD of the three-sphere    %%%%%%%%%%
%%%%%%%%%%%%%%%%%%%%%%%%%%%%%%%%%%%%%%%%%%%%%%%%%%%%%
\section{Generalised T-duality and infinite distance of the three-sphere }\label{sec:NATD_S3}
Having examined the implications of studying different Abelian T-duality frames for the three-sphere with and without three-form flux $\mathpzc{H}$, in this section we would like to go one step further. Again taking the three-sphere as our main example and realise its geometry as the non-Abelian Lie group $SU(2)$, one could consider the T-duality frame realised by a non-Abelian T-duality of the whole $SU(2)$ rather than T-duality of one of its Cartan directions. 

 Non-Abelian T-dual is usually implemented using a Lagrangian multiplier procedure, whereupon the real-valued multiplier becomes one of the coordinates of the background. Missing global information about the non-Abelian T-dual background, this direction becomes \` a priori non-compact. As already discussed in section \ref{sec:generalised_Tdualities}, this is clearly at odds with when one, for example, requires the solution to be again a holographic background. Here however we will choose a different route. Instead of using the Lagrangian multiplier approach to non-Abelian T-duality, we will espouse a more general perspective on non-Abelian T-duality offered by Poisson-Lie T-duality. This is the as of yet most general formulation of T-duality.  Taking this more general approach allows us to circumvent the introduction of an explicitly non-compact coordinate associated to the Lagrangian multiplier. Instead, we will obtain, after T-duality, only compact coordinates, provided we interpret the resulting space as being (globally) non-geometric. This interpretation will turn out to be be critical to obtain a consistent picture on both side of the duality from a point of view of the SDC. In addition, the framework of Poisson-Lie T-duality will equip us with tools to track the momentum/winding exchange of the two dual models \cite{Klimcik:1996nq}.  In appendix \ref{app:commNATD}, we will make a step towards relating of the background considered here with the holographic completion of \cite{Lozano:2019ywa}. Although desirable, it is not obvious how to directly compare, or even possibly match, these two instances of the NATD of the three-sphere.

%%%%%%%%%%%%%%%%%%%%%%%%%%%%%%%%%%%%%%%%%%%
\subsection{Zero modes of non-Abelian T-duals and mass spectrum}
In this section we first identify the zero modes, using the framework provided by Poisson-Lie T-duality and discuss the dependence of the mass spectrum on the scalar field $R$.

\vspace{10pt}

\noindent
\textit{Zero modes}

\vspace{3pt}
\noindent
Generalised T-duality transformations also exchange the zero modes of the dual manifolds. The precise generalised exchange of zero-modes was worked out in \cite{Klimcik:1996nq} and is briefly reviewed in appendix \ref{sec_app:PLnarain}. When applied to the three-sphere and its non-Abelian T-dual with respect to the whole $SU(2)$ the exchange reads 
\begin{align*}
    S^3\cong SU(2): w=0\text{ and } m\in\mathbb Z^{\oplus 3}\quad \longleftrightarrow \quad SU(2)_\mathrm{NATD}: w\in\mathbb Z^{\oplus 3}\text{ and } m=0\,.
\end{align*}
On the left, we have the familiar absence of winding state on the three-sphere together with quantised momenta for each of the three coordinates. On the right, we see that the dual geometry topologically looks like a three-torus. It features three non-contractible cycles hosting three integer valued zero-modes associated to the winding of the string. More surprisingly, but in accordance with the familiar expectations of T-duality that momentum and winding zero modes are exchanged, the dual geometry does not allow for any quantised momentum zero modes. This can be traced back to the scrambling of the (non-Abelian) isometries in the original background due to the non-linear nature of the generalised duality transformation: the dual background has no isometries left.

\vspace{10pt}
\noindent
\textit{Mass spectrum}

\vspace{3pt}
\noindent
Since the three-sphere has already been discussed in the previous section we turn to the NATD of the three-sphere. As this space is no longer maximally symmetric, the task of solving the Laplacian problem becomes an arduous task. The problem of identifying the Laplacian spectrum on the NATD of $S^3$ was already explored in \cite{Lozano:2013oma} within the context of the non-Abelian T-dual of the holographic background $\mathrm{AdS}_3\times S^3\times T^4$. Instead of considering the Laplacian problem head-on, the authors considered a particular limit which is also applicable here: for large values of the radius $R\rightarrow \infty$, the spectrum was shown to remain unchanged compared to the original three-sphere. As result, at least for large values of the radius (which will be the regime we are interested in), the spectrum matches at leading order the one of a conventional three-sphere.

%%%%%%%%%%%%%%%%%%%%%%%%%%%%%%%%%%%%%%%%%%%
%%%%%%%%%%%%%%%%%%%%%%%%%%%%%%%%%%%%%%%%%%%
\subsection{Geometry and non-geometry of the NATD \texorpdfstring{$S^3$}{S3}}\label{sec:nongeom_NATDS3}
We start by stating the background of three-sphere $S^3$ and its non-abelian T-dual, which we will denote by $\hat{T}^3$ for reason that will become clear soon. Both backgrounds are obtained from considering their associated Drinfel'd double and performing a non-Abelian T-dual using the more general setting of Poisson-Lie T-duality, which we briefly review in appendix \ref{sec_app:PL}. We take the three-sphere to be parametrised as follows 
\begin{align}\label{eq:metric_S3}
    \quad \mathrm{d}s^2 = h_{ij}\mathrm dx^i\mathrm d x^j= R^2 \left(\mathrm{d} \eta^2 + \mathrm{d}\xi_1^2 +\mathrm{d}\xi_2^2 + 2 \cos(\eta) \mathrm{d}\xi_1 \mathrm{d}\xi_2 \right)\,,
\end{align}
with $\eta \in  [0,\pi]\,, \xi_1 \in [0, 2\pi)\,, \xi_2 \in [0, 4\pi)$ as before and vanishing Kalb-Ramond two-form. The background metric is completed to a background  admitting a generalised T-dual by the addition of the dilaton \cite{Demulder:2018lmj}
\begin{align}\label{eq:dilaton_S3_NATDsec}
    \Phi_y= \phi_0 + \frac{1}{4}\log\left(\det(h_{ab}) \right) =\frac{1}{2}\log(R^{3})\,,
\end{align}
with $h_{ab}$ carries algebra indices and setting $\phi_0$ to zero.
The dual background, which we will denote by an hat, is given, as dictated by the duality transformations in the appendix, by
\begin{align}\label{eq:background_fields_NATDS3}
    \hat{h} = \frac{1}{R^2 \mathcal{X}}       \begin{pmatrix} R^4+\phi^2 &  \phi \psi &  \theta \phi\\
    \phi \psi & R^4 + \psi^2 & \theta \psi\\
    \theta \phi &  \theta \psi & R^4 + \theta^2
    \end{pmatrix}\,, \quad \hat{B}=
    \frac{1}{\mathcal{X}} \begin{pmatrix} 0 &  -\theta &  \psi\\
   \theta & 0 & -\phi\\
    -\psi &  \phi & 0
    \end{pmatrix}\,,
\end{align} 
together with a non-trivial dilaton $\hat{\Phi}=\log(|\hat{h}|)/4$.  We have defined $\mathcal{X}=R^4+\phi^2+\psi^2+\theta^2$ and $R$ is the radius of the $S^3$ and enters the dual geometry via the dualisation procedure. The range of the coordinates is given by $0 \leq \eta \leq \pi$, $0 \leq \xi_{1,2} < 2 \pi$ and $\{\phi, \psi, \theta \} \in [0,2\pi]$ are angular coordinates parametrising the three independent one-cycles of a topological three-torus.  

Before moving on to the computation of the associated scalar potential, let us first make a crucial observation \cite{Bugden:2019vlj}: insisting that the coordinates of the non-Abelian T-dual background are all compact, this background becomes non-geometric. From the metric geometry in eq. \eqref{eq:background_fields_NATDS3}, the dual geometry is given in terms of three angular coordinates and the background should respect their associated periodicity. We see however that in particular the factor $1/\mathcal{X}$ spoils this periodicity making the (determinant of the) metric and $B$-field are ill-defined.  Plugging in the above expressions for the dual background we find that under a shift of the three coordinates $\varphi^i \to \varphi^i + 2 \pi$ we have that
\begin{align}
    \mathcal{H}_{IJ}(\phi + 2 \pi, \psi + 2 \pi, \theta + 2 \pi) =  M_I^{\ L} \mathcal{H}_{LK}(\phi,\psi,\theta)M^K_{\ J}\,,
\end{align}
with the monodromy matrix given by
\begin{align}
     M^{K}_{\ L} = \begin{pmatrix}
       I_3 & P \\
       0 & I_3
    \end{pmatrix}\,, \qquad P = \begin{pmatrix}
        0 & 2 \pi & - 2 \pi \\
        - 2 \pi & 0 & 2\pi\\
        2 \pi & - 2 \pi & 0 
    \end{pmatrix}\,.
\end{align}
From this we see that the monodromy can not be interpreted as a gauge transformation of the Kalb-Ramond two-form $B$ nor as a coordinate transformation, rendering the background non-geometric. This is a clear indication that the background is in fact non-geometric as reviewed in section \ref{sec:monodromies}. 

This means, and we will see that this observation will be critical for an harmonious interpretation of the distance conjecture, that we should not work in the usual (NSNS) frame given in terms of metric and $B$-field $(g,B)$ but rather in the non-geometric $\beta$-frame (reviewed in section \ref{sec:beta-gravity}) with background given in terms of the pair $(\tilde{h},\beta)$ which in the particular case are easily obtained to be
\begin{align}\label{eq:S3_NATD_beta_fields}
    \tilde{h} = R^{-2} \mathbb{1}_3\,, \quad \beta  =\begin{pmatrix} 0 &  -\theta &  \psi\\
   \theta & 0 & -\phi\\
    -\psi &  \phi & 0
    \end{pmatrix}\,.
\end{align}
We see that in the $\beta$-frame the background tremendously simplifies and is indeed perfectly well-defined under the periodicity of the coordinates. In conclusion, to globally interpret the geometry one is naturally led to glue using a T-duality, leading to a non-geometric space. Adopting the $\beta$-frame formulation of the background fields, the data is globally smooth. This will be additionally motivated by the matching of the metrics on the dual scalar field spaces. In fact, since the NSNS-frame and $\beta$-frame are equivalent reformulation at the level of the action, one could equally study the background and its property directly in the $\beta$-frame.
The main advantage to consider the $\beta$-frame, is that in this frame the background \eqref{eq:S3_NATD_beta_fields} can be recognised to be a three torus $\hat{T}^3$ with radius $\tilde{R}=1/R$ with one unit of ($Q$)-flux. Formally the resulting equations of motion are equivalent to a ``standard'' three torus with one unit of $H$-flux. As discussed also in section \ref{sec:T-duality_chain} as well as in appendix \ref{App:dilute_flux} such a background is expected to have non-conserved momentum states with mass dependence $\propto \tilde{R}^{-1}$ while winding is expected to be conserved and scaling like with its tension, hence $\tilde{R}$. Using that $\tilde{R} \to 0$ corresponds to $R\to \infty$ we see that indeed now the winding states are the ones providing the light tower at infinite radius while there are no (conserved) towers for $R \to 0$ which translates into missing winding modes on the sphere $S^3$.

%%%%%%%%%%%%%%%%%%%%%%%%%%%%%%%%%%%%%%%%%%%%%%%%%%%%
%%%%%%%%%%%%%%%%%%%%%%%%%%%%%%%%%%%%%%%%%%%%%%%%%%%%
\subsection{Scalar potentials and the distance conjecture}
Having specified the metric of the internal manifolds, we now move on to compactify and determine the associated effective fields theories. Following the procedure given in section \ref{sec:reduction_potential}, we first compute the scalar potential of the scalar field space for the three-sphere and subsequently derive the scalar potential of its non-Abelian T-dual. Viewing the latter as a non-geometric space, the coordinates are naively ill-defined, and performing the integration is impeded by the presence of the monodromy. We will see however that moving to the $\beta$-frame  the integration over the internal coordinates can be performed leading to results in line with the expectations set by T-duality.

%%%%%%%%%%%%%%%%%%%%%%%%%%%%%%%%%%%%%%%%%%%
\subsubsection*{Reduced action and scalar potential for \texorpdfstring{$S^3$}{S3}}
Plugging the expression for the dilaton  \eqref{eq:dilaton_S3_NATDsec} together with the explicit form of the internal metric $h$ into the reduced Einstein action (\ref{equ:reduced_NSNS_2}) we obtain ($\alpha=0$)
\begin{multline}
   S = \frac{\widehat{\mathcal{V}}_{int}}{2 \kappa^2} \int \mathrm{d}^d x \sqrt{-g} \Bigl(  \mathcal{R}(g) - \frac{1}{12} \mathpzc{H}_{\mu\nu\lambda} \mathpzc{H}^{\mu\nu\lambda}e^{\frac{-8 \phi}{d-2}}  - \frac{4}{d-2} (\partial \phi)^2- \frac{3}{R^2} (\partial R)^2   - V(\phi,R)\Bigr)\,,
\end{multline}
with $\widehat{\mathcal{V}}_{int} =  16\pi^2$ and the potential given by 
\begin{align}\label{eq:potS3_NATDsec}
    V(\phi,R)= -\widehat{\mathcal{V}}_{int}^{-1} e^{\frac{4\phi}{d-2}} \int \mathrm{d}^3 y \,\Omega\,\mathcal{R}(h)=-\frac{3}{2R^2}e^{\frac{4\phi}{d-2}}\,.
\end{align} 

%%%%%%%%%%%%%%%%%%%%%%%%%%%%%%%%%%%%%%%%%%%
\subsubsection*{Reduced action and scalar potential for \texorpdfstring{$\hat{T}^3$}{hT3} in \texorpdfstring{$\beta$}{beta}-frame}
Identifying the NATD $S^3$ as a globally non-geometric background, we will treat the description of the background data in terms of $(h,B)$ for one given by $(\tilde{h},\tilde{\beta})$, which is given in terms $(h,B)$. As seen earlier, expressions significantly simplify in terms of the the \linebreak{$\beta$-gravity} fields given in eq. \eqref{eq:S3_NATD_beta_fields}. In what follows we will systematically use the tilde to denote background fields in the $\beta$-frame to differentiate those in the NSNS-frame.
In the $\beta$-frame, the dilaton reads
\begin{align}
\tilde{\Phi}_y = \Phi_y + \frac{1}{4}\mathrm{tr}(\mathrm{ln}(I_d - \tilde{\beta} \tilde{g} \tilde{\beta} \tilde{g}))= - \frac{3}{2} \ \mathrm{ln}(R)= \frac{1}{4} \ \mathrm{ln}(|\tilde{h}|) \,.
\end{align}
We can write the background data for our theory in this frame (including the external space) as
\begin{align}
ds^2_\beta&=\tilde{G}_{IJ}\mathrm{d}X^I\mathrm{d}X^J=e^{\frac{4 \phi_x}{d-2}}g_{\mu \nu}(x) \mathrm{d}x^\mu \mathrm{d}x^\nu + \tilde{h}_{ij}(y) \mathrm{d} y^i \mathrm{d} y^j\,, 
\end{align}
and the dilaton given as
\begin{align}
\tilde{\Phi} = \Phi_0 + \phi + \frac{1}{4} \ \mathrm{ln}(|\tilde{h}|)\,,
\end{align}
again denoting $\Phi_x=\phi$. Note that the above quantities are written with a certain abuse of notation since the line element $ds^2_\beta$ and dilaton $\tilde{\Phi}$ contain both a part that is in the usual NSNS-frame as well as in the $\beta$-frame. This has to be remembered also when doing the reduction procedure. In particular we can still use the formula \eqref{equ:reduced_NSNS_2} by carefully substituting the internal quantities by their $\beta$-frame pendant. 
Again plugging in all the expressions into the reduced Einstein-Hilbert action given in \eqref{equ:reduced_NSNS_2} and using the defining relation for the $\beta$-frame in eq. \eqref{eq:SNSNS_to_beta_fields}, we arrive at
\begin{multline}
S=\frac{\widehat{\mathcal{V}}_{\beta, int}}{2 \kappa^2} \int \mathrm{d}^d x  \sqrt{|g|} \Bigl( \mathcal{R}(g) -\frac{1}{2} \mathpzc{H}_{\mu\nu\lambda}\mathpzc{H}^{\mu\nu\lambda} e^{\frac{-8 \phi}{d-2}}  - \frac{4}{d-2} (\partial \phi)^2
- \frac{3}{R^2} (\partial R)^2  - V(\phi,R)\Bigr)\,,
\end{multline}
with $\widehat{\mathcal{V}}_{\beta, int} = (2\pi)^3$. From the review in section \ref{sec:potential_nongeom} and in particular using  eq. \eqref{eq:potential_Q-flux_bkrgd} for the scalar potential in terms of the non-geometric $Q$-flux, we obtain  
\begin{align}\label{eq:NATD_potential_beta}
\widetilde{V}(\tilde{\phi},R)&= -\widehat{\mathcal{V}}_{\beta,int}^{-1} e^{\frac{4\phi}{d-2}} \int \mathrm{d}^3y \tilde{\Omega}\Bigl\{-\frac{1}{4} \mathpzc{Q}^2 - \frac{1}{2} \tilde{h}_{ij} \mathpzc{Q}_k^{\ lj}\mathpzc{Q}_l^{ki} \Bigr\} = -\frac{3}{2 R^2}e^{\frac{4\phi}{d-2}}\,,
\end{align}
where the defining relations for the involved quantities can be found in section \ref{sec:beta-gravity}.
We see that this indeed precisely agrees with the potential derived in the original $S^3$-background in eq. \eqref{eq:potS3_NATDsec} and in this example, the metrics on the moduli spaces of the T-dual internal spaces agree. Critically, this would not have been the case had we used that standard NSNS Lagrangian. We show explicitly in appendix \ref{App:totdal_deriv} how the NSNS and $\beta$-frame match upon using the total derivative mentioned in section \ref{sec:beta-gravity}.

We can furthermore explicitly check that the expressions for the scalar potential of the original and NATD spaces indeed fit into the scaling of the formula \eqref{eq:pot_scaling} expressing the scalar potential in terms of the volume modulus $\rho$ and the external dilaton $\sigma$. For the $S^3$ background and its NATD, we can compare eq. \eqref{eq:pot_scaling} with  eq. \eqref{eq:metric_S3}, respectively \eqref{eq:S3_NATD_beta_fields}, to read off
\begin{equation*}
    \rho_{S3} = R^2\,,\qquad  \tilde{\rho} \equiv \rho_\mathrm{NATD} = R^{-2}\,.
\end{equation*}
Note that the second volume modulus, like stressed already several times, also this quantity is only well-defined in the $\beta$-frame and the large volume limit is well-defined in this duality frame. See \cite{Andriot:2011uh} for a discussion.
Now recalling that for $S^3$ the only non-zero contribution was coming for the Ricci scalar, i.e $V_f$ in eq. \eqref{sec:nongeom_NATDS3} while for the non-Abelian T-dual the potential is sourced from the $Q$-flux. Therefore eq. \eqref{eq:pot_scaling} gives 
\begin{align*}
     V_{S^3} &= \sigma^{-2}\rho^{-1}V_f^0 \sim  R^{-2}\,, \qquad  V_\mathrm{NATD} = \tilde{\sigma}^{-2} \tilde{\rho} V_Q^0 \sim  R^{-2}\,,
\end{align*}
and we retrieve the potential given earlier in section \ref{sec:potential_nongeom}.

%%%%%%%%%%%%%%%%%%%%%%%%%%%%%%%%%%%%%%%%%%%
%%%%%%%%%%%%%%%%%%%%%%%%%%%%%%%%%%%%%%%%%%%
\subsection{Distance conjecture}

The properties of the three-sphere  and its non-Abelian T-dual relevant to the SDC collected in the previous section are summarised in table \ref{tab:NATDS3}. We saw here that starting from the very symmetric three-sphere, the resulting background obtained after a non-Abelian T-duality transformation displays no isometries. The background remains however tractable, and the zero modes and potential can be computed. 
The example considered in the last section already discussed the three-sphere and we will not repeat it here. The new element here is the non-Abelian T-dual of the three sphere. The properties of the non-Abelian T-dual were computed using Poisson-Lie T-duality and by taking the space to be non-compact at the price of turning it into a non-geometric space. The interpretation of the background as being non-geometric was crucial to see a matching of the potentials on the original background and its non-Abelian T-dual. Using Poisson-Lie T-duality we could trace back the exchange of the zero modes under the duality transformation, leading to a background with three sets of winding zero-modes but plagued by an absence of momentum zero modes. The latter is directly related to the lack of isometries in the non-Abelian T-dual background.

From table \ref{tab:NATDS3}, we see that the three-sphere and its non-Abelian T-dual perfectly fit into the expectations set by the SDC but again only \textit{provided}  we take into account the potential which signals a pathological behaviour in certain directions. The potential of the non-geometric non-Abelian T-dual background displays a divergence precisely in accord with the absence of zero modes which could give rise to a tower, here the momentum modes. 

\begingroup
    \setlength{\tabcolsep}{9pt}
    \renewcommand{\arraystretch}{1.7}
\begin{table}[t]
\centering
  \begin{tabular}{c|c|c|c}
  {Background} &  Potential $V$ &   Obstructed modes & Frame \\ \hline \hline
     $S^3$ and $\mathpzc{H}=0$ & $-\frac{3}{2R^2}$ &  absence of winding & NSNS \\
  NATD $S^3$ and $[\mathpzc{H}]\neq 1$ & $-\frac{3\tilde{R}^2}{2}=-\frac{3}{2R^2}$  &  absence of momentum & $\beta$
  \end{tabular}\caption{The basic data of the three-sphere with trivial $H$-flux and its non-Abelian T-dual. The last column refers to the ``frame'' in which the background is globally well-defined: in the conventional NSNS-formulation or the $\beta$-supergravity formulation and where the dual and original radius are related through the usual relation $\tilde{R}=R^{-1}$.}\label{tab:NATDS3}
\end{table}
\endgroup

%%%%%%%%%%%%%%%%%%%%%%%%%%%%%%%%%%%%%%%%%%%%%%%%%%%%%
%%%%%%%%%%%        T-duality chain         %%%%%%%%%%
%%%%%%%%%%%%%%%%%%%%%%%%%%%%%%%%%%%%%%%%%%%%%%%%%%%%%
\section{T-duality, infinite distance and the non-geometry chain}\label{sec:T-duality_chain}
In this section we will consider the SDC for backgrounds with internal manifolds one of the backgrounds in the non-geometric T-duality chain, which were already briefly introduced in section \ref{sec:intro_nongeomchain}. That is, we will take as total space $M_D=M_d\times K_n$ where the compact part $K_n$ will be any of the following backgrounds with fluxes 
\begin{align}\label{eq:H-torus-chain}
   \begin{matrix}
       H\text{-torus}\\
       \mathpzc{H}_{ijk}
   \end{matrix} \quad \xleftrightarrow[]{\; \; y_3\; \; } \quad \begin{matrix}
       \text{tw-torus}\\
       \mathpzc{f}^{i}{}_{jk}
   \end{matrix}\quad   \xleftrightarrow[]{\; \; y_2\;\;  } \quad \begin{matrix}
       Q\text{-flux}\\
       \mathpzc{Q}^{ij}{}_{k}
   \end{matrix}\,\,.
\end{align}In each of these backgrounds we have a $T^2$ conformal field theory fibered over a circle base, for which, upon going once around the base, the fiber picks up a monodromy valued in the T-duality group $O(2, 2; \mathbb Z)$ as outlined in section \ref{sec:monodromies}.
Each background will be treated in detail below. We will not discuss the R-flux background which would arise by performing a last T-duality on the last rug of this sequence.

In this section we will find a tension between the non-geometric T-duality chain and the SDC: the absence of certain zero modes and associated towers cannot be explained, even after including a potential. One has to keep in mind here that we restrict the discussion to the simplest incarnation of the T-duality chain and do no consider any other towers than those coming from momentum and winding modes. We leave these generalisations to future work \cite{nextup}.

\begin{figure}[t]
	\centering
	\includegraphics[scale=0.35]{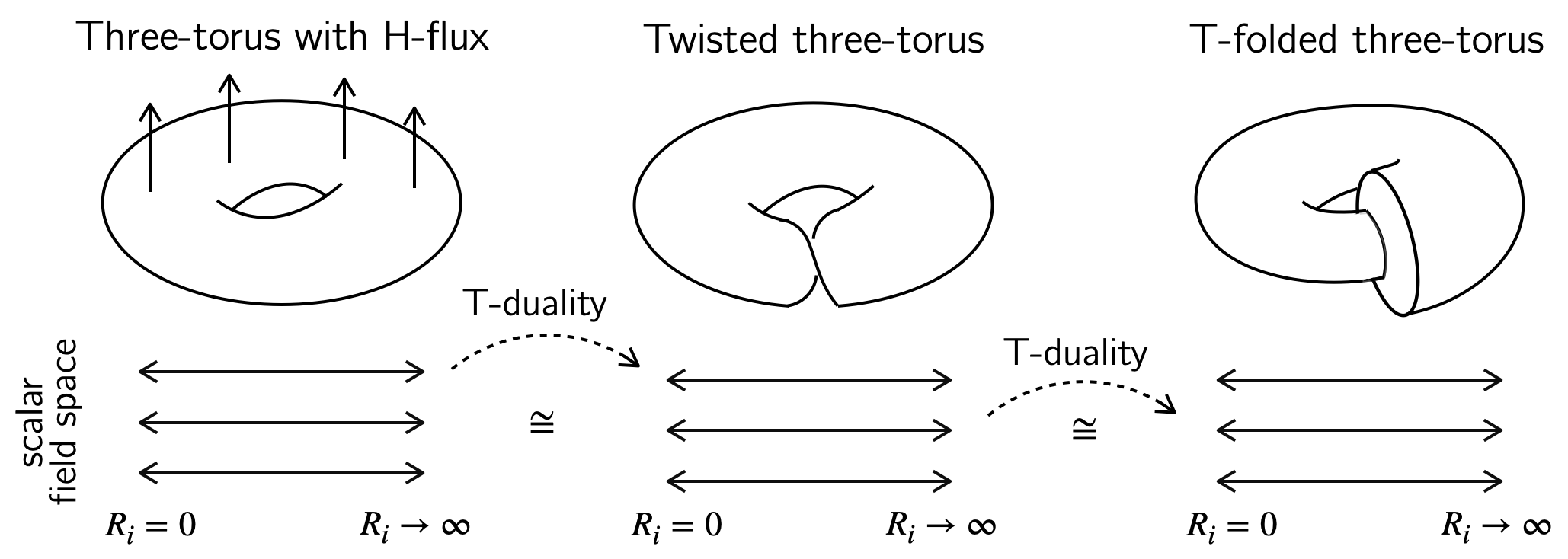}
	\caption{Illustration of the change in global structure after T-dualisation. Although the scalar field space of T-dual space are diffeomorphic, the corresponding target space are not. Note that the the base circle of each (schematically illustrated) background has been suppressed.}\label{fig:moduli_spaces_chain}
\end{figure}

Before moving on, let us comment on the scalar fields following the observations made in section \ref{sec:generalised_Tdualities}. Upon T-dualising a given theory, its target space is mapped together with its scalar field space to a different but (canonically) equivalent background and moduli space respectively. Crucially though, although equivalent, the corresponding target spaces that these describe may differ dramatically featuring different local and global geometrical properties. This phenomenon is succinctly exemplified by the T-duality chain. For each of the T-dual backgrounds, the moduli spaces are diffeomorphic. The target spaces however have different global properties and are not diffeomorphic. After one duality, the cycles will be non-trivially intertwined, see eq. \eqref{eq:twist_twT3}. Applying a second duality, one of the cycle will be locally glued together by an $O(d,d;\mathbb Z)$-transformation. This is illustrated in figure \ref{fig:moduli_spaces_chain}.

%%%%%%%%%%%%%%%%%%%%%%%%%%%%%%%%%%%%%%%%%%%%%%%%%%%%%
%%%%%%%%%%%%%%%%%%%%%%%%%%%%%%%%%%%%%%%%%%%%%%%%%%%%%
\subsection{Zero-modes, mass spectrum and scalar potentials}
The three-dimensional space of the duality chain is embedded  into a higher-dimensional space with some generic external space $M_d$ such that the total metric reads
\begin{align}\label{eq:generalmetric}
ds^2&=G_{IJ}\mathrm{d}X^I\mathrm{d}X^J=e^{\frac{4\phi}{d-2}}g_{\mu \nu}(x) \mathrm{d}x^\mu \mathrm{d}x^\nu + ds_\mathrm{int}^2 \,, 
\end{align}
where $\hat{g}_{\mu \nu}(x)\equiv e^{\frac{4\phi}{d-2}}g_{\mu \nu}(x) $ denotes the metric on the external space $M_d$.  Before delving into the details, let us summarise the resulting potentials in table \ref{tab:Tdualitychain_pots}. Here we have adopted the following notation. The tilde radii $\tilde R_i$ denote the natural radii in each frame. For example, when performing a first T-duality along  the $x_1$ direction, the radius $R_1$ is identified with the new radius $\tilde R_1=1/R_1$. As we will see, in the original moduli the potential remains invariant, in line with what can be expected from T-duality.

%%%%%%%%%%%%%%%%%%%%%%%%%%%%%%%%%%%%%%%%%%%%%%%%%%%%
\subsubsection{Torus with \texorpdfstring{$H$}{H}-flux}
With as starting point the three-torus with three-from flux, denote the three angular coordinates on the torus by $y_i$ for $i\in \{1,2,3\}$. Then the internal metric we  consider is simply 
\begin{align}\label{eq:T_H_ds}
    \mathrm ds_\mathrm{int}^2 = h_{ij}\mathrm{d}y_i y_j = \sum_i R_i^2 (\mathrm{d}y_i)^2\,.
\end{align}
embedded as the internal part of a ten-dimensional background and we fix the choice of $B$-field for the $H$-flux to be
\begin{align}\label{eq:T_H_B}
     B=h y_3\mathrm d y_1\,\wedge \mathrm d y_2\,. 
\end{align}
together with a (internal) dilaton that can at most depend on $y_3$, i.e $\Phi_y=\Phi_3(y_3)$. We choose the full dilaton to be 
\begin{align}
    \Phi =\phi+ \frac{1}{2}\log(R_1R_2R_3)  + \Phi_3\,,
\end{align}
where again $\phi=\phi(x)$ is the dilaton for the external space and the constant part is such that $\sqrt{(\det(h))}e^{\Phi} \propto R^0$.
These fields then enter via the higher dimensional line element \eqref{eq:generalmetric} into the $D$-dimensional Einstein-Hilbert action for the metric $G$ including an H-flux $\mathpzc H$ and dilaton $\Phi$. Substituting into the multivariable analogue of the reduction formula \eqref{equ:reduced_NSNS_multivar} on obtains
\begin{align*}
S&=  \frac{\mathcal{V}_{int}}{2\kappa^2} \int \mathrm{d}^d x \sqrt{-g}\Bigl( \mathcal{R}(g) -\frac{1}{12} \mathpzc{H}_{\mu\nu\lambda} \mathpzc{H}^{\mu\nu\lambda} e^{-\frac{8 \phi}{d-2}}- \frac{4}{d-2}(\partial \phi)^2 - \sum_i R_i^2 (\partial R_i)^2 -V(\phi,R_i)\Bigr)\,,
\end{align*}
with $\mathcal{V}_{int} =(2\pi)^3$ is the volume of the internal space and where the potential is given by 
\begin{align*}
    V(\phi,R_i) =-\widehat{\mathcal{V}}_{int}^{-1} e^{\frac{4\phi}{d-2}} \int \mathrm{d}^3 y \,\Omega\, \left( 4 \partial_i \Phi_3 \partial^i \Phi_3 -\frac{1}{12}\mathpzc{H}_{ijk} \mathpzc{H}^{ijk} \right)\equiv -V(\Phi_3) +\frac{h^2}{2} \frac{1}{R_1^2R_2^2R_3^2}e^{\frac{4\phi}{d-2}}\,.
\end{align*}
Hence yielding a potential that diverges as the scalar fields $R_i$ approach zero.

\bgroup
\def\arraystretch{1.8}
\begin{table}[t]
\centering
\begin{tabular}{c|c|c|c}
\textbf{}                         & {$T^3$ with $H$-flux} & {Twisted three-torus} & { Q-flux three-torus} \\ \hline \hline
 Ricci scalar $\mathcal{R}$                     & $-$                          &  $\checkmark$         & $-$                        \\ \hline
Flux $-\frac{1}{2}\mathpzc{H}^2$                 & $\checkmark$              & $-$                    & $-$                        \\ \hline
Flux $-\frac{1}{2}\mathpzc{Q}^2$                 & $-$                          & $-$                    &  $\checkmark$             \\ \hline \hline
potential $ V(\vec R)$ & $+\frac{h^2}{2} (R_1^2R_2^2R_3^2)^{-1} $            & $  +\frac{h^2}{2} \tilde R_1^2 (R_2^2R_3^2)^{-1}$      & $ +\frac{h^2}{2} \tilde R_1^2 \tilde R_2^2 (R_3^2)^{-1}$
\end{tabular}\caption{Summary of the potentials for the different backgrounds in the T-duality chain. A check mark indicates which fluxes contribute to sourcing the potential. The tilde $R$'s denotes a T-dualised parameter but defined in such a way that it is the natural radii in that particular frame. Note that since the dilaton do not contribute in a significant way to the scalar potential, they are not included in this table.}\label{tab:Tdualitychain_pots}
\end{table}

%%%%%%%%%%%%%%%%%%%%%%%%%%%%%%%%%%%%%%%%%%%%%%%%
\subsubsection{Twisted torus or \texorpdfstring{$f$}{f}-flux background}\label{sec:nilmanifold}
T-dualising the background fields of the three-torus with H-flux given in eqs. \eqref{eq:T_H_ds} and \eqref{eq:T_H_B} along the $x_1$-direction, the internal metric  becomes
\begin{align}\label{eq:metric_tw_torus}
    \mathrm d s^2= \sum_{i=1}^3 (\eta_i)^2=
   \frac{1}{R_1^2} \mathrm{d}y_1^2 + R_2^2\left(1+\frac{h^2y_3^2}{R_1^2 R_2^2}\right)\mathrm{d}y_2^2- 2 \frac{h y_3}{R_1^2}\mathrm{d}y_1 \mathrm{d}y_2+R_3^2 \mathrm{d}y_3^2 \,,
\end{align}
where we have used the one-forms $\eta_1=1/R_1(\mathrm dy_1+hy_3\mathrm d y_2)$, $\eta_2=R_2\mathrm dy_2$, $\eta_3=R_3\mathrm dy_3$. The background is a compact manifold with discrete identification given by
\begin{align}\label{eq:twist_twT3}
y_1\sim y_1+n_1-n_3hy_2\,,\qquad y_2\sim y_2+n_2 \,,\qquad 	y_3\sim y_3+n_3\,, \quad n_{1,2,3}\in \{0,1\} \,.
\end{align}
This is an example of a so-called nilmanifold: it can be identified via this lattice action to be the quotient of a nilpotent group by one of its discrete subgroups. After performing the T-duality transformation, the the $B$-field vanishes while the dilaton remains unchanged. Instead the geometry features a ``geometric''-flux under the form of a non-vanishing  spin connection (solving the Cartan structure equation \textit{with} torsion) \cite{Kachru02}
\begin{align}
    \omega= h \eta_1\wedge \eta_2 \wedge \eta_3 \,.
\end{align}
 The vielbeins verify the Maurer-Cartan equation
\begin{align}
	\mathrm d \eta_3= \mathpzc{f}^1{}_{23}\eta_2\wedge\eta_3\,,
\end{align}
where $\mathpzc{f}^1{}_{23}$ is the sole non-zero component of the geometric flux characterising this background.

Turning to the dilaton, it transforms under the T-duality transformation to 
\begin{align}
    \Phi=\phi + \frac{1}{2}\log\left(\frac{R_2R_3}{R_1}\right) + \Phi_3\,,
\end{align}
such that indeed again $\sqrt{(\det(h))}e^{\Phi} \propto R^0$.
Therefore performing a compactification and using the reduction formula \eqref{equ:reduced_NSNS_2} we obtain the potential, which this time however, in absence of a Kalb-Ramond field, is sourced by the curvature/geometric flux and given by
\begin{align*}
    V(\phi,R_i) &= - \widehat{\mathcal{V}}_{int}^{-1}e^{\frac{4\phi}{d-2}} \int \mathrm{d}^3 y\,\Omega \, \left(\mathcal R + 4 \partial_i \Phi_3 \partial^i \Phi_3\right)=-V(\Phi_3) +\frac{h^2}{2} \frac{ \tilde{R}_1^2}{  R_2^2R_3^2}e^{\frac{4\phi}{d-2}}\nonumber\\
    &=-V(\Phi_3) +\frac{h^2}{2} \frac{ 1}{ R_1^2 R_2^2R_3^2}e^{\frac{4\phi}{d-2}}\,.
\end{align*} 
Here, as before, the tilde stands for the radius in the current frame, which equals via \linebreak{T-duality} to the inverse of the original radius, i.e. $\tilde R_1=1/R_1$. Note that again, the potential is preserved under T-duality.

%%%%%%%%%%%%%%%%%%%%%%%%%%%%%%%%%%%%%%%%%%%%%%%%
\subsubsection{\texorpdfstring{$Q$}{Q}-flux background}
Dualising again but now along the $y_2$-direction, the new line element reads
\begin{gather}
\begin{aligned}
    \mathrm d s^2= \left(1+\frac{h^2y_3^2}{R_1^2 R_2^2}\right)^{-1}&\left(R_1^{-2}\mathrm d y_1^2+R_2^{-2}\mathrm d y_2^2\right) + R_3^2 \mathrm d y_3^2\,,\\
   \quad B=\frac{hy_3}{R_1^2R_2^2+h^2y_3^2}\mathrm d x_1\wedge \mathrm d y_2&\,,\quad  \Phi_y=\Phi_3-\frac{1}{2}\log\left[R_2^2\left(1+\frac{h^2y_3^2}{R_1^2 R_2^2}\right)\right]\,.
\end{aligned}
\end{gather}
The background (in the conventional $B$-frame) is locally not well-defined and features a non-trivial monodromy, as was already briefly reviewed in section \ref{sec:monodromies}. Moving along the fiber $y_3\mapsto y_3+1$, the generalised metric picks up a monodromy mixing the volume density, i.e. determinant of the metric $\sqrt{g}$, and the $B$-field. Alternatively, one can show that the supergravity Lagrangian will yield a total derivative, rendering the whole action not single-valued, signalling a non-geometric background.

To address this ill-definiteness that would hinder one to perform the necessary integrations in the reduction of the Einstein-Hilbert action, we adopt as before, the strategy of moving to the $\beta$-frame.  Transforming to the $\beta$-frame, the metric becomes diagonal with $\beta$-field and dilaton field
\begin{gather}
    \mathrm d \tilde s^2=R_1^{-2}\mathrm{d}y_1^2 +R_2^{-2}\mathrm{d}y_2^2 +R_3^2 \mathrm{d}y_3^2\,,\qquad \beta = -h y_3 \mathrm d y_1\wedge \mathrm d y_2\,,\nonumber\\ \tilde \Phi_y= \Phi_3+\frac{1}{2}\log(\frac{R_3}{R_1R_2})\,.
\end{gather}
Using the expression for the $\mathpzc{Q}\,$-flux in terms of the $\beta$-field in \eqref{eq:Qflux_ito_beta}, the non-vanishing components of the non-geometric Q-flux for this background are
\begin{align}
\mathpzc Q\,_3{}^{12}=-\mathpzc Q\,_3{}^{21}=-h\,.  
\end{align}
These fields now define globally well-defined $\beta$-supergravity Lagrangian, allowing us to perform the reduction with the internal part now expressed in the $\beta$-frame. 
As a result, it is now the $Q$-flux that sources the scalar potential
\begin{align*}
    \tilde{V}(\phi, R_i) &= -\widehat{\mathcal{V}}_{\beta,int}^{-1}e^{\frac{4\phi}{d-2}}  \int \mathrm{d}^3 y \,\tilde{\Omega}\, \left( 4 \partial_i \Phi_3 \partial^i \Phi_3 -\frac{1}{4}\mathpzc{Q}_{\ i}^{\ jk} \mathpzc{Q}^i_{\ jk} \right) =-V(\Phi_3)  +\frac{h^2}{2} \frac{ \tilde R_1^2 \tilde R_2^2}{R_3^2}e^{\frac{4\phi}{d-2}}\nonumber\\
    &=-V(\Phi_3) +\frac{h^2}{2} \frac{ 1}{ R_1^2  R_2^2 R_3^2}e^{\frac{4\phi}{d-2}}\,.
\end{align*} 
We observe here again that the potential remains invariant under the T-duality transformation.

%%%%%%%%%%%%%%%%%%%%%%%%%%%%%%%%%%%%%%%%%%%%%%%%%%%%%
\subsubsection{Zero modes and mass spectra}\label{subsec:zero_modes_duality_chain}
Having computed the potential, we now determine the presence or absence of zero-modes that would source possible towers of light states as well as providing a brief discussion of the corresponding mass spectra.

\newpage
%\vspace{10pt}
\noindent
\textit{Zero modes}

\vspace{3pt}
\noindent
Considering first the three-torus with $H$-flux,  we encounter a situation where due to the presence of a non-trivial $H$-flux, the momentum is found to be conserved only modulo $k$. This phenomenon is very similar to the familiar cyclotron motion of a particle moving in a (constant) magnetic field. In particular, by exploiting the dilute flux approximation\footnote{For large values of the radius we apply the so called dilute flux approximation, that is $\mathpzc{H}/\mathrm{Vol}\ll 1$. In this approximation one can then perform a field expansion in linear order of the flux, effectively treating the three-form flux $\mathpzc{H}$ as a perturbation on a flat space CFT. This enables the use of usual CFT techniques, at least in a perturbative scheme. The resulting perturbative theory was therefore also dubbed CFT$_{\mathpzc{H}}$.}, in \cite{Blumenhagen:2011ph} it was shown that whenever the momenta and the winding modes on the three sphere have parallel components, the momenta are no longer conserved, schematically
\begin{align}\label{eq:SphereLens_nonconserv}
   \dot X \propto \mathpzc{H} \quad \text{whenever}\quad \vec{p}\parallel \vec{w}\,,
\end{align}
 where the indices $i,j$  label the components of the winding and momentum numbers, $w$ and $N$ respectively.  The precise form of  \eqref{eq:SphereLens_nonconserv} and its derivation is given in appendix \ref{App:dilute_flux}, together with a concise review to the dilute flux approximation. We thus expect that the physical momenta are no longer conserved and will form an obstruction to the existence of a corresponding tower of states. 
On the other hand, since the three torus counts three non-contractible cycles of $T^3$,  we have a $\pi_1(T^3)=\mathbb Z^{\oplus 3}$-worth of winding zero modes \cite{Kachru02}.

  Moving on to the twisted torus, in absence of $H$-flux or any other obstruction the momentum zero modes form a $\mathbb Z^{\oplus 3}$-lattice. The twisted torus has a non-trivial homology structure \cite{Kachru02} (and see \cite{Marchesano:2006ns} for a detailed description)
\begin{align}
    H_1(T^3_\mathrm{tw},\mathbb Z)=\mathbb{Z}\oplus \mathbb{Z}\oplus \mathbb{Z}_k\,.
\end{align}
 The presence of the torsion cycle shows that the corresponding cycle is not adequate to realise a tower of states associated to winding modes since quanta vanish after $N$ windings.  Note that this gives us an alternative derivation, without having to resort to a perturbative calculation on the T-dual side, that in the torus with H-flux, one of the momenta is only conserved modulo $k$. A similar phenomena arise when considering D3 brane in twisted $T^6$-background, although its charges are non conserved it is however BPS \cite{Marchesano:2006ns}.

The final background, the $Q$-flux background, is somewhat more arduous to discuss as the topology and differential geometry of $Q$-flux backgrounds, are more generally \linebreak{T-folds}, remain mostly ill-understood. Instead, we will choose to reverse the logic here. Espousing the extension to the SDC proposed in the earlier sections and inferring from it the topological and differential properties of the $Q$-flux background. We thus postpone dealing with the zero-modes and the spectrum of this background to the discussion of the SDC in the last part of this section.

\vspace{10pt}
\noindent
\textit{Mass spectrum}

\vspace{3pt}
\noindent
The spectra of low-energy effective field theories associated to the different backgrounds in the T-duality chain and relating them to the momentum zero modes are deceptively difficult to access. In this section we will mainly point out the different issues one faces. As we will see in the discussion of the SDC, the problem will already lie in the (non-)existence of certain modes.

Starting with the $H$-torus, using a reasoning similar to that of the three-sphere with $H$-flux in section \ref{sec:threesphereH_zeromodes_spectrum_pot}, we can similarly infer that at large values of $R$ one can discard contribution from the $H$-flux. As a result, the KK-spectrum coincides, at large values of the radius, to that of a conventional flat torus.

Turning now to the twisted torus, the manifold is already significantly different in many aspects compared to the torus with $H$-flux. This can in particular be seen in its Laplacian spectrum, which was computed in \cite{Andriot:2018tmb} and where it was shown to display an unusual gap between a number of light states and the rest of the spectrum. As pointed out by the authors, the eigenforms of the Laplacian for the nilmanifold obtained there forms a perfect starting point to compute the spectrum of the low energy effective theory, possibly having to take into account so-called space invaders \cite{Duff:1986hr}. This goes however beyond the scope of this paper, and we will instead assume that the spectra follow the pattern dictated by T-duality, see table \ref{table:T-duality_chain_DC}.

Finally, for the non-geometric $Q$-flux note that in the $\beta$-frame the problem is very similar to that of the $H$-torus. Indeed, there the metric is flat. The presence of the $Q$-flux does however twist the conventional differential to become \cite{Shelton:2006fd,Ihl:2007ah,Andriot:2014uda} $\mathrm d_{\mathpzc Q}$. At large values of the radius, this modification to the differential can be neglected. It would be interesting to investigate if one can equally carry out the identification and determination of the \linebreak{KK-spectrum} in the $\beta$-frame. Here however again, we will use T-duality as guiding principle when discussion the SDC.

%%%%%%%%%%%%%%%%%%%%%%%%%%%%%%%%%%%%%%%%%%%%%%%%%%%%%
%%%%%%%%%%%%%%%%%%%%%%%%%%%%%%%%%%%%%%%%%%%%%%%%%%%%%
\subsection{Distance conjecture}

Having tabulated the presence and absence of zero modes (see table \ref{table:T-duality_chain_DC}),  the infinite distance points in the respective duality frames together with the scalar potentials, we are now in a position to discuss the T-duality properties of this chain of backgrounds in the light of the SDC.

The infinite distance points are located once again at $R_i=0$ and $R_i \rightarrow \infty$ of the respective cycles. At the same time, we saw that the zero modes, and in particular the winding modes, are not always conserved. Looking at the potentials in table \ref{tab:Tdualitychain_pots}, all the backgrounds display a divergent potential along (a single cycle) at zero radius\footnote{From the torus with H-flux the absence of winding modes and the necessity for the divergent potential is no apparent since there is no obvious obstruction for the existence of winding modes, c.f however the discussion in section \ref{subsec:zero_modes_duality_chain}.}. We thus seem to retrieve a similar pattern to that of the previous examples, at least for going to $R\rightarrow 0$. Turning to the other limit, first note that the $H$-flux or $Q$-flux spoil the conservation of at least one set of momentum zero-modes and hence these particular mode cannot give rise to a tower of states. The potential however only protects from the possible absence of winding modes. This is indicated in the figure embedded in table \ref{table:T-duality_chain_DC}. The regular behaviour of the potential at large values of the radii $R_i$ is therefore at odds with the absence of momentum zero modes in $H$-torus. This leads us to conclude that, without supplementing the solution with additional fluxes, the backgrounds in the duality chain cannot serve as the internal spaces for a low-energy effective realisation of a theory of quantum gravity. We plan to consider this scenario in the near future \cite{nextup} as one may expect that the new fluxes will lead to a different scalar potential or the possibility for new towers.

\begingroup
    \begin{figure}[t]
        \centering\hspace{2,3cm}
        \includegraphics[scale=0.312]{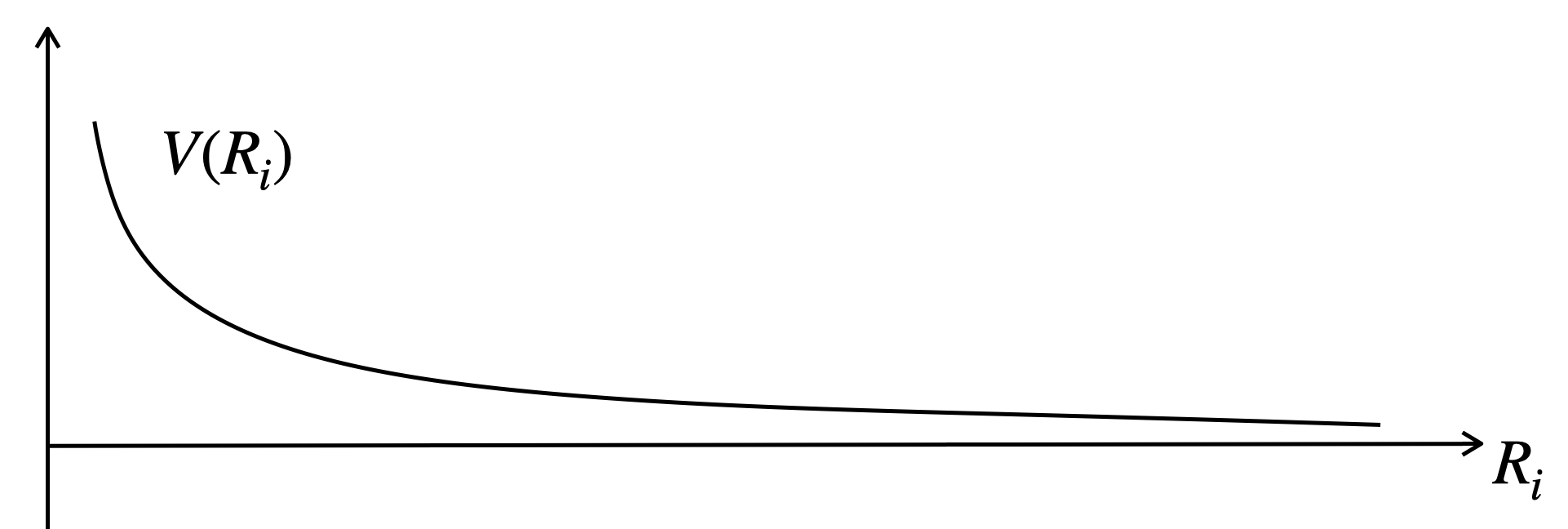}
    \end{figure}
    \setlength{\tabcolsep}{5pt}
    \renewcommand{\arraystretch}{1.3}
\begin{table}[t]
\centering
  \begin{tabular}{c|c|c|c}
    & modes &  $R_i\rightarrow 0$ & $R_i\rightarrow \infty$ \\ \hline \hline
      \multirow{2}{*}{ $\underset{\{R_1,R_2,R_3\}}{T^3_{\mathpzc H}}$, $[\mathpzc H]=k$}  & $w: \,\mathbb Z\oplus\mathbb Z\oplus \mathbb Z$  &$w:$ (light, light, light) &  $w:$ (heavy, heavy, heavy)  \\
     &$p:  \,nc \oplus nc \oplus nc$ & $p:$ ( \cancel{heavy}, \cancel{heavy}, \cancel{heavy}) & $p:$ (\cancel{light}, \cancel{light}, \cancel{light})  \\ \hline 
      \multirow{2}{*}{$\underset{\{R_1,R_2,R_3^{-1}\}}{T^3_{\mathrm{tw}}}$, $[\mathpzc{f}]=k$}    & $w:  \,\mathbb Z\oplus\mathbb Z\oplus \mathbb Z_k$ & $w:$ (light, light,  \cancel{heavy}) &   $w:$ (heavy, heavy, \cancel{light})    \\
         &$p:  \,nc \oplus nc \oplus  \mathbb Z$ & $p:$  (\cancel{heavy}, \cancel{heavy}, light)  & $p:$(\cancel{light}, \cancel{light}, heavy)\\ \hline 
      \multirow{2}{*}{$\underset{\{R_1,R_2^{-1},R_3^{-1}\}}{T^3_{\mathpzc{Q}}}$, $[\mathpzc Q] = k$} &  $w:  \,\mathbb Z \oplus \mathbb Z_k \oplus \mathbb Z_k$    &  $w:$ (light,  \cancel{heavy},  \cancel{heavy}) &  $w:$ (heavy, \cancel{light}, \cancel{light})  \\
         &$p:  \,nc \oplus \mathbb Z \oplus  \mathbb Z$& $p:$  (\cancel{heavy}, light, light)  & $p:$ (\cancel{light}, heavy, heavy)
    \end{tabular}
     \caption{The graph indicates the divergent scalar potential as a function of the scalar fields $R_i$. The limits are to be understood for a single $R_i$ with the remaining two kept fixed. The radii are defined with respect to the three cycles of the torus with H-flux in the first line. The table indicates the zero modes potentially sourcing a tower of light states for the corresponding asymptotic value of the radius $R_i$. Crossed \cancel{expressions} denote non-conserved ($nc$) or only finite number of conserved quantities. For large values of $R_i$, where the momentum modes become very light, we distinguish between the actual $R$ is infinite point and the value approaching that infinite point. For the latter, the theory becomes effectively free, under the torsion contribution vanishes. Note that the $\mathpzc Q$-flux background has to be treated with care, as it has been inferred by T-duality arguments rather than be direct computations. Finally, the masses for the momenta are speculative and would require dedicated computations of the associated spectra, as explained in the main text.}
    \label{table:T-duality_chain_DC}
\end{table}
\endgroup

%%%%%%%%%%%%%%%%%%%%%%%%%%%%%%%%%%%%%%%%%%%%%%%%%%%%%
%%%%%%%%%%%           Discussion           %%%%%%%%%%
%%%%%%%%%%%%%%%%%%%%%%%%%%%%%%%%%%%%%%%%%%%%%%%%%%%%%
\section{Future directions}\label{sec:discussion}
 The main motivation behind this work was the exploration of the relation between the Swampland distance conjecture with T-duality when the internal manifold is curved or carries non-trivial fluxes. Having summarised the main results in the section \ref{sec:summary_results}, we will conclude here with a number of future directions.

\pagebreak
\noindent
\textbf{Tackling more realistic examples}

\vspace{3pt}
\noindent
The examples treated here are relatively simple, and the result drawn from them have to be treated with care. Addressing this apparent impediment requires studying more general and richer curved backgrounds with and without supporting (non-geometric) fluxes, most notably
\begin{itemize}
    \item \textit{KK-monopole and NS5 non-geometry chain.} The more involved sibling of the non-geometric T-duality chain discussed in section \ref{sec:T-duality_chain} is the chain of T-dualities with as starting point the NS5-brane background. The ensuing T-duals share similar but considerably more intricate features: non-conservation of zero modes (which is lifted by instantonic corrections \cite{Gregory:1997te,Harvey05,Tong:2002rq}) and non-geometry.
    \item \textit{The type IIB $T^6/\mathbb Z_2$ orientifold.} Yet another interesting background is the type IIB $T^6/\mathbb Z_2$-background with the choice of solution to the tadpole condition introducing an NSNS flux and its T-duals \cite{Kachru02}. One could wonder what the synergy is between the zero-modes, potential and distance conjecture in this example which does verify the tadpole condition, contrary to the T-duality chain. This (or closely related) backgrounds have recently already been considered in the context of the Swampland Distance Conjecture in \cite{CaboBizet:2019sku} and  tadpole cancellation and moduli stabilisation in \cite{Betzler:2019kon}.
\end{itemize}
We plan to report in these examples and their place within the SDC in the near future \cite{nextup}.

\vspace{10pt}
\noindent
\textbf{Addressing other extensions and puzzles within the SDC}

\vspace{3pt}
\noindent
Notably, in many of the examples presented here, we encountered issues closely related to very recent extensions and puzzles around the SDC
\begin{itemize}
    \item {\textit{Flux variations.}} As recently discussed in \cite{Li:2023gtt}, extracting the metric on moduli space requires a subtle interplay between on- and off-shell arguments. In the particular example of the supergravity background formed by the three-sphere and associated three-form flux, T-duality offered an additional motivation to why it is critical to better understand this phenomenon. It would thus be interesting to see if (generalised) T-duality can offer additional insights when applied the Freund-Rubin backgrounds supported by RR-fluxes considered in \cite{Li:2023gtt} withing this context. 
     
    \item\textit{Higher dimensional scalar field spaces and SUSY.}
     The backgrounds considered here are controlled by a single parameter\footnote{In section \ref{sec:T-duality_chain}, although each cycle carried its own modulus generating a three-dimensional scalar field space that space is flat and the direction are effectively independent.} and are purely bosonic. This begs the question to whether expanding the discussion to higher dimensional parameter space and supersymmetry, can introduce additional challenges to the interpretation of the SDC and its convex hull \cite{Calderon-Infante:2020dhm}.
\end{itemize}

\pagebreak
\noindent
\textbf{Other types of topology changes featured of T-duality}

\vspace{3pt}
\noindent
One of the crucial themes was the change of topology induced under T-duality. One may then wonder if topology change in string theory and its relation to the SDC can be investigated in a more general and comprehensive way.
\begin{itemize}
     \item \textit{Deformations and topology.} Deformations of known backgrounds not only provide examples of tractable theories although being less symmetric, they also enlarge the scalar field space of the theory. It is well-known that deformations can lead to changes in topology when varying the deformation parameters \cite{Giveon:1993fd}. It would be interesting to explore these ideas using generalised T-duality as a tool to explore these properties but also the related framework of integrable deformations known as Yang-Baxter \cite{Klimcik:2002zj} and $\lambda$-deformations \cite{Sfetsos:2013wia}.  
     
     \item \textit{Disconnected moduli space.} T-duality and mirror symmetries have been known for a long time to sometime identify two \textit{disconnected} components of moduli space. This phenomenon is e.g. known to happen for example for compactifications on orbifolds with discrete torsion. The components correspondingly often describe backgrounds with different topologies \cite{Greene:1990ud,Giveon:1994fu}. This is particularly puzzling from a point of view of SDC where the moduli space is expected to be connected.

\end{itemize}

\vspace{10pt}
\noindent
\textbf{Towards a careful treatment of curved internal manifolds within the SDC}

\vspace{3pt}
\noindent
The results presented here enabled us to also draw a clear path towards a number issues one needs to face in the future when considering the SDC and curved internal manifolds.
\begin{itemize}
    \item\textit{Backreaction and curvature corrections.} 
    Considering curved backgrounds strained at the limit of their moduli space, one inevitable encounters highly curved manifolds. A thorough understanding of the implications to the SDC would require to study the corresponding backreaction of the geometry and curvature corrections affecting the moduli stabilising scalar potential. Although extremely challenging, a possible way  towards harnessing these correction would be to use the technology in e.g. \cite{Becker:2002nn,Conlon:2005ki}.

    \item\textit{Other types of towers.} One of the simplifying working assumption taken here, was to consider momentum and winding modes as the only possible stable states giving rise to tower of light state at infinite distances. This is however too simplistic, as e.g. pointed out in \cite{Grimm:2018ohb}, and the appearance of a massless tower might be in fact difficult to trace back to its originating stable state. It would be interesting to investigate the influence of including other types of towers to the examples provided here, and to verify if they would still showcase scalar potentials with a similar behaviour as observed in the above sections.

    \item\textit{Truly non-geometric spaces?}
    Finally, the present results remained constrained to non-geometric space which are dual to conventional manifolds. Ideally, one would like to address truly non-geometric space. However in absence of an approach that enables us to access the target space geometry, rather than one based on asymmetric orbifolds, the path towards putting truly non-geometric spaces under the scrutiny of the SDC remains unclear. Note however that recently in \cite{Gkountoumis:2023fym} an alternative strategy was adopted to explore the role of asymmetric orbifold in quantum gravity.
\end{itemize}

%%%%%%%%%%%%%%%%%%%%%%%%%%%%%%%%%%%%%%%%%%%%%%%%%%%%%
%%%%%%%%%%%%%%%%%%%%%%%%%%%%%%%%%%%%%%%%%%%%%%%%%%%%%
\subsubsection*{Acknowledgements}
We thank  Anayeli M. Ramirez, Andriana Makridou, David Andriot, Christian Northe, Eran Palti, Falk Hassler, Ivano Basile, Jacob Leedom, Laurent Gallot, Yixuan Li for very interesting comments and discussions. We would also like to thank the organisers and participants of the ``First DIP collaboration meeting'' and the ``Integrability, Duality and Deformation workshop 2023'' for many useful discussions  as well as the Banff virtual workshop ``Geometry and Swampland'', which sparked our interest for looking into the Swampland program from a T-duality point of view.

S.D. is grateful to the Azrieli foundation for the award of an Azrieli fellowship. 
The work of S.D. is also partially supported by the Israel Science Foundation (grant No. 1417/21),  by Carole and Marcus Weinstein through the BGU Presidential Faculty Recruitment Fund and the ISF Center of Excellence for theoretical high energy physics. 
The work of D.L. and S.D. is supported by the Origins Excellence Cluster and by the German-Israel-Project (DIP) on Holography and the Swampland.

T.R. and S.D. would like to, respectively, thank the physics departement of Ben Gurion University and the Max-Planck Institute for Physics for hospitality at different stages of this project. SD would like to also thank the theory division of CERN for hospitality during the finalisation of this project.

\newpage
\appendix

%%%%%%%%%%%%%%%%%%%%%%%%%%%%%%%%%%%%%%%%%%%%%%%%%%%%%
%%%%%%%%%%%%%%%%%%%%%%%%%%%%%%%%%%%%%%%%%%%%%%%%%%%%%
%%%%%%%%%%%%%%%%%%%%%%%%%%%%%%%%%%%%%%%%%%%%%%%%%%%%%
\section{Derivation reduction formula}
\label{App:reduction_formula}
In this appendix we derive a general reduction formula for a $D$-dimensional total space with metric $G_{IJ}$ in block diagonal from, which in the special case of only one scalar field reduces to \eqref{equ:reduced_NSNS_2}.  In particular we assume $G$ is of the form
\begin{align}
    G_{IJ}\mathrm{d}X^I\mathrm{d}X^J=g_{\mu\nu}(x,\varphi_a(x))\mathrm{d}x^\mu\mathrm{d}x^\nu + h_{ij}(y,x,\varphi_a(x))\mathrm{d}y^i\mathrm{d}y^j\,,
\end{align}
with $\mu=0,\dots ,d-1$, $i=1,\dots ,n$ and $D=d+n$. Hence we allow for the internal metric $h$ to depend on the external coordinates explicitly as well as implicitly through some scalar fields $\phi_a(x)$. Decomposing the Ricci scalar $\mathcal{R}(G)$ and re-summing properly one can show it splits as
\begin{align}\label{eq:splittingRicciapp}
    \mathcal{R}(G)&=\mathcal{R}(g)+\mathcal{R}(h) +\frac{3}{4}\mathrm{tr}(k_\mu k_\nu)g^{\mu \nu} - \frac{1}{4}\mathrm{tr}(k_\mu)\mathrm{tr}(k_\nu)g^{\mu \nu}- \mathrm{tr}(h^{-1}\square h)\,,
\end{align}
with 
\begin{align}
    k_\mu=h^{-1}\partial_\mu h\,.
\end{align}
We want to use the relation in eq. \eqref{eq:splittingRicciapp} to derive the reduced action starting from the 10D expression 
\begin{align}
S = \frac{1}{2 \kappa^2_0} \int \mathrm{d}^D X \sqrt{-G} e^{-2 \Phi} \left( \mathcal{R}(G) - \frac{1}{12}\mathpzc{H}_{IJK}\mathpzc{H}^{IJK} + 4 \partial_I \Phi \partial^I \Phi \right)\,,
\end{align}
with metric 
\begin{align}
ds^2&=G_{IJ}\mathrm{d}X^I\mathrm{d}X^J=e^{\frac{4\Phi_x}{d-2}}\mathcal{F}^{\frac{-2\alpha}{d-2}}(\varphi_a)g_{\mu \nu}(x) \mathrm{d}x^\mu \mathrm{d}x^\nu + h_{ij}(y,x,\varphi_a) \mathrm{d} y^i \mathrm{d} y^j\,,
\end{align}
and $\mathcal{F}(\varphi_a)$ a yet undetermined function of the moduli. We express the determinant $\det(h)$ as
\begin{align}
   \sqrt{ \det(h)} = \mathcal{D}(\varphi_a) \Omega(y)\,,
\end{align}
where we absorbed all moduli dependence in the function $ \mathcal{D}(\varphi_a)$. We later us, we will denote the external metric with all its factors by
\begin{align}
    \hat g_{\mu\nu}(x)\equiv e^{\frac{4\Phi_x}{d-2}}\mathcal{F}^{\frac{-2\alpha}{d-2}}(\varphi_a)g_{\mu \nu}(x)\,.
\end{align}
We can thus, using the above decomposition of the action, write the total action as (hats denoting contraction with $\hat{g}$)
\begin{multline}
S= \frac{1}{2 \kappa^2} \int \mathrm{d}^d x \mathrm{d}^n y \sqrt{-g}\, \Omega\, e^{\frac{-4 \Phi_x}{d-2}} e^{-2 \Phi_y}  \mathcal{D} \mathcal{F}^{\frac{-d \alpha}{d-2}}\Bigl( \mathcal{R}(\hat{g}) + \mathcal{R}(h)- \mathrm{tr}(h^{-1}\hat{\square}h) \\
 +\frac{3}{4}\mathrm{tr}(k_\mu k_\nu)\hat{g}^{\mu \nu} - \frac{1}{4}\mathrm{tr}(k_\mu)\mathrm{tr}(k_\nu)\hat{g}^{\mu \nu}-\frac{1}{12} \mathpzc{H}_{\mu\nu\lambda} \mathpzc{H}^{\hat{\mu}\hat{\nu}\hat{\lambda}} -\frac{1}{12} \mathpzc{H}_{ijk} \mathpzc{H}^{ijk} -\frac{1}{4} \mathpzc{H}_{\mu jk} \mathpzc{H}^{\hat{\mu} jk}\\
+ 4 \partial_\mu \Phi \partial^{\hat{\mu}} \Phi + 4 \partial_i \Phi_y \partial^i \Phi_y  \Bigr)\,.
\end{multline}
Now using the transformation rule of the Ricci scalar under a Weyl rescaling we can express $\mathcal{R}(\hat{g})$ as
\begin{align}\label{eq:Sigma}
    \mathcal{R}(\hat{g}) &= e^{\frac{-4 \Phi_x}{d-2}} \mathcal{F}^{\frac{-2 \alpha}{d-2}}\Bigl( \mathcal{R}(g) -2(d-1)e^{\frac{-2\Phi_x}{d-2}}\mathcal{F}^{\frac{\alpha}{d-2}}\square(e^{\frac{2\Phi_x}{d-2}}\mathcal{F}^{\frac{-\alpha}{d-2}})\nonumber\\
    &\qquad \qquad \qquad -(d-1)(d-4)e^{\frac{-4\Phi_x}{d-2}}\mathcal{F}^{\frac{2\alpha}{d-2}}(\partial(e^{\frac{2\Phi_x}{d-2}}\mathcal{F}^{\frac{-\alpha}{d-2}}))^2\Bigr)\nonumber\\
    &\equiv e^{\frac{-4 \Phi_x}{d-2}} \mathcal{F}^{\frac{-2 \alpha}{d-2}}\Bigl( \mathcal{R}(g) - \Sigma(\Phi_x,\mathcal{F})\Bigr)\,.
\end{align}
We arrive at 
\begin{multline}\label{eq:intermediateaction}
S= \frac{1}{2 \kappa^2} \int \mathrm{d}^d x \mathrm{d}^n y \sqrt{-g}\, \Omega\,   e^{-2 \Phi_y} \mathcal{D} \mathcal{F}^{-\alpha}\Bigl( \mathcal{R}(g) - \frac{1}{12}\mathpzc{H}_{\mu\nu\lambda} \mathpzc{H}^{\mu\nu\lambda}e^{\frac{-8 \Phi_x}{d-2}}\mathcal{F}^{\frac{4\alpha}{d-2}}+4\partial_\mu \Phi_x \partial^\mu \Phi_x\\
    + \left(\mathcal{R}(h)-\frac{1}{12} \mathpzc{H}_{ijk} \mathpzc{H}^{ijk} + 4 \partial_i \Phi_y \partial^i \Phi_y  \right) e^{\frac{4\Phi_x}{d-2}}\mathcal{F}^{\frac{-2\alpha}{d-2}}\\
    - \mathrm{tr}(h^{-1}\hat{\square}h)e^{\frac{4\Phi_x}{d-2}}\mathcal{F}^{\frac{-2\alpha}{d-2}}+\frac{3}{4}\mathrm{tr}(k_\mu k_\nu)g^{\mu \nu} - \frac{1}{4}\mathrm{tr}(k_\mu)\mathrm{tr}(k_\nu)g^{\mu \nu}\\
     -\frac{1}{4} \mathpzc{H}_{\mu jk} \mathpzc{H}^{\mu jk}+ 8 \partial_\mu \Phi_x \partial^\mu \Phi_y+ 4 \partial_\mu \Phi_y \partial^\mu \Phi_y - \Sigma(\Phi_x,\mathcal{F})\Bigr)\,.
\end{multline}
The purpose of introducing $\mathcal{F}$ becomes now evident, as it allows us to easily put the action in the Einstein frame
\begin{align}
    e^{-2 \Phi_y} \mathcal{D} \mathcal{F}^{-\alpha} = F(y)\,,
\end{align}
hence the factor in front of $\mathcal{R}(g)$ can be at most a function of the internal coordinates $y$ but neither $\varphi_i(x)$ nor $x^\mu$ explicitly. 

Two terms in the action \eqref{eq:intermediateaction} can be further simplified by imposing this condition as well as partially integrating\footnote{This relation is only true if the total derivative terms are identically zero. This vanishing has to be checked for each case at hand, and one can indeed verify that this is the case for the examples considered in the main text.}
\begin{multline}
    \int \mathrm{d}^d x \mathrm{d}^n y \sqrt{-g}\,\Omega\, F \left( - \mathrm{tr}(h^{-1}\hat{\square}h)e^{\frac{4\Phi_x}{d-2}}\mathcal{F}^{\frac{-2\alpha}{d-2}}\right) =
    \int \mathrm{d}^d x \mathrm{d}^n y \sqrt{-g}\,\Omega\, F \Bigl( - \mathrm{tr}(k_\mu k_\nu)g^{\mu\nu}\\-2\mathrm{tr}(k_\mu)\partial^\mu \Phi_x + \alpha \mathcal{F}^{-1}\mathrm{tr}(k_\mu)\partial^\mu \mathcal{F}\Bigr)
\end{multline}
and 
\begin{multline}
    \int \mathrm{d}^d x \mathrm{d}^n y \sqrt{-g}\,\Omega\, F \left(\Sigma(\Phi_x,\mathcal{F})\right) =
    \int \mathrm{d}^d x \mathrm{d}^n y \sqrt{-g}\,\Omega\, F \Bigl(\alpha^2 \frac{d-1}{d-2}\mathcal{F}^{-2}\partial_\mu \mathcal{F} \partial^\mu \mathcal{F}\\ - 4 \alpha \frac{d-1}{d-2} \mathcal{F}^{-1}\partial_\mu \mathcal{F} \partial^\mu \Phi_x + 4  \frac{d-1}{d-2} \partial_\mu \Phi_x \partial^\mu \Phi_x \Bigr)\,,
\end{multline}
one arrives at 
\begin{multline}
S= \frac{1}{2 \kappa^2} \int \mathrm{d}^d x \mathrm{d}^n y \sqrt{-g}\,\Omega\, F \Bigl\{ \mathcal{R}(g) - \frac{1}{12}\mathpzc{H}_{\mu\nu\lambda} \mathpzc{H}^{\mu\nu\lambda}e^{\frac{-8 \Phi_x}{d-2}}\mathcal{F}^{\frac{4\alpha}{d-2}}-\frac{4}{d-2}\partial_\mu \Phi_x \partial^\mu \Phi_x\\
    + \left(\mathcal{R}(h)-\frac{1}{12} \mathpzc{H}_{ijk} \mathpzc{H}^{ijk} + 4 \partial_i \Phi_y \partial^i \Phi_y  \right) e^{\frac{4\Phi_x}{d-2}}\mathcal{F}^{\frac{-2\alpha}{d-2}}\\
  -\frac{1}{4}\mathrm{tr}(k_\mu k_\nu)g^{\mu \nu} - \frac{1}{4}\mathrm{tr}(k_\mu)\mathrm{tr}(k_\nu)g^{\mu \nu} -2\mathrm{tr}(k_\mu)\partial^\mu \Phi_x + \alpha \mathcal{F}^{-1}\mathrm{tr}(k_\mu)\partial^\mu \mathcal{F}\\
  -\alpha^2 \frac{d-1}{d-2}\mathcal{F}^{-2}\partial_\mu \mathcal{F} \partial^\mu \mathcal{F} + 4 \alpha \frac{d-1}{d-2} \mathcal{F}^{-1}\partial_\mu \mathcal{F} \partial^\mu \Phi_x -\frac{1}{4} \mathpzc{H}_{\mu jk} \mathpzc{H}^{\mu jk}+ 8 \partial_\mu \Phi_x \partial^\mu \Phi_y+ 4 \partial_\mu \Phi_y \partial^\mu \Phi_y\Bigr\}\,.
\end{multline}
This can be further simplified using the identity
\begin{align}
    \partial(\det(A))=\det(A) \mathrm{tr}(A^{-1}\partial A)
\end{align}
as well as $\det(h)=\Omega^2 (e^{2 \Phi_y}F \mathcal{F}^\alpha)^2$ in order to obtain the following simplifying expressions
\begin{align}
    \mathrm{tr}(k_\mu) = 4 \partial_\mu \Phi_y + 2\alpha \mathcal{F}^{-1} \partial_\mu \mathcal{F}\,.
\end{align}
This in turn gives
\begin{multline}\label{equ:reduced_NSNS_multivar}
S= \frac{1}{2 \kappa^2} \int \mathrm{d}^d x \mathrm{d}^n y \sqrt{-g}\,\Omega F \Bigl\{ \mathcal{R}(g) - \frac{1}{12}\mathpzc{H}_{\mu\nu\lambda} \mathpzc{H}^{\mu\nu\lambda}e^{\frac{-8 \Phi_x}{d-2}}\mathcal{F}^{\frac{4\alpha}{d-2}}-\frac{4}{d-2}\partial_\mu \Phi_x \partial^\mu \Phi_x\\
    + \left(\mathcal{R}(h)-\frac{1}{12} \mathpzc{H}_{ijk} \mathpzc{H}^{ijk} + 4 \partial_i \Phi_y \partial^i \Phi_y  \right) e^{\frac{4\Phi_x}{d-2}}\mathcal{F}^{\frac{-2\alpha}{d-2}}\\
   -\frac{1}{4}\mathrm{tr}(k_\mu k_\nu)g^{\mu \nu}
  - \frac{\alpha^2}{d-2}\mathcal{F}^{-2}\partial_\mu \mathcal{F} \partial^\mu \mathcal{F} + \frac{ 4 \alpha}{d-2} \mathcal{F}^{-1}\partial_\mu \mathcal{F} \partial^\mu \Phi_x -\frac{1}{4} \mathpzc{H}_{\mu jk} \mathpzc{H}^{\mu jk}\Bigr\}\,.
\end{multline}
Finally we work out the explicit moduli dependence by ``splitting'' the derivative \linebreak ${\partial_\mu(\cdot) = \partial_\mu ( \cdot) \vert_{\varphi =\mathrm{const.}} +  (\partial_\mu \varphi_a)\partial_a(\cdot)}$ where $\partial_a \equiv \frac{\partial}{\partial \varphi_a}$. This gives, with now all partials $\partial_\mu$ understood as  $\partial_\mu ( \cdot) \vert_{\varphi =\mathrm{const.}}$, the result\footnote{We assume $\Phi_x=\Phi_x(x)$ but not a function of the scalar fields $\varphi_a(x)$. This is just for convenience to keep expressions more compact. It can be restored easily be replacing $\partial_\mu \Phi_x \to \partial_\mu \Phi_x \vert_{\varphi=\mathrm{const.}} + (\partial_a \Phi_x) \partial_\mu \varphi_a $.}
\begin{multline}
S  = \frac{\widehat{\mathcal{V}}_{int}}{2 \kappa^2} \int \mathrm{d}^d x \sqrt{-g}\,  \Bigl\{ \mathcal{R}(g) - \frac{1}{12}\mathpzc{H}_{\mu\nu\lambda} \mathpzc{H}^{\mu\nu\lambda}e^{\frac{-8 \Phi_x}{d-2}}\mathcal{F}^{\frac{4\alpha}{d-2}}-\frac{4}{d-2}\partial_\mu \Phi_x \partial^\mu \Phi_x\\
 + \kappa_a \partial_\mu \varphi_a \partial^\mu \Phi_x + \kappa_\mu \partial^\mu \Phi_x  -\gamma_{ab}\partial_\mu \varphi_a \partial^\mu \varphi_b  -\gamma_{a\mu} \partial^\mu \varphi_a  -V(\varphi_a,\Phi_x)\Bigr\}\,,
\end{multline}
where we defined
\begin{align}
    \gamma_{ab} &= \widehat{\mathcal{V}}_{int}^{-1} \int    \mathrm{d}^n y\,\Omega F \Bigl\{\frac{1}{4}\mathrm{tr}(k_a k_b) + \frac{\alpha^2}{d-2}\mathcal{F}^{-2} \partial_a \mathcal{F} \partial_b \mathcal{F}  +\frac{1}{4} \partial_a B_{jk}\partial_b B_{lm}h^{jl}h^{km}\Bigr\}\,,\nonumber \\
    \gamma_{a \mu} &=\widehat{\mathcal{V}}_{int}^{-1} \int    \mathrm{d}^n y\,\Omega F \Bigl\{ \frac{1}{2}\mathrm{tr}(k_a k_\mu) +\frac{2 \alpha^2}{d-2}\mathcal{F}^{-2} \partial_a \mathcal{F} \partial_\mu \mathcal{F}   +\frac{1}{2} \partial_\mu B_{jk}\partial_a B_{lm}h^{jl}h^{km}\Bigr\}\,,\nonumber \\
   \kappa_a &=\widehat{\mathcal{V}}_{int}^{-1} \int    \mathrm{d}^n y\,\Omega F \Bigl\{\frac{ 4 \alpha}{d-2} \mathcal{F}^{-1} \partial_a \mathcal{F}\Bigr\}\,,\nonumber \\
   \kappa_\mu &= \widehat{\mathcal{V}}_{int}^{-1} \int    \mathrm{d}^n y\,\Omega F \Bigl\{\frac{ 4 \alpha}{d-2} \mathcal{F}^{-1} \partial_\mu \mathcal{F} \Bigr\}\,,\nonumber\\
   \widehat{\mathcal{V}}_{int}&=\int    \mathrm{d}^n y\, \Omega(y)F(y)\,.
\end{align}
and the potential reads
\begin{align}
V(\varphi_a,\Phi_x)=-\widehat{\mathcal{V}}_{int}^{-1}\int    \mathrm{d}^n y\,\Omega F \Bigl\{ \left(\mathcal{R}(h)-\frac{1}{12} \mathpzc{H}_{ijk} \mathpzc{H}^{ijk} + 4 \partial_i \Phi_y \partial^i \Phi_y  \right) e^{\frac{4\Phi_x}{d-2}}\mathcal{F}^{\frac{-2\alpha}{d-2}}\Bigr\} + \mathcal{C}\,,
\end{align}
where 
\begin{align}
    \mathcal{C} &= \widehat{\mathcal{V}}_{int}^{-1} \int    \mathrm{d}^n y\,\Omega F \Bigl\{ \frac{1}{4}\mathrm{tr}(k_\mu k^\mu)   + \frac{\alpha^2}{d-2}\mathcal{F}^{-2} \partial_\mu \mathcal{F} \partial^\mu \mathcal{F}  +\frac{1}{4} \partial_\mu B_{jk}\partial^\mu B_{lm}h^{jl}h^{km}\Bigr\}\,.
\end{align}
Note that
\begin{align}
    \mathrm{tr}(k_a k_b)\partial_\mu \varphi_a \partial^\mu \varphi_b  = h^{ij} h^{kl}\frac{\partial h_{jk}}{\partial \varphi_a} \frac{\partial h_{li}}{\partial \varphi_b} \partial_\mu \varphi_a \partial^\mu \varphi_b 
\end{align}
is nothing else then the multivariable version of the 
DeWitt-like metric for scalar field variations \cite{DeWitt:1967yk,Gil-Medrano:1991ncm}. Furthermore the argument around equation \eqref{eq:Odd_expr_metric} straightforwardly generalises to the metric for multiple scalar fields, i.e to the $\gamma_{ab}$, showing that also in this case the metric is invariant under T-duality.

Lastly let us point out that due to the fact that we allowed for an explicit $x$ dependence of the internal metric $h$, Kalb-Ramond field  $B$ as well as $\mathcal{F}$ there can appear terms that are not quadratic in the scalar field derivative like  $\gamma_{a\mu} \partial^\mu \varphi_a$ and therefore seem to be challenging to interpret. We will leave this here as an observation but refer to \cite{Li:2023gtt} for one possible resolution of this issue. 
%%%%%%%%%%%%%%%%%%%%%%%%%%%%%%%%%%%%%%%%%%%%%%%%%%%%%
%%%%%%%%%%%           Appendix             %%%%%%%%%%
%%%%%%%%%%%%%%%%%%%%%%%%%%%%%%%%%%%%%%%%%%%%%%%%%%%%%
\section{A quick guide to generalised T-duality}
\label{App:gen_T-duality}
This appendix is meant as a minimalist's guide to the main definitions and properties of Poisson-Lie T-duality and has to be mostly considered as a reference guide. And in particular we seek to address two challenges generalised T-duality give rise to.  One of the of these conundrums is how non-Abelian T-duality, when performed via a Lagrangian multiplier method, results in backgrounds often lacking any isometry. Without the latter, the transformation back to the original background could not be performed. Leaving one enable to perform the inverse transformation, non-Abelian T-duality could not, at least using this method, be truly consider as a \textit{duality} transformation. We will discuss briefly in the following how Poisson-Lie T-duality alleviates that problem. The second point relates to identify of and the exchange of zero modes. This problem already arose for Abelian \linebreak{T-duality} on curved spaced performed using the Lagrange multiplier method. There, by virtue of the Lagrangian multiplier, the connection is flat, it can display non-trivial monodromies around non-trivial cycles of the geometry \cite{Alvarez:1993qi}. In the case of Abelian T-duality, this problem was addressed by turning the Lagrangian intro periodic coordinates. This then lead to the identification of the winding numbers. For generalised T-dualities a different identification was proposed using the algebraic structure of Poisson-Lie T-duality \cite{Klimcik:1996nq}, which will be succinctly reviewed in  section \ref{sec_app:PLnarain}.

%%%%%%%%%%%%%%%%%%%%%%%%%%%%%%%%%%%
%%%%%%%%%%%%%%%%%%%%%%%%%%%%%%%%%%%
\subsection{An algebraic approach using the Drinfel'd double}\label{sec_app:PL}
Non-Abelian T-duality generalises conventional T-duality to isometry groups of the background fields forming a non-Abelian group. Poisson-Lie T-duality in turn not only starts with a non-Abelian Lie group manifold, it also weakens the requirement that these vector generate an isometry of the background fields. In particular, denoting $E_{ij}=G_{ij}+B_{ij}$, where $G_{ij}$ and $B_{ij}$ are respectively the metric and Kalb-Ramond field, on no long requires that $ L_{v_a}E_{ij}=0$ but instead the weaker condition \cite{Klimcik:1995ux,Klimcik:1996nq}
\begin{align}\label{eq:PL_symmetry}
     L_{v_a}E_{ij}=\tilde f^{bc}{}_a v_b{}^lE_{il}E_{kj}v_c{}^j\,,
\end{align}
where the letters $a,b,c,\dots$ denote algebra indices while the letters $i,j,k,\dots$ are `curved' indices and $\tilde f^{bc}{}_a$ are some constants, which will play a crucial role.
When this condition is satisfied for a set of vectors $v_a$ generating a Lie algebra $\mathfrak g$ with Lie group $G$, the background fields are called Poisson-Lie T-dualisable. How the corresponding T-dual background is obtained rather distinct from Abelian and non-Abelian T-duality. In fact, the consistency relation $[L_{v_a},L_{v_b}]=L_{[v_a,v_b]}$ implies that the constants $\tilde f^{bc}{}_a$ are structure constants of an algebra turning $\mathfrak g$ into a bialgebra. A bilagebra is a simply a Lie algebra $\mathfrak g$ for which there exists a second Lie algebra $\tilde{\mathfrak{g}}$ which verifies a mixed-Jacobi identity. Dually, the Lie algebra  $\tilde{\mathfrak{g}}$ is also the bialgebra for  $\mathfrak{g}$. The direct sum of the Lie algebra forming a bialgebra pair $\mathfrak d=\mathfrak g\oplus \tilde{\mathfrak{g}}$ is called the Drinfel'd double. As was shown \cite{Klimcik:1995ux,Klimcik:1996nq}, the Lie group $\widetilde G$ of the Lie algebra $\mathfrak g$ naturally encodes the Poisson-Lie T-dual background. Details on how to obtain the background fields can be found in the reviews \cite{Klimcik:2021bjy,Demulder:2019bha,Thompson:2019ipl,Quevedo:1997jb}.

\begin{table}[t]
\centering
  \begin{tabular}{c|c|c|c}
   T-duality & Vector fields & Symmetry cond. & Conservation eq. \\ \hline\hline
   Abelian & $[v_a,v_b]=0$ & $ L_{v_a}E_{ij}=0$& $\mathrm d\star J_a=0$  \\
  non-Abelian  &$[v_a,v_b]=f_{ab}{}^c v_c$ & $ L_{v_a}E_{ij}=0$ & $\mathrm d\star J_a=0$\\
  Poison-Lie  &  $[v_a,v_b]=f_{ab}{}^c v_c$& \eqref{eq:PL_symmetry} & $\mathrm d\star J_a=\tilde f^{bc}{}_aJ_b\wedge J_c$
  \end{tabular}
  \caption{Summary of T-duality conditions and its non-Abelian generalisations. See text for more details.}\label{table:generalising_Tduality}
\end{table}

In Poisson-Lie T-duality, the duality has been lifted to an algebraic problem of understanding the Drinfel'd double $\mathbb D=\exp(\mathfrak d)$ and how it decomposes into a sum of bialgebra pairs
\begin{align}\label{eq:dinrfelddecompt}
    \mathbb D=G\times \widetilde G\,.
\end{align}
Poisson-Lie symmetric sigma-models for a group $G$ take on the form
\begin{align}
    S=\int_\Sigma L^a(E_0^{-1}+\Pi(g))^{-1}_{ab}L^b\,,
\end{align}
where $\Sigma$ is the worlsheet, $L=g^{-1}\mathrm d g$ are the left-invariant Maurer-Cartan forms in a basis of generators $\{T_a\}$ of the Lie algebra $\mathfrak g$ of the Lie group $G$ for a group element $g\in G$ with coordinates $\{x^i\}$, $E_0$ is a constant matrix and $\Pi$ is the Poisson-Lie structure for the Lie group $G$ in terms of a group elemeng $g\in G$ obtained from the $r$-matrix solving its modified classical Yang-Baxter equation. The T-dual metric and $B$-field are then  readily obtained by writing down the Poisson-Lie T-dual sigma model
\begin{align}\label{eq:PLdual}
    S_\mathrm{PL-dual}=\int_{\Sigma}\tilde L_a[(E_0^{-1}+\tilde \Pi(\tilde g))^{-1}]^{ab}\tilde L_b\,.
\end{align}
This time the Maurer-Cartan form and Poisson-Lie structure are those of the algebra paired up to $G$ in its Drinfel'd double, i.e. $\tilde G$ and its corresponding algebra in eq. \eqref{eq:dinrfelddecompt}.

As is similarly the case for conventional flat space T-duality, the dilaton transformation rules under duality requires separate consideration and were derived in \cite{Demulder:2018lmj} (see also \cite{Jurco:2017gii} for a different approach). In order to have a PL-symmetric background the dilaton needs to be given by
\begin{align}
    \Phi = \phi_0 + \frac{1}{4}\log\left(\det(G_{ab}) \right)\,,
\end{align}
where again the symbol $G_{ab}$ (with indices) denotes the metric here in the basis of the Maurer-Cartan forms $L_a$.
The dilaton on the PL T-dual background can in turn be computed using the formula
\begin{align}
    \tilde \Phi= \Phi + \frac{1}{4} \log\left(\frac{\det(\tilde G_{ab})}{\det(G_{ab})}\right)\,.
\end{align}
In the limiting case of Abelian T-duality, the double is simple
\begin{align}\label{eq:DD_AbelianT}
     \mathbb D=U(1)^{2n}=U(1)^n\times\widetilde{ U(1)^n}\,,
\end{align}
where we have kept the tilde notation to distinguish one tori from its T-dual.
One of Poisson-Lie T-dualities first victory was to explain how non-Abelian T-duality is still an invertible symmetry despite what might infer from the Lagrangian procedure. How this is consistent with the problem posed in the special case of non-Abelian T-duality becomes apparent when considering the relevant Drinfel'd double, which can be shown to be
\begin{align}\label{eq:DD_NAbelianT}
    \mathbb D=G\times U(1)^{\mathrm{dim}\,G}\,.
\end{align}
We see that as expected the dual background $\widetilde G=U(1)^{\mathrm{dim}\,G}$ feature no (non-Abelian) symmetry. Within Poisson-Lie T-duality this no longer forms a problem to invert transformation as the relevant Drinfel'd double together with additional data pin-pointing the particular background under scrutiny is sufficient to go from one background to its dual. One can in addition readily check that the Drinfel'd double for Abelian T-duality in eq. \eqref{eq:DD_AbelianT} and for non-Abelian T-duality eq. \eqref{eq:DD_NAbelianT}, reduces the Poisson-Lie symmetry condition to the familiar isometry conditions in table \ref{table:generalising_Tduality}.
On both case one can also see that the dual Poisson-Lie action in eq. \eqref{eq:PLdual} reduces to the known Abelian or non-Abelian T-dual background.

%%%%%%%%%%%%%%%%%%%%%%%%%%%%%%%%%%%
%%%%%%%%%%%%%%%%%%%%%%%%%%%%%%%%%%%
\subsection{Generalised Narain lattice}\label{sec_app:PLnarain}
In this section, we aim to review the motivation and reasoning behind the generalisation of the winding/momentum-exchange put forward in \cite{Klimcik:1996nq}. As transpires from above, backgrounds admitting a Poisson-Lie T-duality need not feature isometries and as a result the conservation laws one would have for an Abelian T-dualisable model are generically absent. The symmetry condition \eqref{eq:PL_symmetry} inevitable comes with a constraint. The insight of \cite{Klimcik:1996nq} to understand this apparent drawback as having to deal with constrained system as in fact providing information about the models at the quantum level, ultimately providing the quantisation condition necessary to identify the winding and momentum quantum modes of Poisson-Lie symmetric models. In this appendix we will sketch the reason in order to motivate and introduce the formula used to determine the momentum and winding exchange used in section \ref{sec:NATD_S3}.

Defining the first currents in terms of the background fields $E_{ij}$ by the expression
\begin{align}\label{eq:currentPL}
  J=e_a{}^jE_{ij}\partial_+X^i\mathrm d \sigma^++e_a{}^jE_{ij}\partial_+X^i\mathrm d \sigma^-\,,   
\end{align}
where we have used the vielbeins $g^{-1}\mathrm d g=T^ae^a{}_j\mathrm dx^j$ for a group element $g\in G$, these currents $J$ are \textit{not conserved} but are instead $\tilde G$-flat\footnote{Note that it is flat with respect to the Poisson-Lie T-dual target space $\tilde G$.}.
In particular this implies that the charges in Poisson-Lie models are no longer numbers but group valued, here in $\widetilde G$, which in turn are not conserved but flat.
This flatness relation can be shown to imply that the current has to verify the following unit-monodromy condition
\begin{align}\label{eq:unit_monodromy}
     \mathcal P\exp \int_\gamma  J_\sigma(\tilde g)=\tilde e\in \tilde G\,.
\end{align}
In fact a very similar relation also pops-up when quantising the free-boson sigma-model in the form of its quantisation condition
\begin{align}\label{eq:unit_monodromy_constr}
   \mathcal  P\exp\oint_\gamma  J=\tilde e\in \widetilde {U(1)}\,,
\end{align}
where the integral evaluates to the unit element of the T-dual $\widetilde {U(1)}$ isometry group.  Extrapolating to Poisson-Lie manifolds, instead of tori, led the authors in \cite{Klimcik:1996nq} to propose that the relation \eqref{eq:unit_monodromy} effectively serves as quantisation condition on the Poisson-Lie currents $J$ and to propose the fundamental group of the Drinfel'd double as generalised Narain lattice
   $ \Lambda_\mathrm{PL}=\pi_1(\mathbb D)\,.$
Taking into account that for all cases considered here the decomposition is perfect, i.e. $\mathbb D=G\widetilde G$, the above equation can be shown to break down into 
\begin{align}\label{eq:PL_lattice}
  \Lambda_\mathrm{PL}=  \pi_1(\mathbb D)=\pi_1(\mathbb D/ G)\oplus \frac{\pi_1(G)}{\pi_2(\mathbb D/G)}\,.
\end{align}
Applying this relation to the Abelian case we retrieve the familiar Narain lattice for tori
    $\pi_1(U(1)^{2D})=\mathbb Z^{\oplus D}\oplus \mathbb Z^{\oplus D}$, we retrieve the conventional Narain lattice of the free boson.

%%%%%%%%%%%%%%%%%%%%%%%%%%%%%%%%%%%%%%%%%%%%%%%%%%%%%
%%%%%%%%%%%%%%%%%%%%%%%%%%%%%%%%%%%%%%%%%%%%%%%%%%%%%
%%%%%%%%%%%%%%%%%%%%%%%%%%%%%%%%%%%%%%%%%%%%%%%%%%%%%
\section{Dilute flux and cyclotron effect}
\label{App:dilute_flux}
The dilute flux approximation forms a valuable approximation scheme in the context backgrounds of the three-torus when supported by a three-form flux. The approximations assumes a large volume limit of the geometry essentially diluting the flux, i.e. $h/\mathrm{Vol}(T^3)\ll 1$, where here $h$ denotes schematically the strength of the $H$-flux but will be specified more precisely later. 
Indeed, the three-form flux spoils the conformality of the three-torus background, which can however can be approximately recover in the limit of very weak three-form flux.  The obvious advantage of the limit, at leading order in $h$, the background becomes an approximate fixed point of the renormalisation group equations and CFT techniques can be applied. For that reason, the corresponding theory was also called CFT$_{\mathpzc H}$. See \cite{Blumenhagen:2010hj,Blumenhagen:2011ph,Bakas:2013jwa}.

Consider the three-torus with $H$-flux as described in the usual  NSNS-frame. Recall that the metric and $B$-field in this setup are given by
\begin{align}
    G = \mathrm{diag}( R_1^2,R_2^2,R_3^2)\,, \qquad B=h X^3\mathrm d X_1\wedge \mathrm dX_2\,.
\end{align}
Then the equations of motion of the sigma model for $H$-flux three torus read
\begin{align}\label{eq:eomssdf}
    G_{\kappa \nu } \partial^2 X^\nu =h \epsilon_{\kappa \mu \nu} \partial_\sigma X^\mu \partial_\tau X^\nu\,.
\end{align}
In the limit of large radii one can solve the equations of motion perturbatively. In fact assuming the usual boundary conditions for the torus (allowing for winding) 
\begin{align}
    X^\mu(\tau,\sigma +2\pi) = X^\mu(\tau,\sigma) + 2 \pi N^\mu
\end{align}
as well as truncating to zero modes it is easy to see that a solution is given by\footnote{For more details, see \cite{Andriot12a}.}
\begin{align}
    X^\mu(\tau, \sigma)= x^\mu_0 + \frac{h}{R_\mu^2} x^\mu_H + p^\mu_0 \tau + \frac{h}{R_\mu^2} p^\mu_H \tau +  N^\mu \sigma - \sum_{\nu \rho}\frac{h}{2 R_\mu^2} \epsilon^\mu_{\nu \rho} p_0^\rho N^\nu \tau^2\,,
\end{align}
where repeated $\mu$ indices should not be summed over.
First we note that since $B$ is linear in the coordinates (in particular it does not involve $\dot{X}^\mu$) the canonical momentum is still given by 
\begin{align}
    \Pi_\mu = \frac{1}{2 \pi}(G_{\mu \nu} \dot{X}^\nu + B_{\mu \nu} X^{\prime \nu})\,.
\end{align}
We can obtain $\dot{X},X^\prime$ from the mode expansion of the string. For $X^\prime$ we get the standard expression of the free string  $X^{\prime \mu} = N^\mu$
while for $\dot{X^\mu}$ we obtain
\begin{align}\label{}
   \dot{X}^\mu = p_0^\mu + \frac{h }{R^2_\mu} p^\mu_H - \sum_{\nu \rho}\frac{h }{R^2_\mu} \epsilon^\mu_{\nu \rho} p_0^\rho N^\nu \tau\,,
\end{align}
where again we do not sum over repeated $\mu$'s.
This is a rather odd expression since it will give a center of mass momentum that depends on the worldsheet parameter $\tau$ and therefore is not conserved anymore.
Collecting the momentum and winding numbers into vectors $\Vec{p}$ and $\Vec{N}$ this is nothing else than 
\begin{align}\label{eq:obstruction_conser_mom}
    \Vec{p}_0 \times \Vec{N} = 0 \Longleftrightarrow \Vec{p}_0 \parallel \Vec{N}\,,
\end{align}
hence momentum conservation is only restored if the winding and momentum vector are parallel! Since on the three torus there is winding, this implies that generically momentum is not conserved.

%%%%%%%%%%%%%%%%%%%%%%%%%%%%%%%%%%%%%%%%%%%
%%%%%%%%%%%%%%%%%%%%%%%%%%%%%%%%%%%%%%%%%%%
\section{On the NATD dual of the three-sphere}\label{app:commNATD}
In this appendix, we will touch on the possible relation of the dual geometry discussed above with that of  \cite{Lozano:2019ywa}. In the latter, the non-Abelian T-dual metric is obtained by a gauging procedure where Lagrangian multiplier are promoted to the T-dual coordinates. As such, even if we start for a compact Lie group, the resulting geometry will trade the compact directions for as many non-compact ones.
The non-compactness of the background is however at odds with the existence of an holographic CFT. Addressing this point and be realising that this NATD backgrounds belongs to a particular class of supergravity solutions known as Gaiotto-Maldacena backgrounds, \cite{Lozano16,Lozano:2019ywa} put forward a compact and well-behaved completion to the NATD background. In contrast, the metric considered here, is obtained using Poisson-Lie T-duality and seen as a compact at the cost of being non-geometric (see also \cite{Bugden:2019vlj}). This leaves ones with the obvious question on how the two backgrounds are related.

At fault of being able to provide a more direct relation between the two backgrounds, the holographic central charge offers a first order comparison.
 For a given compactification background one can compute the associated central charge following \cite{Macpherson:2014eza}. The central charge on the gravity side is essentially related to the volume of the internal manifold. Under non-Abelian T-duality the resulting dual background will have a different volume and thus central charge. The original and dual volumes differ however only by a multiplicative constant in the form of the Faddeev-Popov determinant that can be accordingly traced back to performing the transformation at the level of the path integral \cite{Quevedo:1997jb}. 

The starting point is the generalised T-dualisable background fields for the three-sphere. The dilaton compatible with the existence of a generalised T-dual takes on the form \cite{Demulder:2018lmj}
\begin{align}
    \Phi_y = \phi_0 + \frac{1}{4}\log\left(\det(h_{ab}) \right) =  \phi_0 + \frac{1}{2}\log\left(R^3 \right)\,.
\end{align}
While in the main text it was convenient to keep the dilaton constant, here we set it to zero by the appropriate choice of $\phi_0$ in order to have the same initial configuration as considered in  \cite{Lozano:2019ywa}. 
In this case the Einstein frame after reduction corresponds to setting the exponent to $\alpha=3$ in \eqref{eq:metric_Ansatz}. The dilaton for the generalised T-dual background is then again computed using the transformation rule given in appendix \ref{App:gen_T-duality} and after going to the $\beta$-frame reads
\begin{align}
\tilde{\Phi}_y =  - \frac{1}{2}\log(R^{-6})\,.
\end{align}
Having identified the dilaton in the $\beta$-frame of the NATD background, we can now see the implications for the reduced theory. Where, upon setting $\alpha=3$, the metric on moduli space now reads
\begin{align}
    \gamma_{RR} = \left( \frac{9}{d-2}+3\right)R^{-2}(\partial R)^2 = 3 \frac{d-1}{d-2} R^{-2}(\partial R)^2 \,.
\end{align}
We see that the only two changes compared to the original background are a) the additional $d$-dependent factor in front of the kinetic term and b) the appearance of a mixed term contribution
\begin{align}
    \frac{4 \alpha}{d-2} R^{-1} (\partial \phi \partial R) =  \frac{12}{d-2} R^{-1} (\partial \phi \partial R)\,.
\end{align}
We can now attempt compare the central charges associated to the NATD $S^3$ considered in the main text with completion of the NATD $S^3$ of \cite{Lozano:2019ywa} within a $AdS_3 \times S^3 \times CY_2$ solution. Note that here, in contrast to  \cite{Lozano:2019ywa}, the dual background was not studied as a full-fledged ten-dimensional supergravity solution. To move forward, and keeping a deeper analysis of the solution for later work, we will assume that the NATD considered in the main text can be also embedded in the same ten-dimensional Ansatz. That is starting from the supergravity background\footnote{Note that we are using a different normalisation accounting for a missing factor of 4 in front of the $S^3$-line element. This difference can however be reabsorbed by taking a different normalisation of the $SU(2)$-generators.} 
\begin{align}\label{eq:10d_dual_original}
    \mathrm{d} s^2 = 4 L^2 \mathrm{d} s^2(AdS_3) + M^2 \mathrm{d}s^2(\mathrm{CY}_2) + L^2 \mathrm{d} s^2(S^3)\,,
\end{align}
with vanishing dilaton and Kalb-Ramond field and has to be supplemented by $F_3$ and $F_7$ flux to support the supergravity solution. The unit sphere for the line element $\mathrm{d} s^2(S^3)$ has volume $V_{S^3}=2\pi^2$. Performing the duality, the background metric and dilaton read
\begin{align}\label{eq:10d_dual}
    \widetilde{\mathrm{d} s^2} &= 4 R^2 \mathrm{d} s^2(AdS_3) + M^2 \mathrm{d}s^2(CY_2) +  \delta^2 R^{-2}\mathrm{d}\tilde{s}^2(\hat{T}^3)\,,\qquad  e^{2 \tilde{\phi}} = R^{-6}\,,
\end{align}
which again provide a solution to the supergravity equations when  supplemented with the dual expressions for the fluxes.
Note that that the line element has to be treated as a schematic expression: the third factor as well as the dilaton are given in the $\beta$-frame, while the $AdS_3 \times CY_2$ part is in the usual NSNS-frame. The addition scaling factor $\delta^2$ stresses the change of volume of the dualised part of the geometries and will enable us to track down the difference in central charges.

Using the formula for the holographic central charge \cite{Klebanov:2007ws} (setting $G_N = 8 \pi^6$ following the conventions of \cite{Lozano:2019ywa}), we see that in this particular case the  holographic central charge is becomes simply the square root of the volume of the internal space.  Hence the central charge for the original background given by the metric \eqref{eq:10d_dual_original} reads 
\begin{align}
   c &= \frac{16 \pi^2 R^4 M^4 \mathrm{Vol}_{\mathrm{CY}_{2}}}{G_N} = \frac{2 R^4 M^4 \mathrm{Vol}_{\mathrm{CY}_{2}}}{\pi^4}\,,
\end{align}
while the central charge for the generalised dual background \eqref{eq:10d_dual} is given by
\begin{align}
    \tilde{c} &= {\delta^3} \frac{8 \pi^3 R^4 M^4 \mathrm{Vol}_{\mathrm{CY}_{2}}}{G_N} ={\delta^3} \frac{R^4 M^4 \mathrm{Vol}_{\mathrm{CY}_{2}}}{\pi^3}\,,
\end{align}
hence the two expression match when setting $\delta^3=\frac{\pi}{2}$. This is not a surprise: under generalised T-duality the invariant measure changes according to $e^{-2\Phi}\sqrt{\det g}\,\Delta_\mathrm{F.P.}=e^{-2\tilde \Phi}\sqrt{\det\tilde g}$, we can identify $\tilde{c}=c \Delta_\mathrm{F.P.} =c \frac{\pi}{2}$. This is not a surprise since, the path integrals of non-Abelian T-dual theories are known to differ up to a proportionality constant which is the Faddeev-Popov determinant \cite{Quevedo:1997jb}. With the expression for the central charge of the generalised T-dual background of the three-sphere used in section \ref{sec:NATD_S3} one can try to make a first step towards comparing with the completion of the NATD of the three-sphere constructed in \cite{Lozano:2019ywa}.

%%%%%%%%%%%%%%%%%%%%%%%%%%%%%%%%%%%%%%%%%%%
%%%%%%%%%%%%%%%%%%%%%%%%%%%%%%%%%%%%%%%%%%%
\section{\texorpdfstring{$\beta$}{beta}-gravity and the total derivative}\label{App:totdal_deriv}
As outlined in section \ref{sec:beta-gravity}, the Lagrangians of the standard NSNS frame and the $\beta$-frame agree up to a total derivative. Using as a Lagrangian for the $\beta$-frame the one given by \eqref{eq:beta_Lagra_w_check_R}, hence the DFT Lagrangian of \cite{Andriot12, Andriot:2013xca} with section condition $\tilde{\partial}^i(\cdot)=0$, the total derivative reads
\begin{multline}\label{eq:app_total_deriv}
    \mathcal L_\mathrm{NSNS} = \mathcal L_\beta +  \partial_m\Bigl\{  e^{-2d}\Bigl( \tilde{h}^{mn}\tilde{h}^{pq}\partial_n \tilde{h}_{pq} - h^{mn}h^{pq}\partial_n h_{pq} + \partial_n(\tilde{h}^{mn}-h^{mn})\Bigr)\\
    + \Bigl(\frac{e^{-2d}}{|\tilde{h}|}\partial_n(\tilde{h}_{pq}\beta^{pm}\beta^{qn}|\tilde{h}|)-4 e^{-2d} \beta^{pm}\tilde{h}_{pq}(\mathpzc{Q}_{\ p}^{\ q p} - \frac{1}{2}\beta^{qm}\tilde{h}_{pq}\partial_m \tilde{h}^{pq}) \bigr)\Bigr\}\,,
\end{multline}
where the generalised dilaton  is defined as $e^{-2d}\equiv e^{-2 \tilde{\Phi}_y}\sqrt{|\tilde{h}|}$.
Let us demonstrate this explicitly for the example of the NATD of $S^3$ of section \ref{sec:NATD_S3}. Recall first that the metric $\tilde{h}$  in the $\beta$-frame is very simple, and in fact constant $\tilde{h}= R^{-2} \mathbb{1}_3$. Furthermore the generalised dilaton $d$ is also constant and one can check that $\mathpzc{Q}_{\ l}^{\ kl}=0$ such that the expression simplifies to
\begin{align}
    \mathcal L_\mathrm{NSNS} = \mathcal L_\beta + e^{-2d}\partial_m \Bigl(- \partial_n h^{mn} -h^{mn}h^{pq}\partial_n h_{pq}  + \tilde{h}_{pq}\beta^{qn}\partial_n\beta^{pm} \Bigr)\,.
\end{align}
Plugging in the explicit expressions for the metric in NSNS frame $h$ ($\equiv \hat{h}$ from section \ref{sec:NATD_S3}) as well as the $\beta$-frame quantities $\tilde{h},\beta$ this can be evaluated to give
\begin{align}
    \partial_k(\dots) =  4 R^2 \mathcal{X}^{-2} \left( 3R^4 +\phi^2 +\psi^2 +\theta^2\right)\,,
\end{align}
where $\mathcal{X}=R^4+\phi^2 +\psi^2 +\theta^2$.
We can evaluate $\mathcal{L}_\mathrm{NSNS}$ which is given by
\begin{align}
    \mathcal{L}_\mathrm{NSNS} = \mathcal{R}(h) - \frac{1}{12}\mathpzc{H}^2 + 4(\partial \Phi_y)^2 = \frac{3}{2R^2} + 4 R^2 \mathcal{X}^{-2} \left( 3R^4 +\phi^2 +\psi^2 +\theta^2\right)\,.
\end{align}
Lastly recall that we already calculated $\mathcal{L}_{\beta}$, which is nothing else than the expression given in eq. \eqref{eq:NATD_potential_beta} (stripped off the factor $e^{\frac{4\phi}{d-2}}$)
\begin{align}
    \mathcal{L}_{\beta} = \frac{3}{2R^2}\,.
\end{align}
This indeed verifies relation \eqref{eq:app_total_deriv}.

Let us close this section with another remark. When introducing the $\beta$-frame Lagrangian \eqref{eq:beta_Lagra_w_check_R} in section \ref{sec:beta-gravity} we were careful to use a $\beta$-frame Lagrangian  that only contains covariant expressions.
Up to total derivatives it is equivalent to other expressions given in the literature, which in turn are also equivalent to the standard NSNS Lagrangian, however with a different total derivative. In order to obtain matching potentials for the reduction on $S^3$ and its generalised T-dual $T^3$ (written in $\beta$-frame) is was however important to use this covariant expression, as otherwise we would have obtained a wrong sign for the potential after dualizing and going to the $\beta$-frame.

%%%%%%%%%%%%%%%%%%%%%%%%%%%%%%%%%%%%%%%%%%%
%%%%%%%%%%%%    Bibliography     %%%%%%%%%%	
%%%%%%%%%%%%%%%%%%%%%%%%%%%%%%%%%%%%%%%%%%%
\newpage
\bibliographystyle{JHEP}
\bibliography{bib_DC_T-duality.bib}

\end{document}